# Competition between mirror symmetry breaking and translation symmetry breaking in ferroelectric liquid crystals with increasing lateral substitution.

Grant J. Strachan,* Ewa Górecka, and Damian Pociecha.
Faculty of Chemistry, University of Warsaw, ul. Pasteura 1, 02-093 Warsaw, Poland
*Author for correspondence: g.strachan@chem.uw.edu.pl

## Abstract

The recently discovered heliconical ferroelectric nematic ($N_{TBF}$) phase is a unique example of spontaneous chiral symmetry breaking in a proper ferroelectric fluid. In this study, we investigate four homologous series of mesogenic compounds, differing in the degree of fluorination of the mesogenic core and bearing lateral alkoxy substituents of varying lengths, to understand how molecular architecture influences the formation and stability of the $N_{TBF}$ phase. Increasing the length of the lateral chain lowers the phase transition temperatures and suppresses smectic layer formation, enabling the emergence of the $N_{TBF}$ phase which replaces the orthogonal ferroelectric smectic A ($SmA_F$) phase. This indicates a competition between lamellar and heliconical polar ordering, driven by the interplay of strong molecular dipoles and the self-segregation of chemically incompatible molecular segments that typically favor layered structures. Notably, the $N_{TBF}$ phase in these compounds exhibits exceptionally short helical pitch lengths, on the order of a few hundred nanometers, as revealed by selective light reflection and atomic force microscopy (AFM). Furthermore, for one of the studied compounds AFM imaging of one compound revealed a regular array of screw dislocations within the $N_{TBF}$ phase, suggesting a possible link to more complex modulated or twist-grain-boundary-like structures.

## Introduction

The emergence of chiral structures from achiral building blocks is an intriguing phenomenon that has caught the interest of those working in both fundamental research and also the development of new technologies. A newly discovered liquid crystal phase, the heliconical ferroelectric nematic phase ($N_{TBF}$), is a prime example of this.[1] The $N_{TBF}$ phase is a member of the recently discovered, and rapidly growing, family of proper ferroelectric liquid crystals, which began with the 2017 discovery of the ferroelectric nematic ($N_F$) phase,[2–4] and now includes orthogonal $SmA_F$, tilted $SmC_F$, and heliconical $SmC_P^H SmC^P{}_H$ layered phases as well as phases with antiferroelectric order ($N_X$/$SmZ_A$/$M_{AF}$/$N_S$ and $SmA_{AF}$).[1,5,6] Sketches of these phases are given in Figure 1. In the $N_{TBF}$ phase, the director follows the heliconical trajectory, being tilted with respect to the helical axis, and the helical pitch is typically reported to be on the micron scale, comparable to the wavelength of visible light. As the breaking of mirror symmetry occurs spontaneously, both left- and right-handed helices are formed (Figure 1).

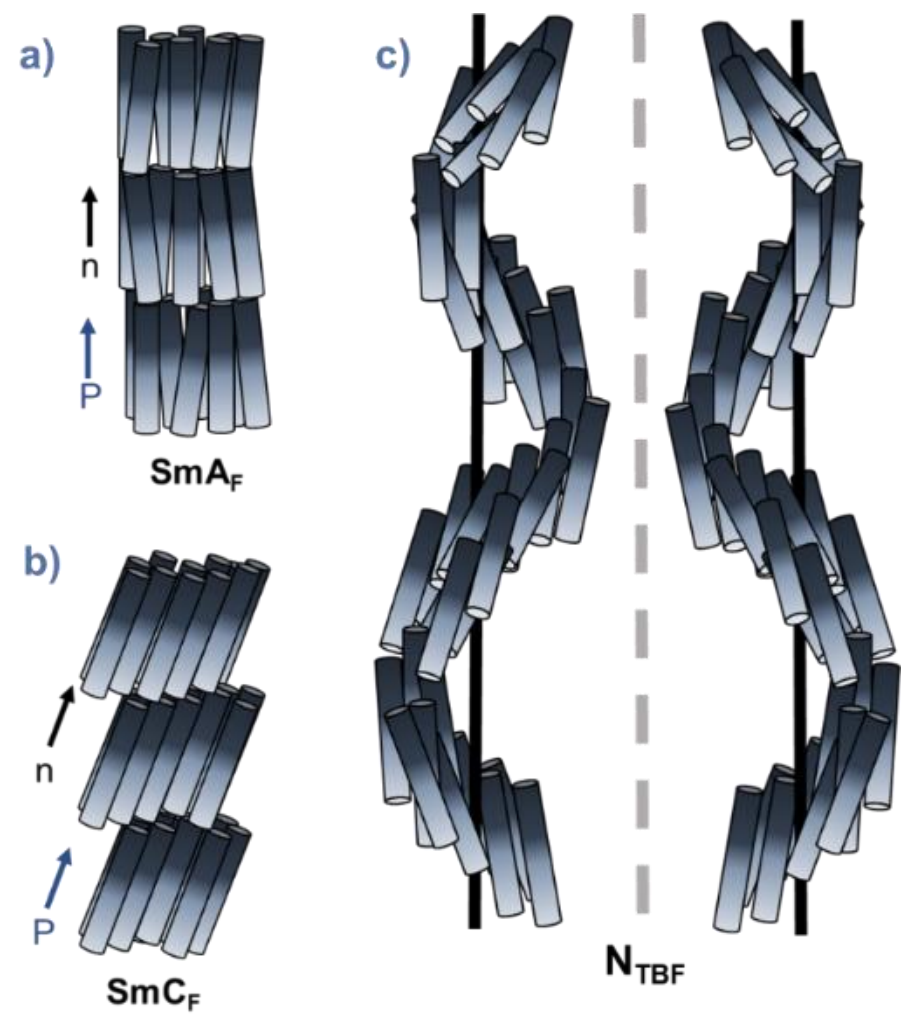

Figure 1. Sketches of the ferroelectric smectic A (SmA$_F$) phase (a); the ferroelectric smectic C (SmC$_F$) phase (b); and the heliconical ferroelectric nematic (N$_{TBF}$) phase (c) with helices of opposite handedness shown. Black arrows represent the orientation of the director (n) and blue arrows the polarization direction (P).

In these recently discovered longitudinally polar phases, ferroelectricity arises directly from the interactions of molecular dipoles. This differs from previously known ferroelectric LCs in which ferroelectricity occurs as a secondary effect, for example induced by molecular chirality or the close packing of bent-core molecules. These new proper ferroelectric phases, despite their fluid nature, exhibit very high spontaneous polarization reaching values comparable to solid, crystalline ferroelectrics. There has been a great deal of attention devoted to these new ferroelectric liquid crystal phases, focusing both on their fundamental properties as well as their potential applications. These systems offer unique insight into the interplay between electrostatic and elastic forces in soft materials. While the properties of paraelectric liquid crystal phases are often considered in terms of elastic energies and deformations, the proper ferroelectric mesophases bring the role of electrostatic interactions to the fore. This has led to several interesting discoveries, including the apparent 'tendency to twist' found for highly polar molecules within fluid phases. While chiral ground states have been reported in both the N$_F$[7] and SmA$_F$[8] phases, the clearest example of this is still the N$_{TBF}$ phase. The chiral symmetry breaking observed in the N$_{TBF}$ phase is believed to be driven by interactions between the strong electric dipoles of the molecules, which promote a non-colinear arrangement and lead to the formation of heliconical structure.[1] This has been compared to the Dzyaloshinskii-Moriya interaction observed in magnetic systems.[9]

However, the N$_{TBF}$ phase is still extremely rare, with only a few examples reported to date, and the precise molecular features required for its formation are not understood. To study this unusual new phase, and to develop its application potential, it is vital to understand both the influence of molecular architecture on the formation of the phase, and to expand the range of molecules showing this phase.

Our previous studies[10] showed that an uneven variation in the electron density along the long molecular axis would promote the formation of the polar smectic phases (SmA$_F$ and SmC$_F$) due to the self-segregation effect of chemically non-compatible units; while a more uniform distribution of electron density favors the N$_F$ phase. For a single material, which seems to be intermediate between these two extremes we observed the N$_{TBF}$ phase. This is somewhat similar to trends seen for two homologous series of polar liquid crystals, both with a terminal alkyl chain, reported by Karcz *et al.*,[1] and by Nishikawa *et al.*[11] In both cases, the N$_{TBF}$ phase (referred to as the $^{HC}$N$_F$ phase in ref. 11) was observed at an intermediate chain length, with shorter homologues only forming the ferroelectric nematic phase, while at longer chain lengths, the $N_{TBF}$ phase was replaced by smectic behavior.

Considering these observations, we hypothesized that mesogens with molecular features expected to suppress smectic ordering may promote the formation of the $N_{TBF}$ phase. To test this, we have selected four structures (Figure 2) forming either $SmA_F$ or $N_{TBF}$ phases and increased the length of their lateral alkoxy substituents. This structural change decreases the length-to-breadth ratio of the molecules and disrupts side-to-side interactions between neighboring molecules, and both effects are expected to decrease the tendency for layer formation. Each compound is given a code, **X-Y-Z**, where **X** represents the length of the lateral alkoxy chain: **M** (methoxy) $OCH_3$, **E** (ethoxy) $OC_2H_5$, **P** (propyloxy) $OC_3H_7$, **B** (butyloxy) $OC_4H_9$, **Q** (pentyloxy) $OC_5H_{11}$. **Y** and **Z** are the number of fluorine substituents on the second and third aromatic rings, respectively.

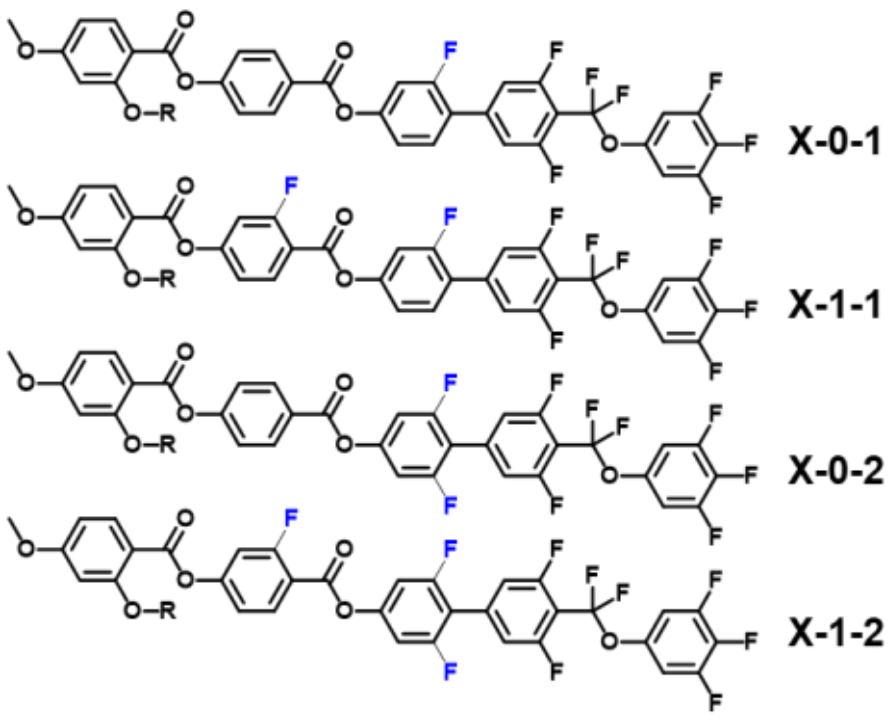

Figure 2. The general molecular structure of the 4 homologous series reported here.

## Methods

The synthetic methods and structural characterization of the reported compounds, and full experimental details are given in the accompanying SI.

*Assignment of liquid crystal phases:*

Preliminary phase assignments were carried out using polarized-light optical microscopy, and phases were identified based on the observation of characteristic optical textures (Figure 3). The nematic phase, N, formed uniform textures in thin cells treated for planar alignment, and in the $N_F$ phase, the uniform texture was accompanied by characteristic conical defects anchored at the cell spacers.[12] In the $N_{TBF}$ phase a striped texture was observed. The $SmA_F$ phase produced a mostly uniform texture, with mosaic-like regions, while the textures of the $SmC_F$ phase were heavily dependent on the preceding phase. Following the $SmA_F$ phase, small striped domains developed from the mosaic texture, while non-characteristic and strongly scattering textures appeared on cooling from the $N_{TBF}$ phase.

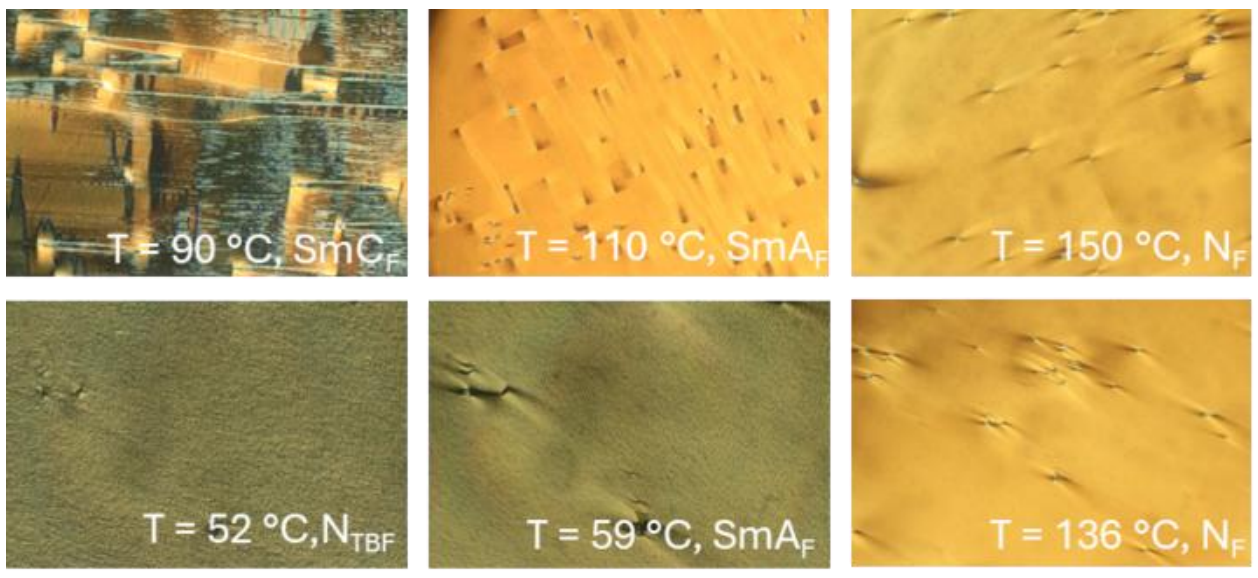

Figure 3. Representative textures for LC phases formed by **E-1-2** (top) and **P-1-2** (bottom) in cells treated for planar alignment and having parallel rubbing on both surfaces.

The phase assignments were supported by measurements of the temperature dependence of the optical birefringence (Figure S1). The classification of nematic or smectic phases were confirmed by X-ray diffraction measurements. For all the phases reported here, the wide-angle X-ray diffraction signal was diffuse, confirming their liquid-like nature. For the nematic phases, the small-angle signal was also diffuse, while in the smectic phases, this signal sharpened and was limited only by instrumental broadening, evidencing true 1D positional order of molecules.

## Results and Discussion

The transition temperatures and phase sequences are given in Table S1, along with the values for the corresponding methoxy-substituted compounds reported previously, and the phase diagrams of the four homologous series studied are presented in Figure 4.

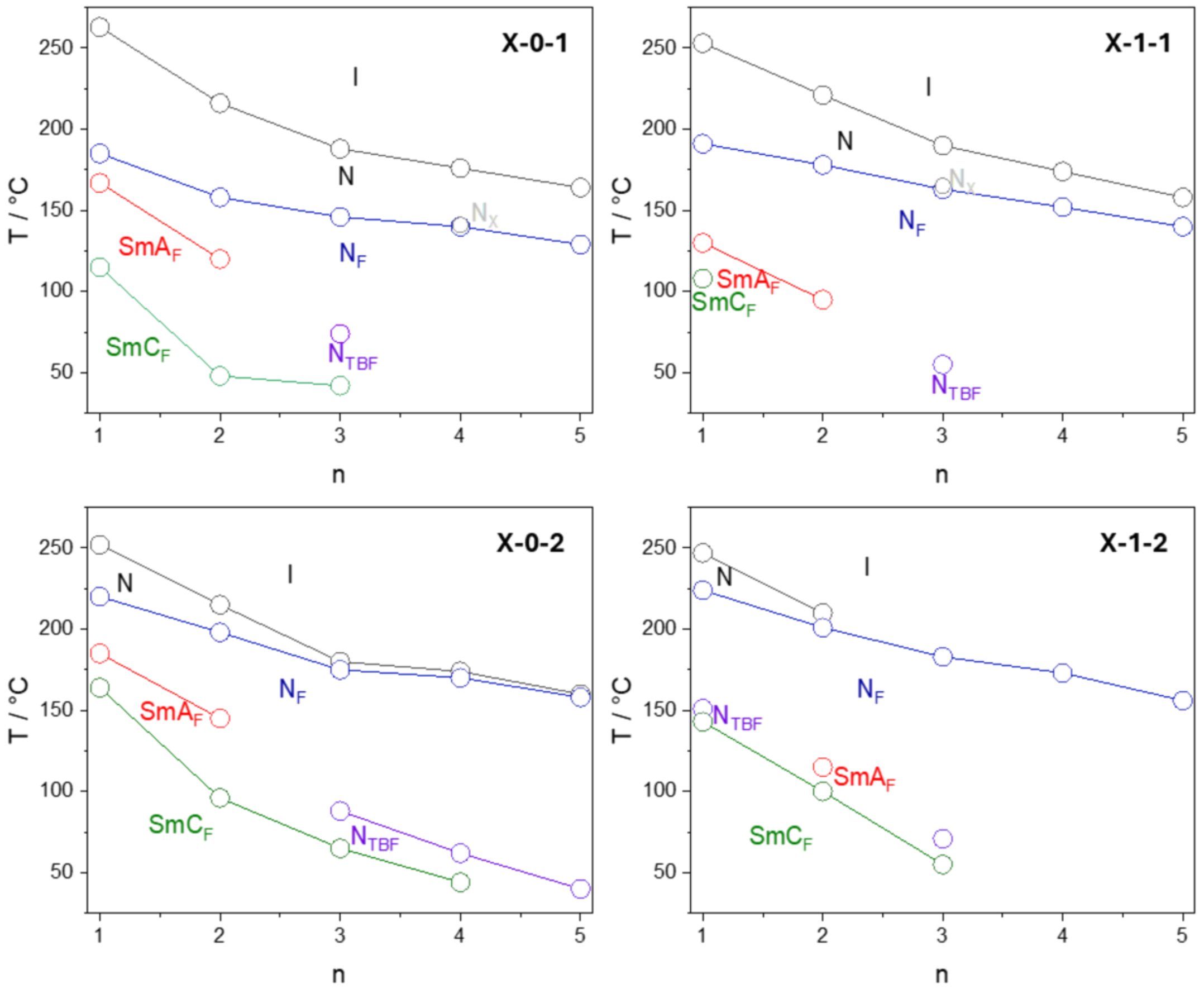

Figure 4 The change in phase sequence seen with increasing lateral chain length for the four series reported here. n is the number of carbons in the lateral chain.

All the compounds reported here formed the ferroelectric nematic phase, and all except the longer homologues of group **X-1-2** formed the paraelectric nematic phase. Mesogens which formed additional liquid crystal phases could be split into two groups, showing on cooling either a sequence of $N_F$ – $SmA_F$ – $SmC_F$ or $N_F$ – $N_{TBF}$ – $SmC_F$. Measurements of the temperature dependence of the optical birefringence showed clear differences between the two phase sequences (Figure 5). In both cases, the N – $N_F$ phase transition was accompanied by a step-like increase in Δn due to the growth of orientational order parameter associated with the onset of polar order. For several of the materials, a pronounced dip in the optical birefringence was additionally seen at the N-$N_F$ transition (Figure S1), which has been previously ascribed to strong fluctuations in the orientational order at the onset of polar order.[13,14] The following transition to $SmA_F$ phase was also accompanied by an increase in Δn due to the coupling of orientational and lamellar order (Figure 5), while in contrast the transition to the $N_{TBF}$ phase was associated with a decrease in Δn due to the tilting of the molecules with respect to the helical axis within the heliconical structure.

The transition between orthogonal and tilted smectic phases was confirmed by change in layer spacing (Figure 6). In the $SmA_F$ phase, the layer spacing was essentially constant, and corresponded to approximately one full molecular length (ca. 31Å). On entering the $SmC_F$ phase it continually decreased due to the tilting of molecules. In both phases, the wide-angle signal remains diffuse, indicating the lack of long-range positional correlations of molecules within the smectic planes.

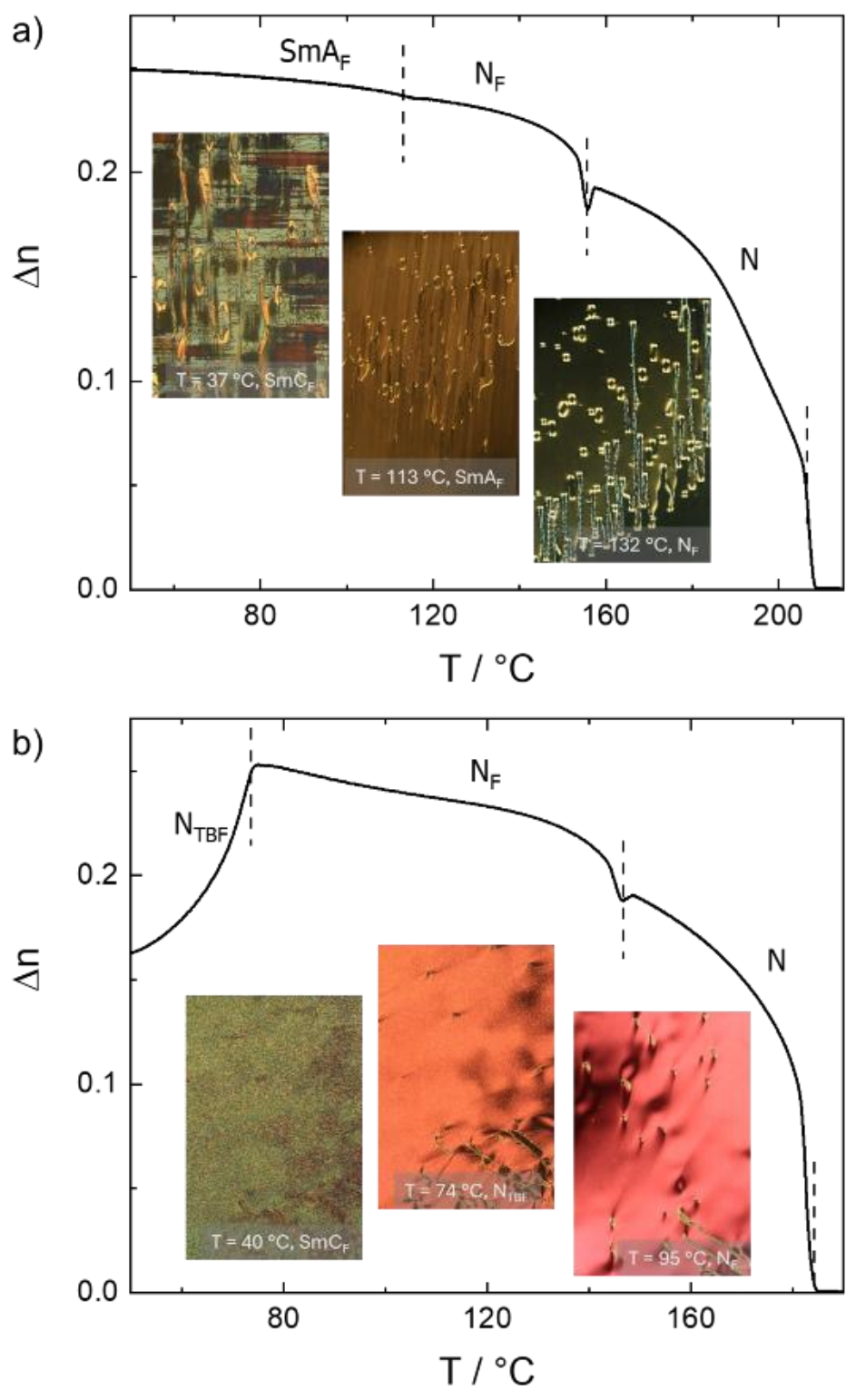

Figure 5. The temperature dependence of the optical birefringence of **E-0-1** (a) and **P-0-1** (b) measured with green light, note that the SmC$_F$ phase was not reached due to sample recrystallization during slow cooling. Inset: example POM textures seen in cells treated for planar alignment.

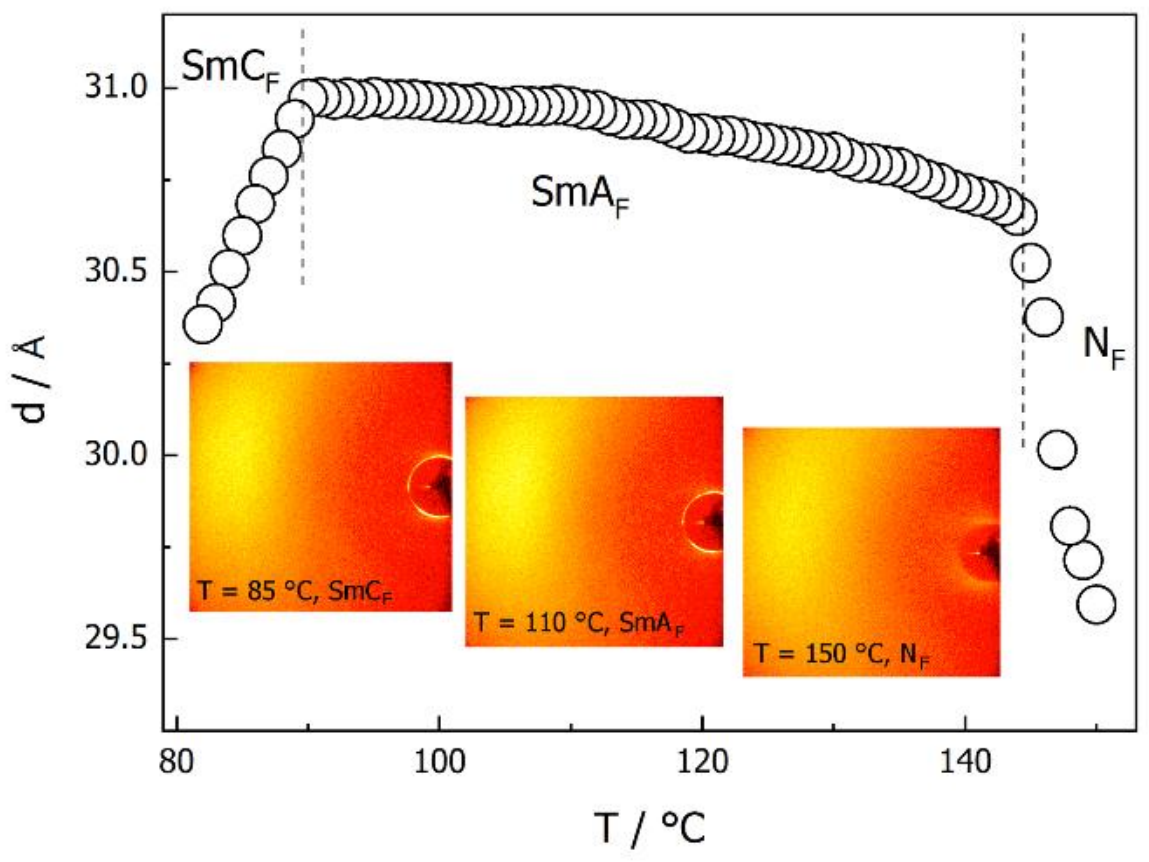

Figure 6.The temperature dependence of the layer spacing measured for **E-0-2**, with the X-ray diffraction patterns taken in the N$_F$, SmA$_F$, and SmC$_F$ phases.

The polar character of liquid crystal phases was established through dielectric spectroscopy and polarization switching measurements. Although the precise physical interpretation of measured permittivity is highly complex for such phases, qualitative trends can clearly be seen. A very strong relaxation mode is detected in the N$_F$ phase in all studied materials, the intensity of which drops off sharply on entering the SmA$_F$ phase, before rebuilding on transition to the SmC$_F$ phase. In contrast, the changes at N$_F$ – N$_{TBF}$ phase transition are much more subtle, with a gradual decrease seen in the strength and relaxation frequency of the mode.

Unfortunately, all the materials studied crystallized before reaching the $N_{TBF}$-$SmC_F$ transition when performing dielectric spectroscopy experiments. These observations are consistent with the behavior of previously studied materials, and representative data is given for **E-0-1** and **P-0-1** in Figure S2.

All polar phases gave clear switching current peaks associated with polarization reversal upon application of ac electric field, and representative examples are given in Figure 7. In the $N_{TBF}$ and $SmA_F$ phases there is a single current peak per half cycle of applied triangular-wave voltage, while in the $SmC_F$ phase, an additional small peak appears that is related to reduction/restoration of the tilt, this signal grows on cooling with growing tilt angle.[10,15]

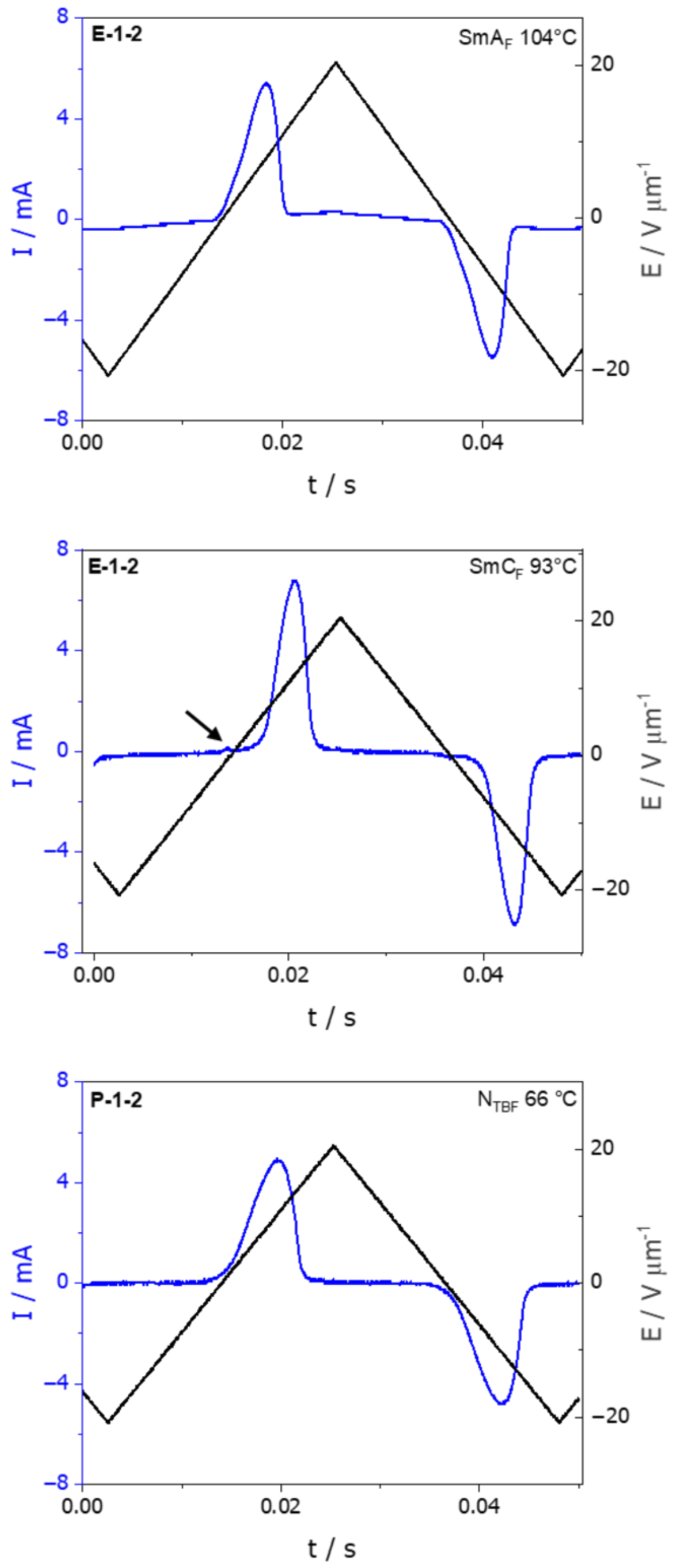

Figure 7. Switching current recorded in the SmA$_F$, N$_{TBF}$, and SmC$_F$ phases under the application of triangular-wave voltage, observed peaks correspond to polarization reversal. Arrow indicates additional current peak present in the SmC$_F$ phase, attributed to removal/restoration of molecular tilt.

## Comparing trends in transition temperatures on increasing lateral chain length

The phase diagrams for the four series are shown in Figure 4. The same trends are seen in the first three series, **X-0-1**, **X-0-2**, and **X-1-1**, the shortest, methoxy-substituted homologues of which showed the same phase sequence: N – N$_F$ – SmA$_F$ – SmC$_F$. Increasing the length of the lateral alkoxy chain led to a decrease in both $T_{NI}$ and $T_{NFN}$, with the former decreasing more rapidly. These trends are in good agreement with those previously reported for ferroelectric nematic liquid crystals and are related to the decreasing molecular shape anisotropy.[16] The temperature of the $N_F$-$SmA_F$ and $SmA_F$-$SmC_F$ transitions also decreased strongly with replacing the methoxy chain with ethoxy one, to such an extent that for **E-1-1** the $SmC_F$ phase was not seen at all prior to crystallization. Interestingly, increasing the length of the lateral chain to three carbons led to the loss of the $SmA_F$ phase, which was replaced by the emergence of the $N_{TBF}$ phase in all three series. For **P-0-1** and **P-0-2**, this then transitioned to the $SmC_F$ phase on further cooling, while in **P-1-1**, the sample crystallized without forming any other LC phases.

Extending the chain length further in series **X-0-1** and **X-1-1** produced only nematic and ferroelectric nematic phases in the butoxy and pentoxy homologues (n = 4 and 5). This presumably reflects a further decrease in the transition temperatures to more ordered phases, such that they are not reached before the sample crystallizes. In contrast, **B-0-2** forms both the $N_{TBF}$ and $SmC_F$ phase, and **Q-0-2** forms the $N_{TBF}$ phase, which could be supercooled to room temperature. The transition temperatures of the ferroelectric phases formed by the **X-0-2** series are consistently higher than those of the **X-0-1** and **X-1-1** series. This highlights the sensitivity of the relationship between fluorination pattern and phase behavior in these materials.

The phase behavior of the final series of compounds, **X-1-2**, was more complex. The parent methoxy-substituted **M-1-2** formed the $N_{TBF}$ phase, rather than the $SmA_F$ phase seen in the other series. From the trends discussed above, it may be expected that increasing the length of lateral substituent would simply lead to a decrease in the $N_F$-$N_{TBF}$ transition temperature, until crystallization prevented its observation. However, the ethoxy-substituted **E-1-2** unexpectedly formed the $SmA_F$ phase. Upon further increasing the length of the substituent to three carbons in **P-1-2**, the $N_{TBF}$ phase reappeared and the $SmA_F$ phase was again lost. Additionally, the N phase was also not seen for this and longer homologues, with **P-1-2**, **B-1-2** and **Q-1-2** all showing direct isotropic-ferroelectric nematic transitions.

### Exploring the unexpected crossover between $SmA_F$ and $N_{TBF}$ phases

To further investigate the apparent competition between the $N_{TBF}$ and $SmA_F$ phases seen in the studied materials, binary mixtures were prepared for members of the **X-1-2** series, and these are shown in Figure 8. The first set of mixtures combined **M-1-2** and **P-1-2** (n = 1 and 3, respectively). While both pure materials showed the same phase sequence N – $N_F$ – $N_{TBF}$ – $SmC_F$, in mixtures the $N_{TBF}$ phase is replaced by the $SmA_F$ phase, as was seen for the intermediate homologue **E-1-2** (n = 2), which can be considered as an analogue of the equimolar mixture. Similarly, mixing of **E-1-2** (n = 2, sequence N –$N_F$ – $SmA_F$ – $SmC_F$) and **B-1-2** (n = 4, only $N_F$ phase seen) induced the $N_{TBF}$ phase, as seen in pure **P-1-2** (n = 3). These observations reinforce the idea that there is a competition between the tendencies for formation of lamellar and heliconical polar structures, however the precise physical mechanisms underpinning this are not clear. We have previously proposed that, in this family of materials, self-segregation between fluorinated and non-fluorinated parts of the molecule drives the formation of layered polar phases.[10] For the $N_{TBF}$ phase, it has been suggested that the spontaneous chiral symmetry breaking occurs in order to partially compensate the large molecular dipoles through the formation of heliconical structure. We have shown here that, for the series **X-0-1**, **X-1-1**, and **X-0-2**, extending the length of lateral substituents destabilizes smectic layer formation, and allows for the $N_{TBF}$ phase to emerge. A partial explanation, then, for the competition between heliconical and lamellar order may be down to their molecular structure: the large dipole moments required for the formation of the $N_{TBF}$ phase require a high degree of fluorination, typically concentrated on one end of the long molecular axis. However, such a concentration of fluorine substituents tends to favor formation of lamellar structure due to self-segregation of non-compatible molecular units. These two factors may contribute to the observed competition between the $N_{TBF}$ and $SmA_F$ phases.

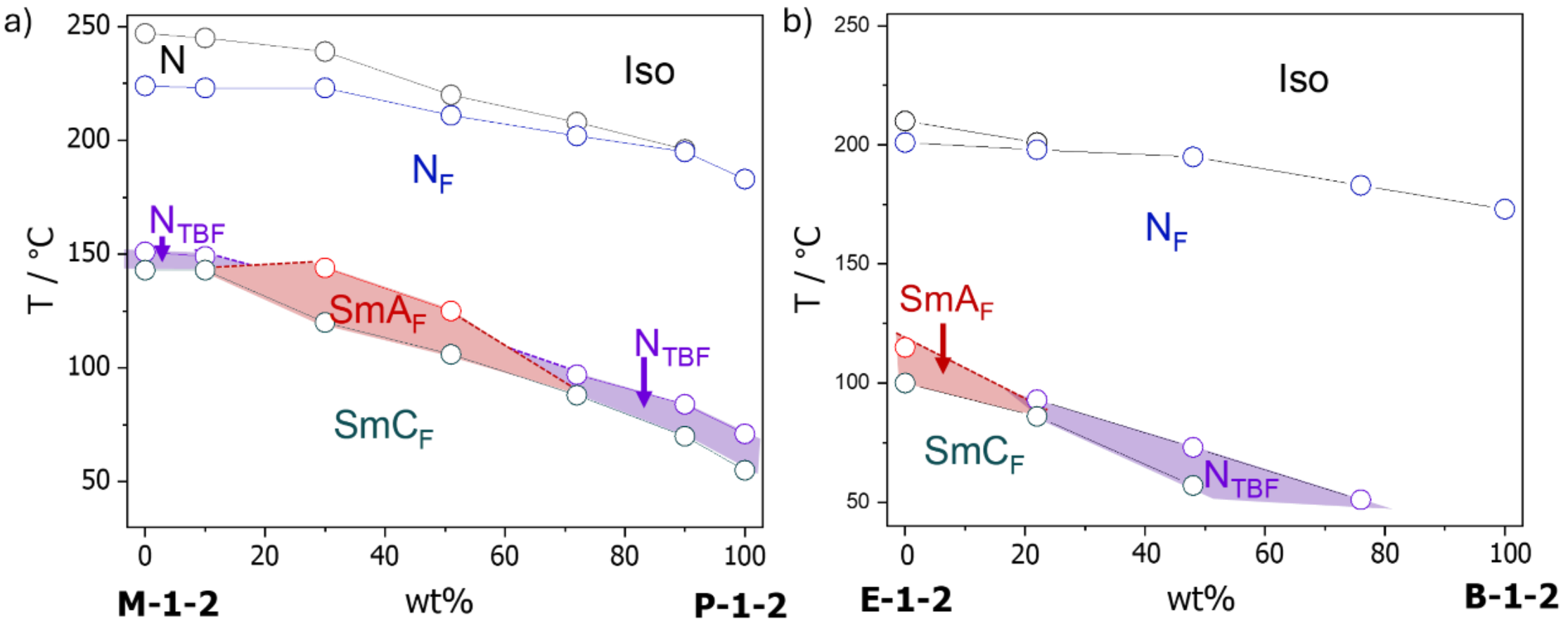

Figure 8. Phase diagrams for binary mixtures of **M-1-2** and **P-1-2** (a) and **E-1-2** and **B-1-2** (b).

## Helical pitch in the $N_{TBF}$ phase

Nearly half of the mesogens discussed here show the $N_{TBF}$ phase. We have previously reported the methoxy-terminated **M-1-2,** for which the pitch of the helix in the $N_{TBF}$ phase changed in a non-monotonic way with temperature. On cooling from the $N_F$ phase it initially decreased, reaching a minimum of ca. 900 nm, before unwinding again on approach to the $SmC_F$ phase. Attempts to measure the helical pitch by laser light diffraction for any of the new materials reported here proved challenging, as the helical pitch in these appears to be shorter and winds more quickly, going out of detection limit ca. 1 or 2 K below the $N_F$-$N_{TBF}$ phase transition. This decrease in the helical pitch compared to **M-1-2** can be attributed to the lower transition temperatures of the $N_{TBF}$ phase in the new compounds. To investigate this more closely, the helical structure of compound **P-0-2** was studied by measuring selective light reflection (Figure 9). The wavelength of selective reflection ($\lambda_{sel}$) can be related to the helical pitch (p) using the relation $\lambda_{sel}$=np, where n is the average refractive index. Assuming a value of 1.5 for n, this indicates that the pitch of the helix decreases from approximately 430 to 265 nm within a few degrees below the $N_F$-$N_{TBF}$ transition.

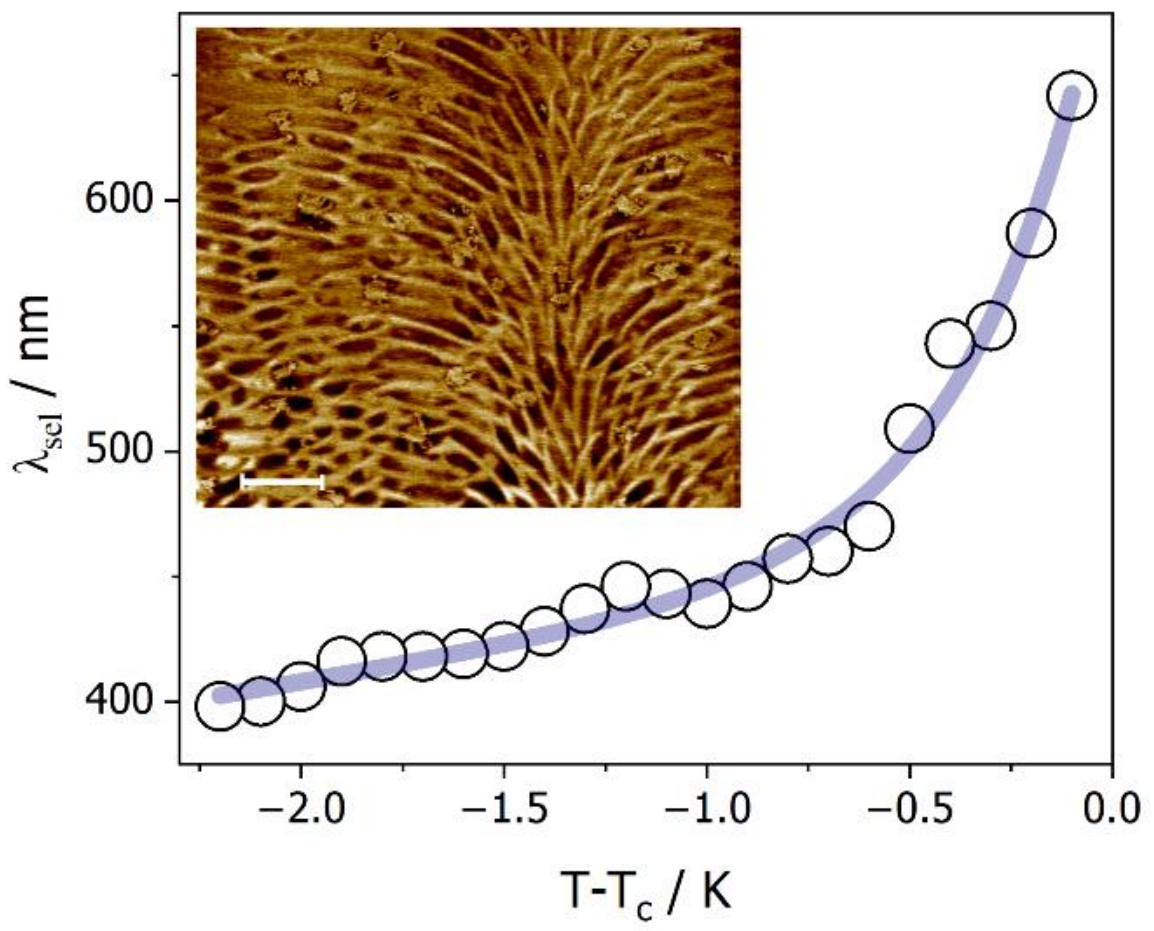

Figure 9. The temperature dependence of the wavelength of selective reflection measured in the $N_{TBF}$ phase of **P-0-2** on cooling. Inset: AFM image recorded for **Q-0-2**.

For two mesogens, the pentyloxy-substituted **Q-0-2** and the propyloxy-substituted **P-1-1,** the $N_{TBF}$ phase could be supercooled to room temperature, which enabled it to be studied with atomic force microscopy (AFM). AFM imaging (inset in Figure 9 for **Q-0-2** and Figure S3 for **P-1-1**) revealed a structural periodicity of approximately 250 nm at room temperature. This periodicity corresponds to subtle height variations on the surface, which reflect a regular stack of equivalent slabs or "pseudo-layers," each one approximately equal in thickness to a single helical pitch. However, in contrast to previously studied materials,[17] where these pseudo-layers appeared largely defect-free over wide areas with only a few edge dislocations, **Q-0-2** exhibits a number of nearly equally spaced, horizontally oriented screw dislocations. It remains unclear whether these screw dislocations arise during the initial nucleation and growth of the pseudo-layers of the $N_{TBF}$ phase, or whether they are an intrinsic feature of a more complex structure — for example, a twist-grain-boundary phase derived from a helical $N_{TBF}$ phase. Apparently, for the studied system, screw dislocations that cause a continuous shift of the layers along the defect line are relatively easy to form.

## Conclusions

We have reported the synthesis and liquid crystalline properties of four homologous series of mesogens with lateral alkoxy substituents. The length of the lateral chains were increased up to five carbon atoms to establish how this would affect the phase behavior. For all materials, increasing the lateral chain length decreased the clearing temperature and the onset of both polar and lamellar order. For the three series **X-0-1**, **X-0-2**, and **X-1-1**, the inclusion of a three-carbon chain leads to the extinction of the $SmA_F$ phase, which is replaced by the heliconical $N_{TBF}$ phase The replacement of the $SmA_F$ phase by the $N_{TBF}$ phase highlights the apparent competition between the driving forces contributing to the appearance of lamellar order and heliconical structure. This is reinforced by the unexpected behavior seen in series **X-1-2**, in which the phase sequence alternates as the length of the lateral chain increases: **M-1-2** forms the $N_{TBF}$ phase, **E-1-2** forms the $SmA_F$ phase, and **P-1-2** again forms the $N_{TBF}$ phase. This underscores the extremely sensitive relationships between molecular structure and the intermolecular interactions producing either lamellar or heliconical structure in these polar LC phases. For the materials studied here, the helical pitch in the $N_{TBF}$ phase is very short, only a few hundred nm, compared to the micron-scale pitch previously reported.[1,6,10] This is attributed to the lower temperatures at which the $N_{TBF}$ phase is observed for these new compounds. In addition, AFM images of **Q-0-2** reveal a structured network of screw dislocations, which may indicate a potential link to more complex phase structures in this family of materials.

## Acknowledgements

This research was supported by the National Science Centre (Poland) under the grant no. 2021/43/B/ST5/00240.

Supporting information

## Experimental Methods

Transition temperatures and the associated enthalpy changes were measured by differential scanning calorimetry using a TA DSC Q200 instrument. Measurements were performed under a nitrogen atmosphere with a heating/cooling rate of 10 K min$^{-1}$, unless otherwise specified.

Observations of optical textures of liquid crystalline phases was carried out by polarised-light optical microscopy using a Zeiss AxioImager.A2m microscope equipped with a Linkam heating stage.

Optical birefringence was measured with a setup based on a photoelastic modulator (PEM-90, Hinds) working at a modulation frequency *f* = 50 kHz; as a light source a halogen lamp (Hamamatsu LC8) equipped with narrow bandpass filters was used. The transmitted light intensity was monitored with a photodiode (FLC Electronics PIN-20) and the signal was deconvoluted with a lock-in amplifier (EG&G 7265) into 1*f* and 2*f* components to yield a retardation induced by the sample. Knowing the sample thickness, the retardation was recalculated into optical birefringence. Samples were prepared in 1.6-µm-thick cells with planar anchoring. The alignment quality was checked prior to measurement by inspection under the polarised-light optical microscope.

X-ray diffraction measurements of samples in liquid crystalline phases were carried out using a Bruker D8 GADDS system, equipped with micro-focus-type X-ray source with Cu anode and dedicated optics and VANTEC2000 area detector. Small angle diffraction experiments were performed on a Bruker Nanostar system (IµS microfocus source with copper target, MRI heating stage, Vantec 2000 area detector).

The complex dielectric permittivity, ε*, was measured using a Solartron 1260 impedance analyser, in the 1 Hz −10 MHz frequency range, and a probe voltage of 50 mV. The material was placed in a 5- or 10-µm-thick glass cell with gold electrodes. Cells without polymer aligning layers were used, as the presence of the thin (~10 nm) polyimide layers at the cell surfaces acts as an additional high capacitance capacitor in a series circuit with the capacitor filled with the LC sample, which for materials with very high values of permittivity, may strongly affect the measured permittivity of the LC phases. Lack of a surfactant layer resulted in a random configuration of the director in the LC phases.

To determine the helical pitch length in $N_{TBF}$ phase selective light reflection studies were carried out with the material placed on a glass substrate without a top cover. Light transmission/reflection was monitored with a fiber-coupled spectrometer (Filmetrics F20-UV) mounted onto the Zeiss Axio Imager A2m microscope.

**Additional data**

Table S1. The transition temperatures (in °C) of the compounds reported here.

| Compound | Melt | SmC$_F$ | | SmA$_F$ | | N$_{TBF}$ | | N$_F$ | | N$_X$ | | N | | Iso |
|---|---|---|---|---|---|---|---|---|---|---|---|---|---|---|
| **M-0-1**[1] | 141 | • | 115 | • | 167 | | | • | 185 | | | • | 263 | • |
| **E-0-1** | 158 | | 48[c] | • | 120 | | | • | 158 | | | • | 216 | • |
| **P-0-1** | 154 | | 42[c] | | | • | 74[a] | • | 146 | | | • | 188 | • |
| **B-0-1** | 118 | | | | | | | • | 140 | • | 141 | • | 176 | • |
| **Q-0-1** | 124 | | | | | | | • | 129 | | | • | 164 | • |
| **M-1-1** [1] | 139 | • | 108[b] | • | 130[a] | | | • | 191 | | | • | 253 | • |
| **E-1-1** | 132 | | | • | 95[d] | | | • | 178 | | | • | 221 | • |
| **P-1-1** | 130 | | | | | • | 55[d] | • | 163 | • | 165[c] | • | 190 | • |
| **B-1-1** | 117 | | | | | | | • | 152 | | | • | 174 | • |
| **Q-1-1** | 119 | | | | | | | • | 140 | | | • | 158 | • |
| **M-0-2** [1] | 138 | • | 164[a] | • | 185 | | | • | 220 | | | • | 252 | • |
| **E-0-2** | 143 | • | 96[c] | • | 145[a] | | | • | 198 | | | • | 215 | • |
| **P-0-2** | 133 | • | 79[c] | | | • | 100[c] | • | 175 | | | • | 180 | • |
| **B-0-2** | 129 | • | 44[c] | | | • | 62[a] | • | 170 | | | • | 174 | • |
| **Q-0-2** | 112 | | | | | • | 42[a] | • | 158 | | | • | 160 | • |
| **M-1-2** [1] | 141 | • | 143 | | | • | 151[a] | • | 224 | | | • | 247 | • |
| **E-1-2** | 145 | • | 100[c] | • | 115[c] | | | • | 201 | | | • | 210 | • |
| **P-1-2** | 115 | • | 55[c] | | | • | 71[a] | • | 183 | | | | | • |
| **B-1-2** | 102 | | | | | | | • | 173 | | | | | • |
| **Q-1-2** | 100 | | | | | | | • | 156 | | | | | • |

[a] Taken from birefringence measurement. [b] Taken from SAXS. [c] Taken from POM. [d] Taken from dielectric spectroscopy.

Table S2. Representative POM textures in thin cells treated for planar alignment.

| | $\mathrm{SmC_F}$ | $\mathrm{SmA_F}$ | $\mathrm{N_{TBF}}$ | $\mathrm{N_F}$ | $\mathrm{N_X}$ |
|---|---|---|---|---|---|
| **E-0-1** | 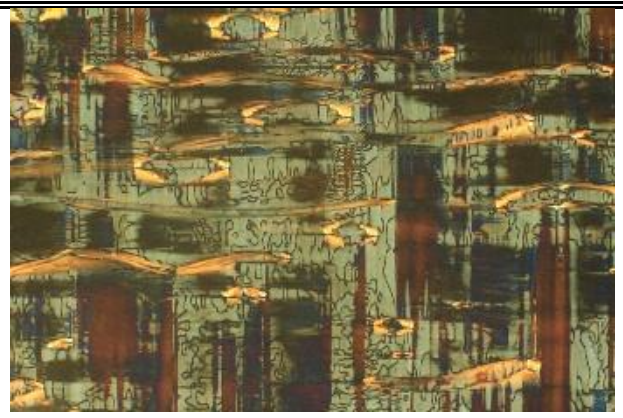 HG Cell, 1.8 μm, 37 °C | 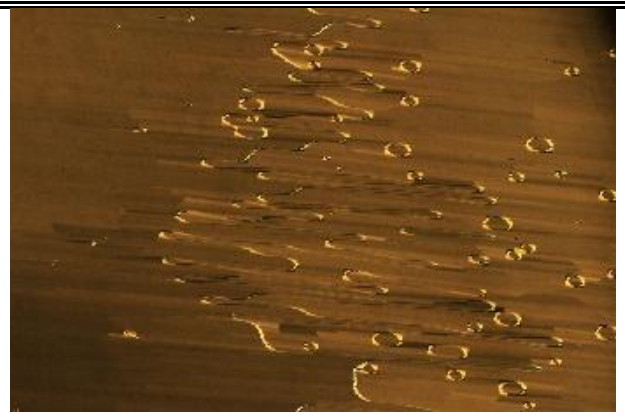 HG Cell, 1.8 μm, 121 °C | | 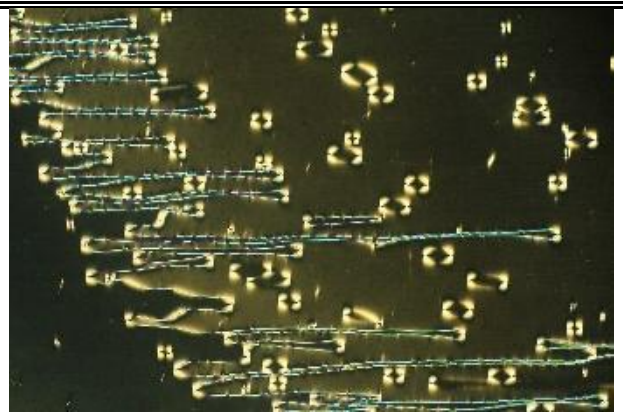 HG Cell, 1.8 μm, 132 °C | |
| **P-0-1** | 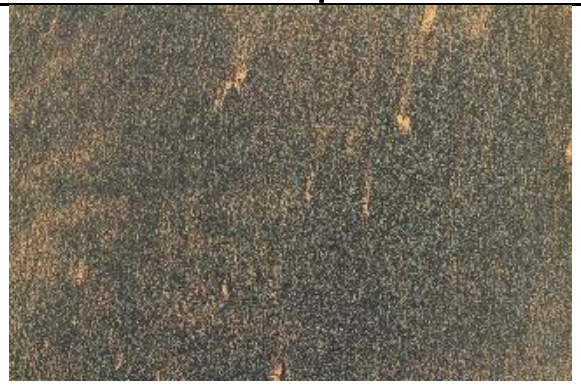 HG Cell, 1.7 μm, 42 °C | | 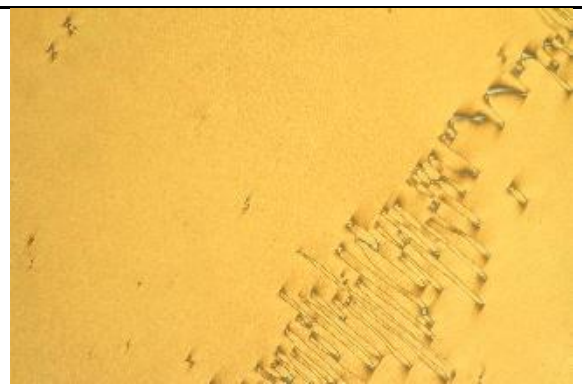 HG Cell, 1.7 μm, 74 °C | 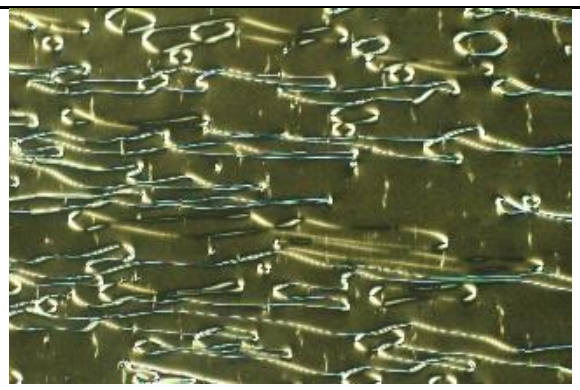 HG Cell, 1.7 μm, 132 °C | |
| **B-0-1** | | | | 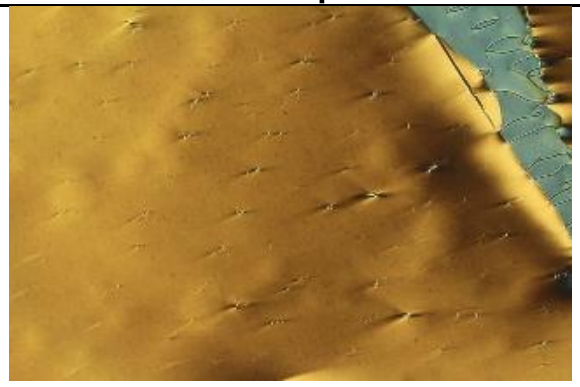 HG Cell, 1.8 μm, 120 °C | 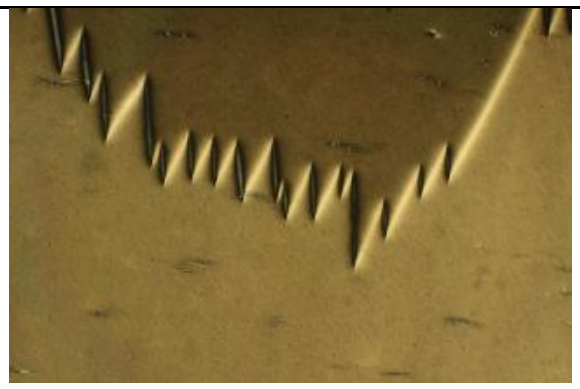 HG Cell, 1.8 μm, 139 °C |
| **Q-0-1** | | | | 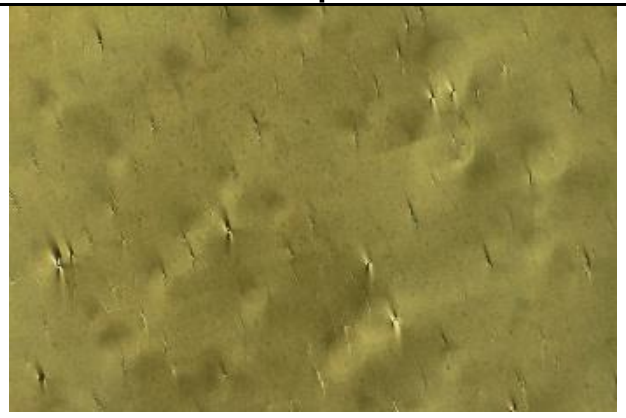 HG Cell, 1.6 μm, 128 °C | |
| **E-1-1** | | 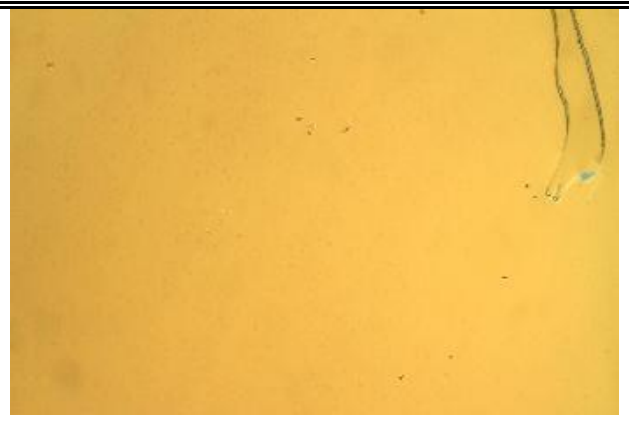 HG Cell, 1.5 µm, 92°C | | 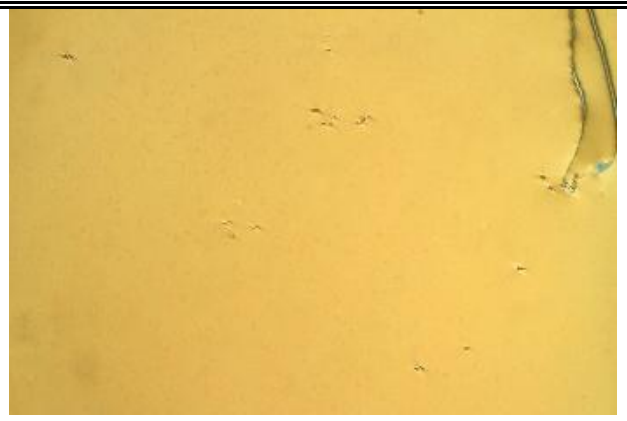 HG Cell, 1.5 µm, 139 °C | |
| **P-1-1** | 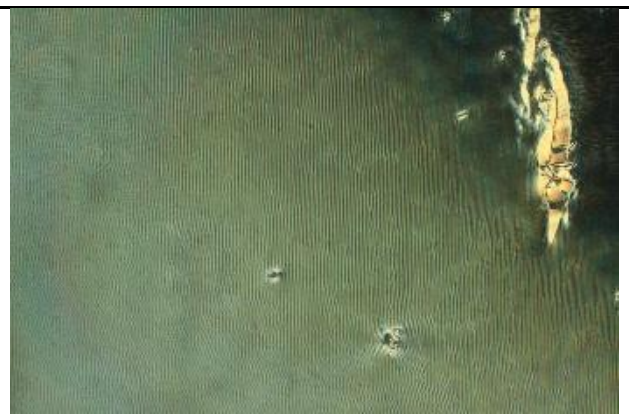 HG 1.6 µm, 32 °C x50 | | 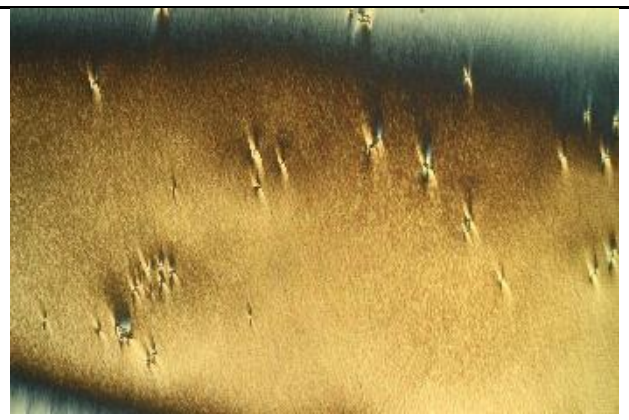 HG 1.6 µm, 54 °C | 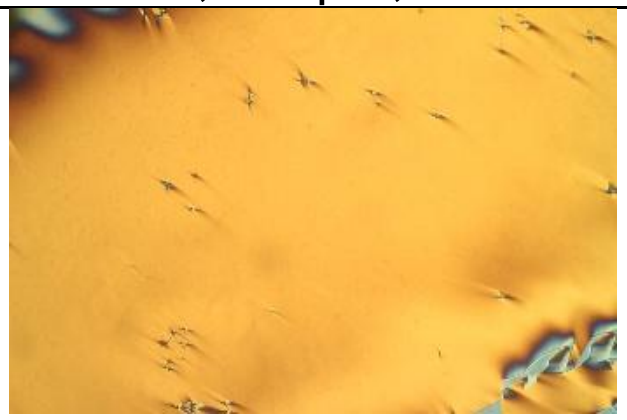 HG 1.6 µm, 97 °C | 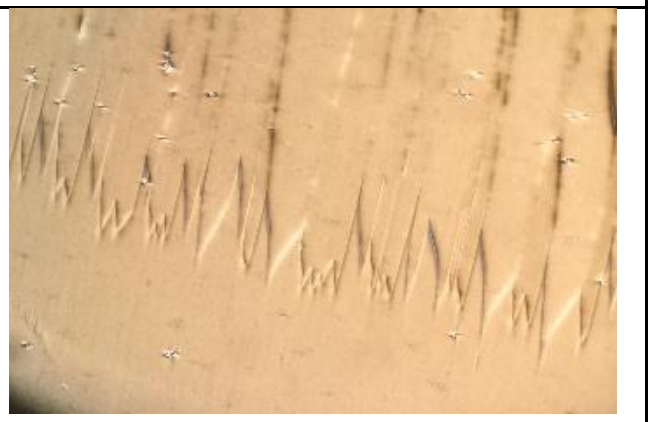 HG 1.6 µm, 165 °C ? |
| **B-1-1** | | | | 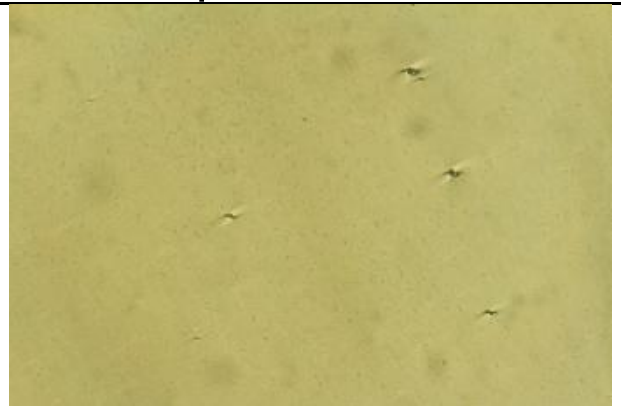 HG 1.4 µm, 145 °C | |
| **Q-1-1** | | | | 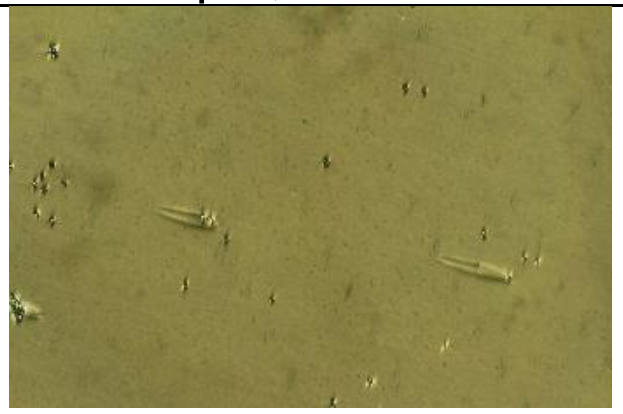 HG 1.6 µm, 141 °C | |
| **E-0-2** | 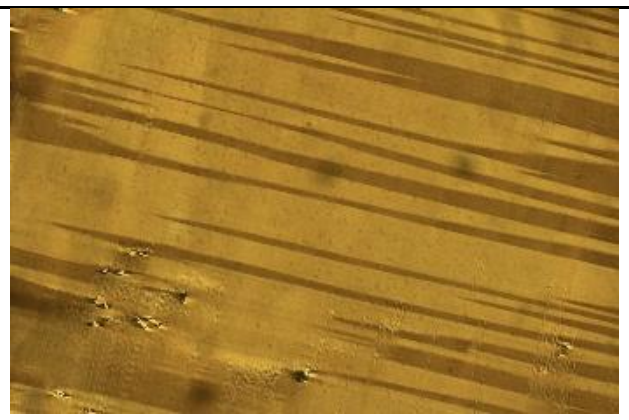 HG 1.6 µm, 90 °C | 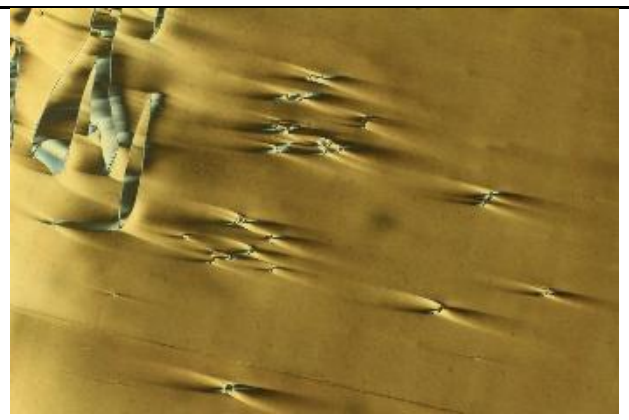 HG 1.6 µm, 134 °C | | 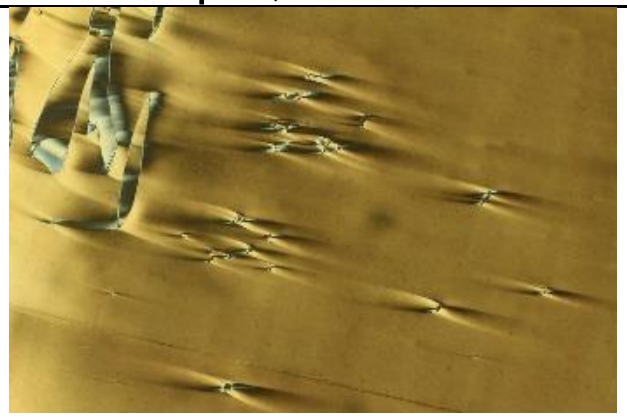 HG 1.6 µm, 158 °C | |
| **P-0-2** | 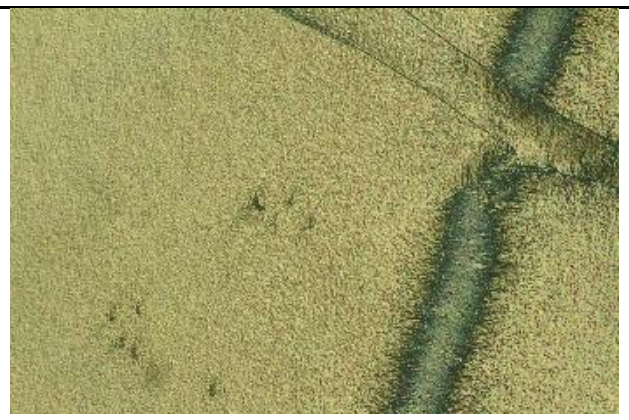<br>HG 1.8 µm, 66 °C | | 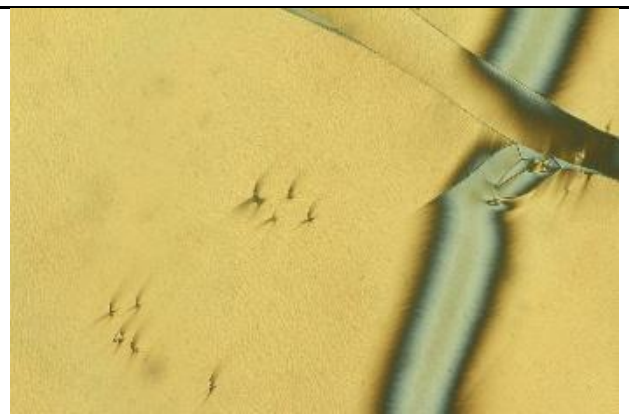<br>HG 1.8 µm, 88 °C | 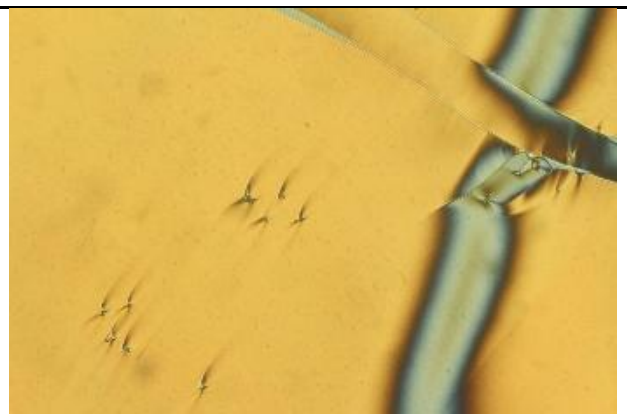<br>HG 1.8 µm, 105 °C | |
| **B-0-2** | 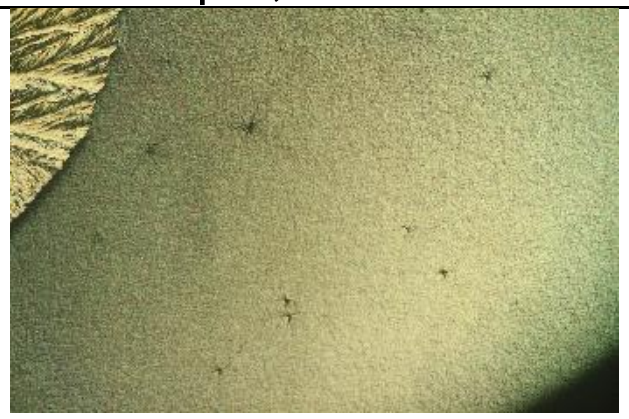<br>HG 1.8 µm, 42 °C + Cr | | 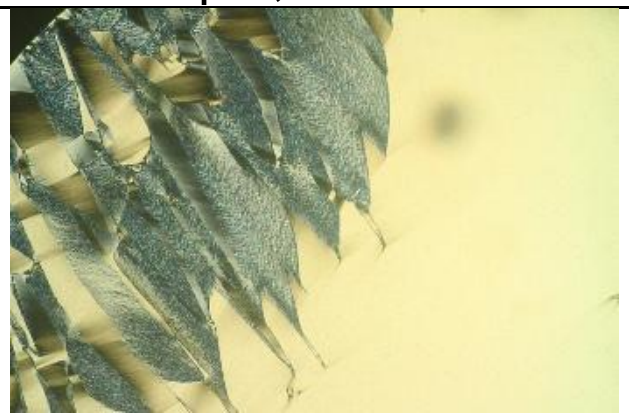<br>HG 1.8 µm, 54 °C | 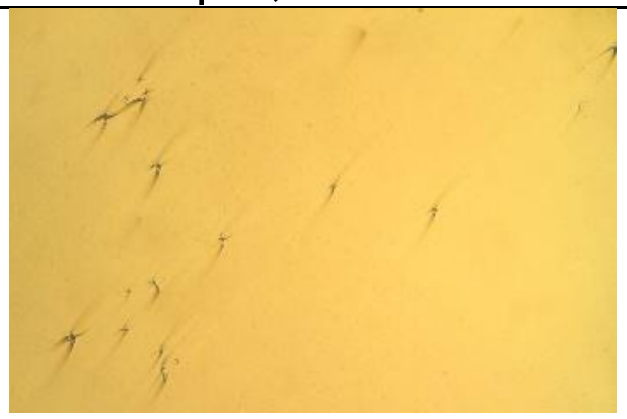<br>HG 1.8 µm, 120 °C | |
| **Q-0-2** | | | 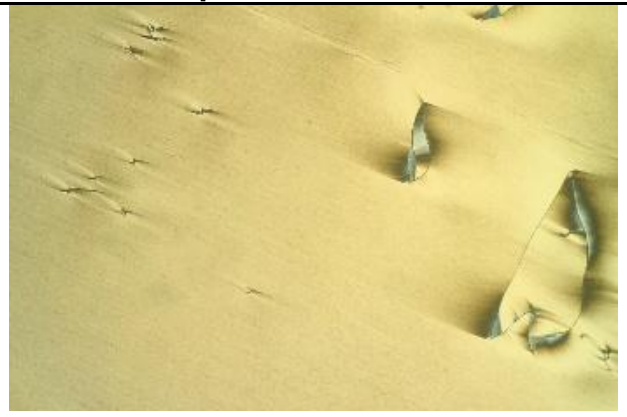<br>HG 1.5 µm, 35 °C | 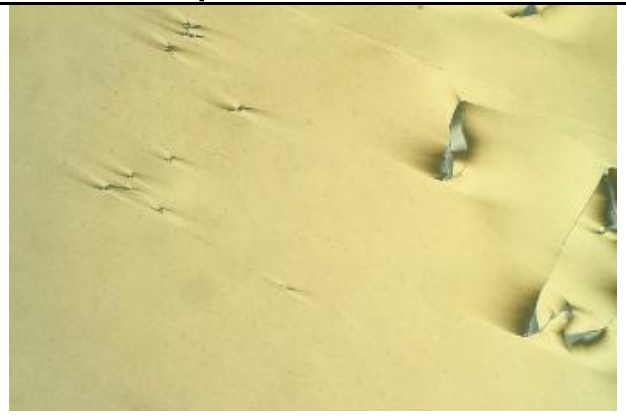<br>HG 1.5 µm, 136 °C | |
| **E-1-2** | 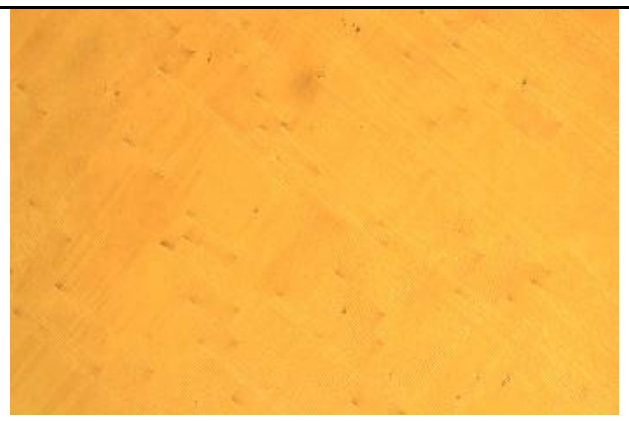<br>HG 1.5 µm, 90 °C | 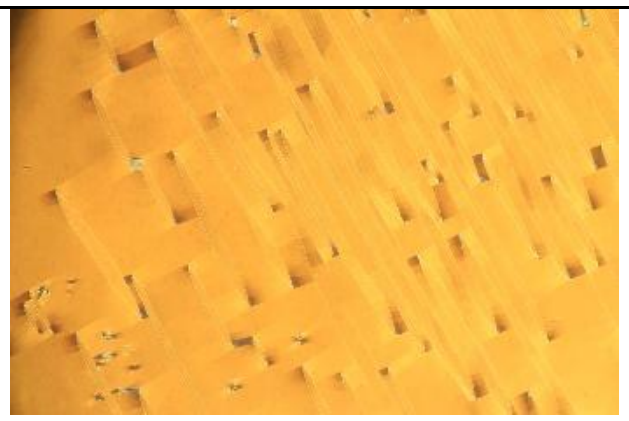<br>HG 1.5 µm, 110 °C | | 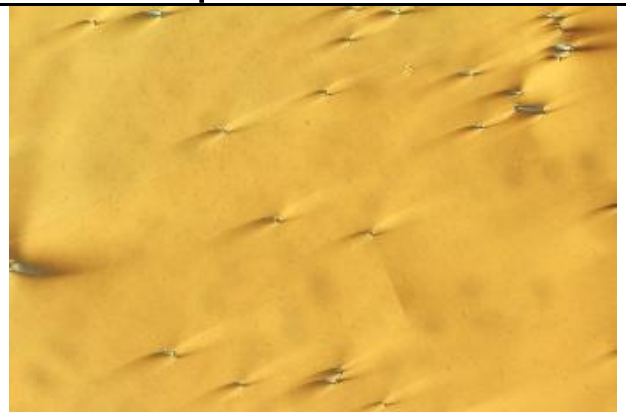<br>HG 1.6 µm, 150 °C | |
| **P-1-2** | 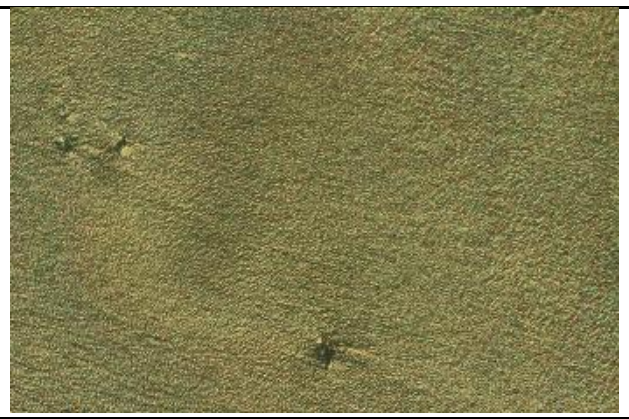 | | 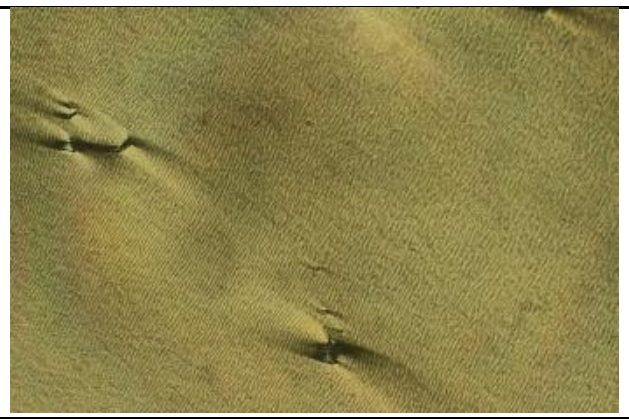 | 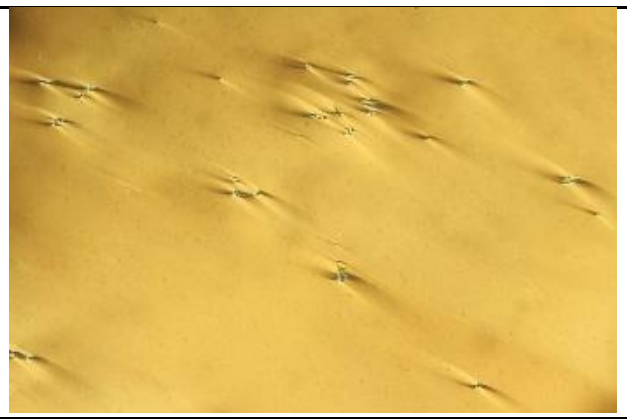 | |

| | HG 1.8 µm, 52 °C x50 | | HG 1.8 µm, 59 °C x50 | HG 1.8 µm, 136 °C | |
|---|---|---|---|---|---|
| **B-1-2** | | | | 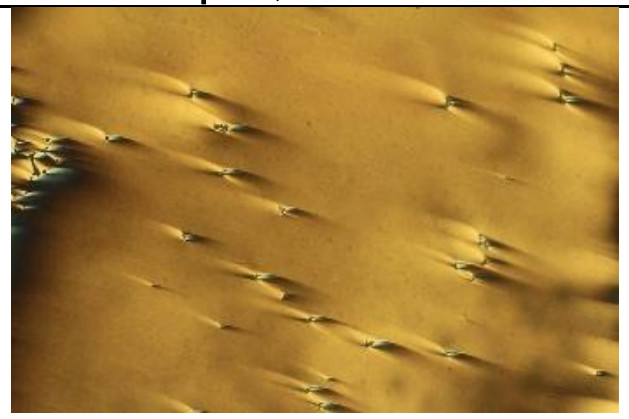 HG 1.7 µm, 84 °C | |
| **Q-1-2** | | | | 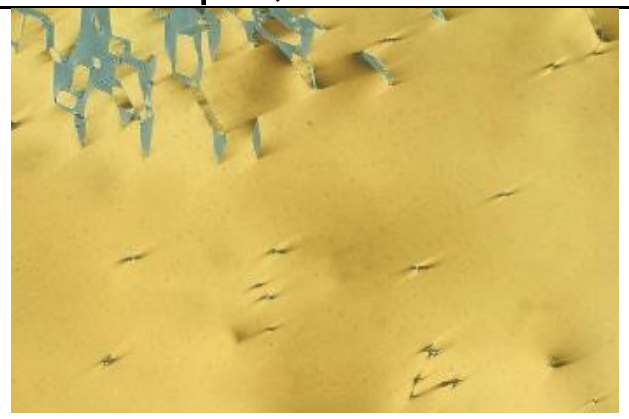 HG 1.6 µm, 116 °C | |

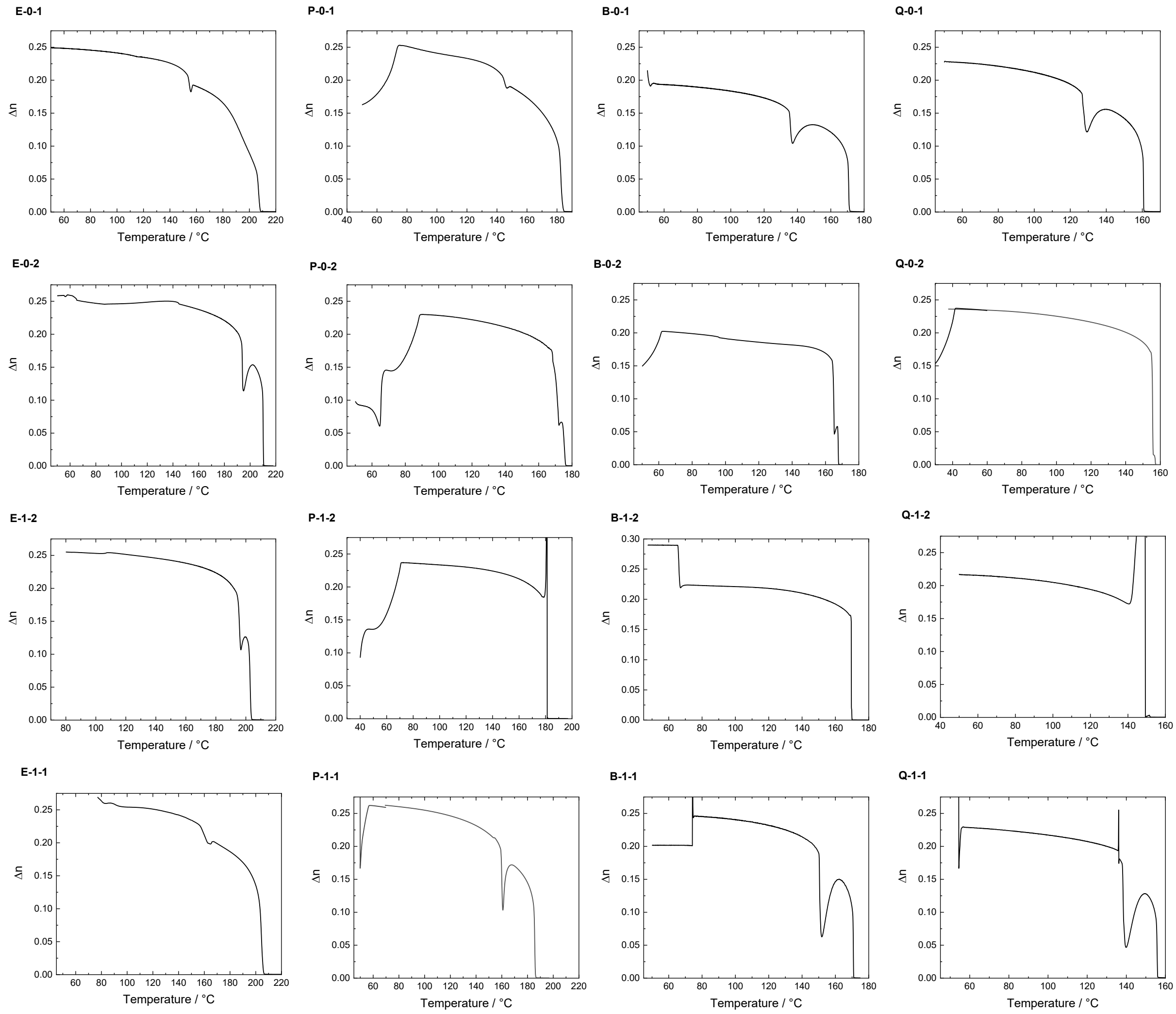

Figure S1.The temperature dependence of the optical birefringence measured with green light in thin cells treated for planar alignment.

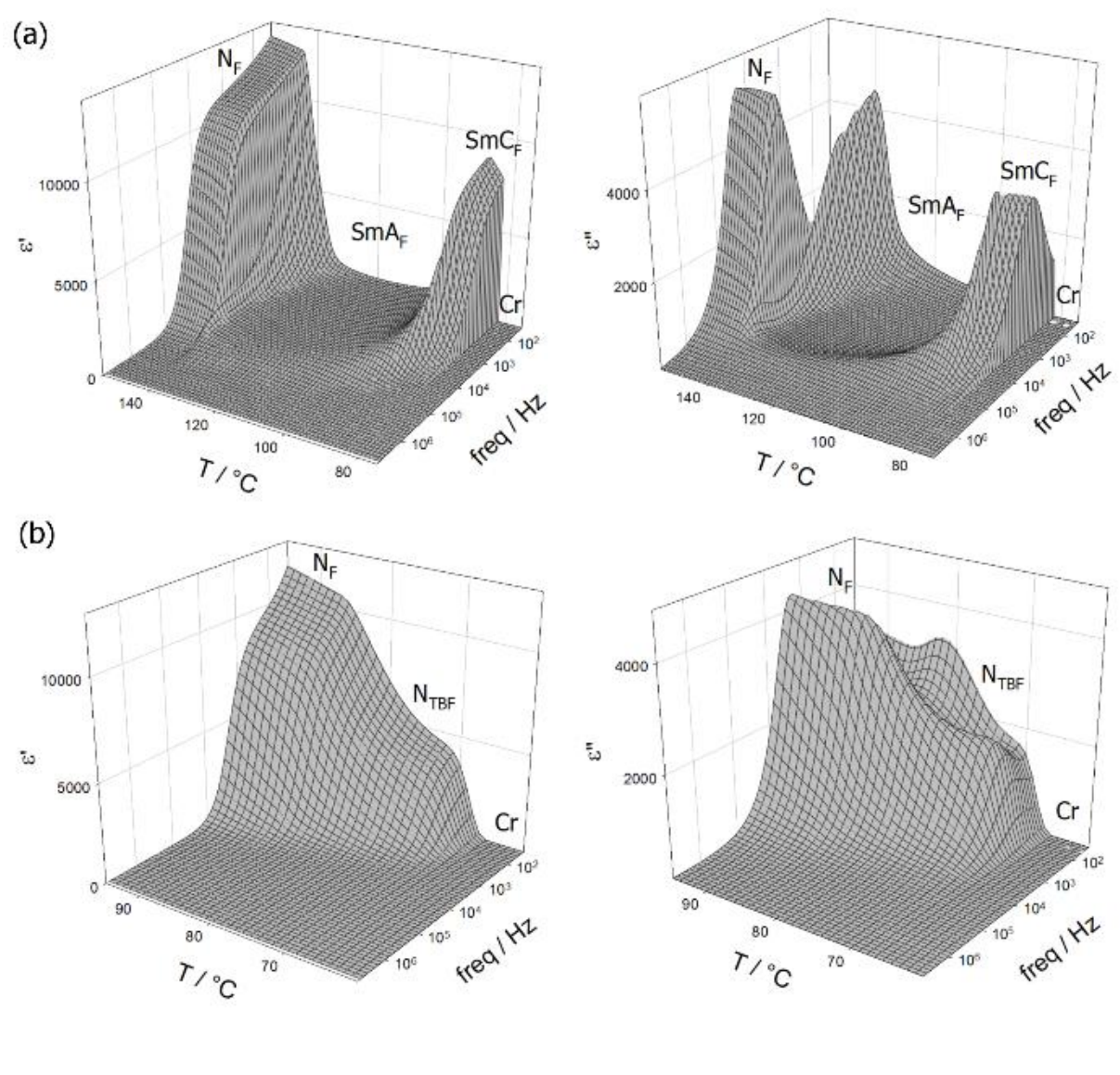

Figure S2. The real (ε′) and imaginary (ε″) parts of apparent dielectric permittivity of (a) **E-0-2** and (b) **P-0-2** compounds, measured in 10-µm-thick cells with gold electrodes.

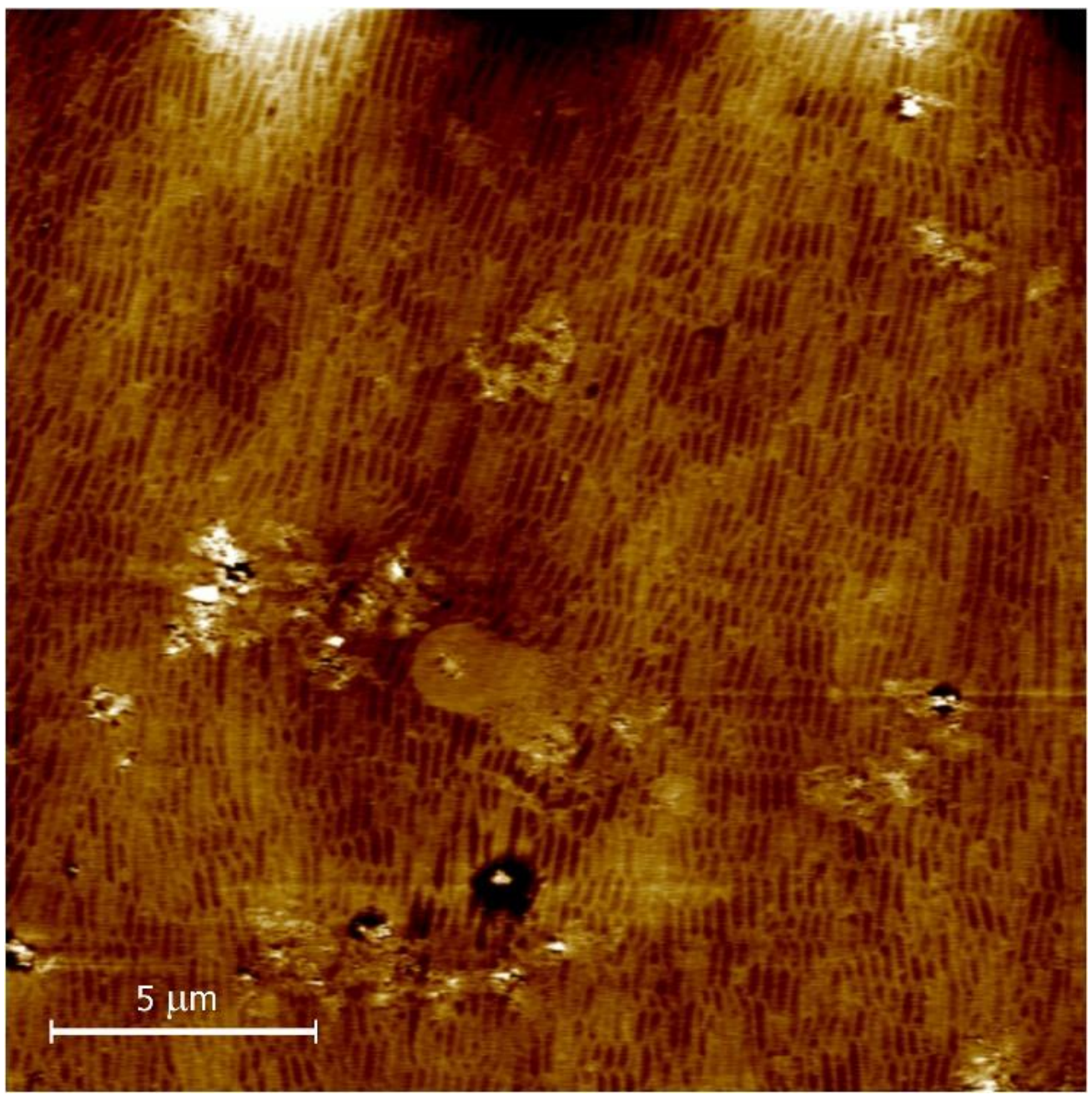

Figure S3. AFM image recorded for **P-1-1**.

**Synthetic Procedures and Structural Characterisation**

Unless otherwise stated, all materials were obtained from commercial sources and used without further purification. Reactions were monitored using thin layer chromatography (TLC) using aluminium-backed plates with a coating of Merck Kieselgel 60 F254 silica and an appropriate solvent system. Spots were visualised using UV light (254 nm). Flash column chromatography was carried out using silica grade 60 Å 40-63 micron. $^{1}H$, $^{19}F$, and $^{13}C$ NMR spectra were recorded on a 400 MHz Agilent NMR spectrometer using $CDCl_3$ as solvent and using residual non-deuterated trace solvents as reference. Chemical shifts (δ) are given in ppm relative to TMS (δ = 0.00 ppm). Due to limited sample availability, solubility and the effect of $^{19}F$ coupling some $^{13}C$ multiplets are very weak, and spectra are reported as fingerprints for comparison. Mass spectroscopy was conducted on a Micromass LCT instrument.

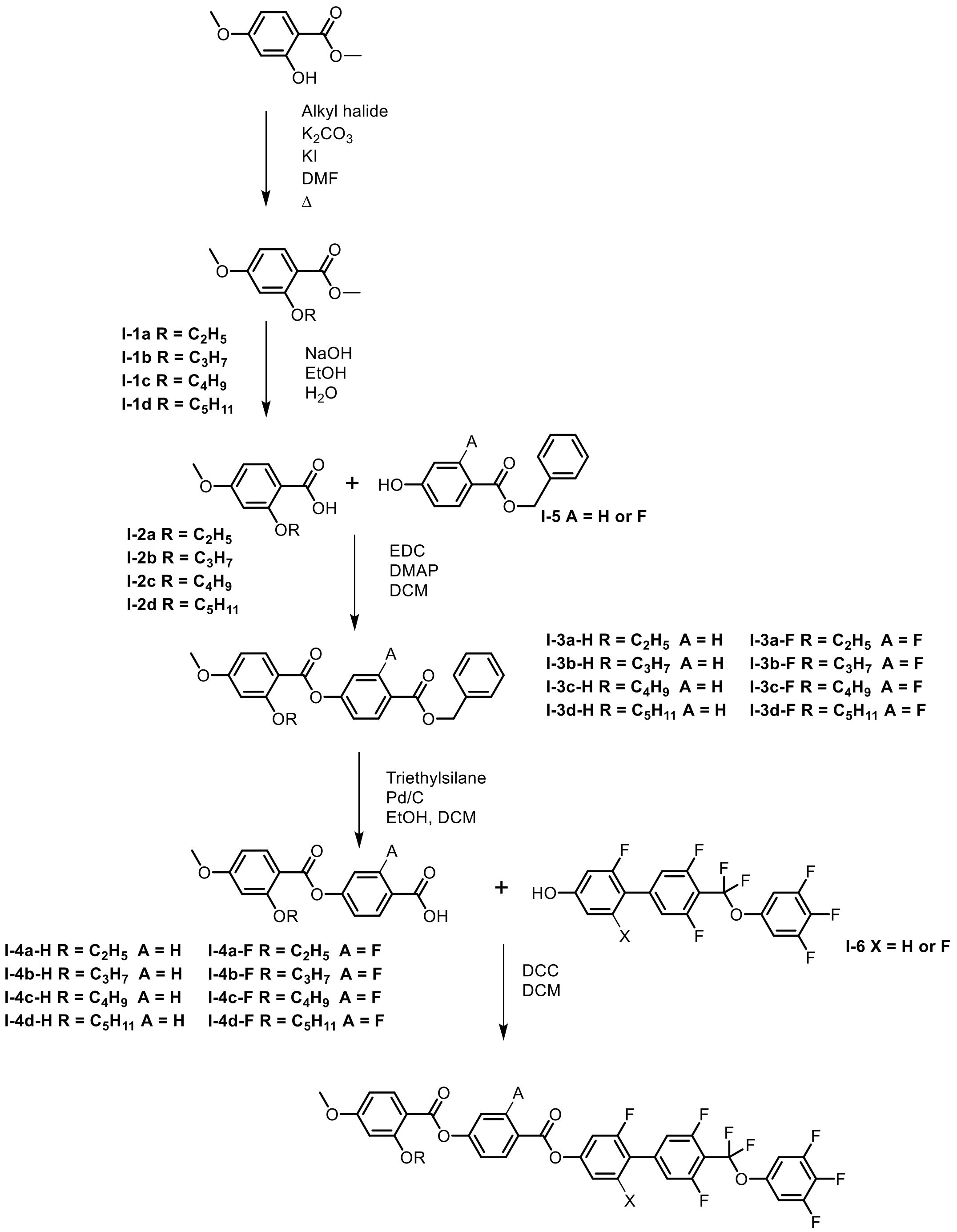

| | A | X | R | | A | X | R | | A | X | R | | A | X | R |
|---|---|---|---|---|---|---|---|---|---|---|---|---|---|---|---|
| E-0-1 | H | H | $C_2H_5$ | P-0-1 | H | H | $C_3H_7$ | B-0-1 | H | H | $C_4H_9$ | Q-0-1 | H | H | $C_5H_{11}$ |
| E-0-2 | H | F | $C_2H_5$ | P-0-2 | H | F | $C_3H_7$ | B-0-2 | H | F | $C_4H_9$ | Q-0-2 | H | F | $C_5H_{11}$ |
| E-1-1 | F | H | $C_2H_5$ | P-1-1 | F | H | $C_3H_7$ | B-1-1 | F | H | $C_3H_7$ | Q-1-1 | F | H | $C_5H_{11}$ |
| E-1-2 | F | F | $C_2H_5$ | P-1-2 | F | F | $C_3H_7$ | B-1-2 | F | F | $C_4H_9$ | Q-1-2 | F | F | $C_5H_{11}$ |

Scheme 1: Synthetic route to the new materials reported here.

The synthesis of intermediate compounds **I-1(a-d), I-2(a-d), I-3(a-d)-H** and **I-4(a-d)-H**, as well as the benzyl protected phenols **I-5** and the fluoroethers **I-6** has been reported previously. [1,2]

**Esterification I-3x-F**

The appropriate benzoic acid and EDC.HCl were dissolved in DCM and stirred for 10 minutes. Benzyl 2-fluoro-4-hydroxybenzoate and 4-dimethylaminopyridine (DMAP) were added and the reaction was left stirring at room temperature overnight. The reaction was washed 3x with water and the solvent removed *in vacuo*. The crude product was recrystallised from ethanol to yield the product as a white solid.

**I-3a-F**

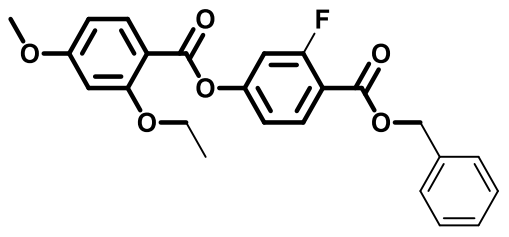

| | | | |
|---|---|---|---|
| 2-fluoro-4-hydroxybenzyl benzoate | 245 mg, | 1.0 mmol, | 1 eq. |
| 2-ethoxy-4-methoxybenzoic acid | 217 mg, | 1.1 mmol, | 1.1 eq. |
| EDC.HCl | 389 mg | 2.0 mmol, | 2 eq. |
| DMAP | 15 mg, | 0.1 mmol, | 0.1 eq. |
| DCM | 10 ml | | |

Recrystallised from ethanol, yield 138 mg.

$^1$H NMR (400 MHz, $CDCl_3$) δ = 8.02 (overlapping multiplets, 2H), 7.49 – 7.30 (m, 5H), 7.09 (d, *J*=9.8, 2H), 6.55 (dd, *J*=8.9, 2.2, 1H), 6.51 (d, *J*=2.2, 1H), 5.39 (s, 2H), 4.12 (q, *J*=7.1, 2H), 3.88 (s, 3H), 1.48 (t, *J*=7.1, 3H).

**I-3b-F**

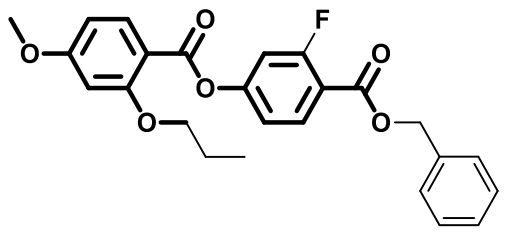

| | | | |
|---|---|---|---|
| 2-fluoro-4-hydroxybenzyl benzoate | 246 mg, | 1.0 mmol, | 1 eq. |
| 2-propoxy-4-methoxybenzoic acid | 239 mg, | 1.1 mmol, | 1.1 eq. |
| EDC.HCl | 387 mg | 2.0 mmol, | 2 eq. |
| DMAP | 12 mg, | 0.1 mmol, | 0.1 eq. |
| DCM | 10 ml | | |

Recrystalised from ethanol, yield 145 mg.

$^1$H NMR (400 MHz, $CDCl_3$) δ = 8.02 (overlapping multiplets, 2H), 7.49 – 7.32 (m, 5H), 7.08 (d, *J*=10.0, 2H), 6.54 (d, *J*=9.8, 1H), 6.50 (s, 1H), 5.39 (s, 2H), 4.07 – 3.97 (m, 2H), 3.88 (s, 3H), 1.87 (q, *J*=7.3, 2H), 1.06 (t, *J*=7.3, 3H).

**I-3c-F**

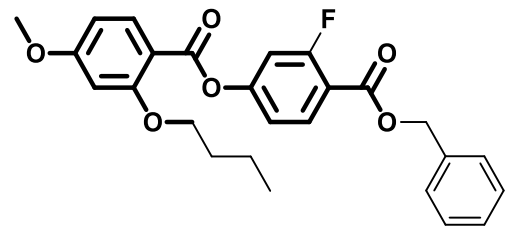

| | | | |
|---|---|---|---|
| 2-fluoro-4-hydroxybenzyl benzoate | 999 mg, | 4.0 mmol, | 1 eq. |
| 2-propoxy-4-methoxybenzoic acid | 1.008 g, | 4.5 mmol, | 1.1 eq. |
| EDC.HCl | 1.604 g, | 8.4 mmol, | 2.1 eq. |
| DMAP | 53 mg, | 0.4 mmol, | 0.1 eq. |
| DCM | 30 ml | | |

Recrystalised from ethanol, yield 954 mg

$^1$H NMR (400 MHz, $CDCl_3$) δ = 8.09 – 7.98 (m, 2H), 7.50 – 7.30 (m, 5H), 7.12 – 7.03 (m, 2H), 6.54 (dd, *J*=8.8, 2.3, 1H), 6.50 (d, *J*=2.3, 1H), 5.39 (s, 2H), 4.05 (t, *J*=6.4, 2H), 3.90 – 3.86 (m, 3H), 1.82 (dq, *J*=8.2, 6.4, 2H), 1.59 – 1.45 (m, 2H), 0.94 (t, *J*=7.4, 3H).

**I-3d-F**

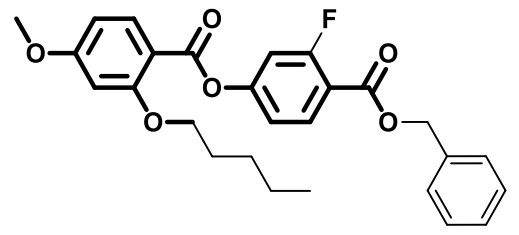

| | | | |
|---|---|---|---|
| 2-fluoro-4-hydroxybenzyl benzoate | 335 mg, | 1.4 mmol, | 1 eq. |
| 2-propoxy-4-methoxybenzoic acid | 379 mg, | 1.6 mmol, | 1.1 eq. |
| EDC.HCl | 569 mg, | 3.0 mmol, | 2.1 eq. |
| DMAP | 18 mg, | 0.4 mmol, | 0.1 eq. |
| DCM | 10 ml | | |

Recrystalised from ethanol, yield 148 mg

$^1$H NMR (400 MHz, $CDCl_3$) δ = 8.07 – 7.98 (m, 2H), 7.50 – 7.30 (m, 5H), 7.12 – 7.04 (m, 2H), 6.54 (dd, *J*=8.8, 2.3, 1H), 6.50 (d, *J*=2.3, 1H), 5.39 (s, 2H), 4.04 (t, *J*=6.5, 2H), 3.88 (s, 3H), 1.90 – 1.78 (m, 2H), 1.53 – 1.41 (m, 2H), 1.35 (h, *J*=7.2, 2H), 0.88 (t, *J*=7.2, 3H).

**Deprotection I-4x**

Under an argon atmosphere triethylsilane was added dropwise to a stirred solution of **I-3x** and 5 % Pd/C in a 1:1 mix of ethanol and DCM. The reaction was stirred for 5 minutes after addition was complete, then filtered through celite and the solvent removed *in vacuo*. The crude product was washed with hexane to yield the product as a white powder.

**I-4a-F**

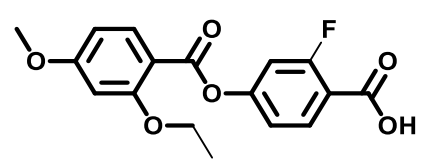

**I-3a-F** 138 mg, 0.3 mmol

5% Pd/C 30 mg

Triethylsilane 0.1 ml

DCM 3 ml

EtOH 3 ml

Recrystalised from ethanol, yield 103 mg.

$^1$H NMR (400 MHz, $CDCl_3$) δ = 8.13 – 8.05 (m, 1H), 8.04 (d, *J*=8.8, 1H), 7.15 (s, 1H), 7.12 (d, *J*=3.6, 1H), 6.58 – 6.49 (m, 2H), 4.13 (q, *J*=7.1, 2H), 3.89 (s, 3H), 1.49 (t, *J*=7.1, 3H).

**I-4b-F**

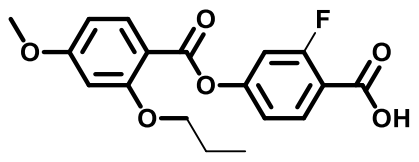

**I-3b-F** 160 mg, 0.36 mmol

5% Pd/C 40 mg

Triethylsilane 0.15 ml

DCM 3 ml

EtOH 3 ml

Crude product washed with hexane, yield 97 mg.

$^1$H NMR (400 MHz, $CDCl_3$) δ = 8.09 (t, *J*=8.5, 1H), 8.03 (d, *J*=8.8, 1H), 7.17 – 7.10 (m, 2H), 6.56 (dd, *J*=8.8, 2.3, 1H), 6.51 (d, *J*=2.3, 1H), 4.02 (t, *J*=6.4, 2H), 3.89 (s, 3H), 1.88 (h, *J*=7.0, 2H), 1.07 (t, *J*=7.4, 3H).

**I-4c-F**

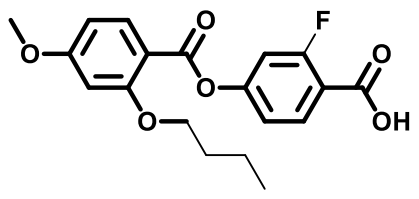

**I-3c-F** 239 mg, 0.5 mmol

5% Pd/C 50 mg

Triethylsilane 0.25 ml

DCM 3 ml

EtOH 3 ml

Crude product washed with hexane, yield 159 mg.

$^1$H NMR (400 MHz, $CDCl_3$) δ = 8.08 (t, *J*=8.5, 1H), 8.02 (d, *J*=8.8, 1H), 7.15 – 7.07 (m, 2H), 6.54 (dd, *J*=8.8, 2.3, 1H), 6.50 (d, *J*=2.3, 1H), 4.05 (t, *J*=6.4, 2H), 3.88 (s, 3H), 1.88 – 1.77 (m, 2H), 1.57 – 1.47 (m, 2H), 0.95 (t, *J*=7.4, 3H).

**I-4d-F**

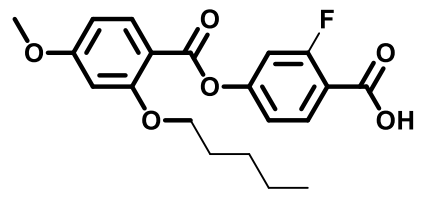

| | |
|---|---|
| **I-3d-F** | 210 mg, 0.45 mmol |
| 5% Pd/C | 50 mg |
| Triethylsilane | 0.2 ml |
| DCM | 3 ml |
| EtOH | 3 ml |

Crude product washed with hexane, yield 137 mg.

$^1$H NMR (400 MHz, $CDCl_3$) δ = 8.08 (t, *J*=8.5, 1H), 8.02 (d, *J*=8.8, 1H), 7.16 – 7.08 (m, 2H), 6.55 (d, *J*=9.5, 1H), 6.51 (s, 1H), 4.05 (t, *J*=6.6, 2H), 3.89 (s, 3H), 1.91 – 1.81 (m, 2H), 1.46 (q, *J*=7.7, 2H), 1.36 (q, *J*=7.3, 2H), 0.89 (t, *J*=7.2, 3H).

**Esterification**

Method A

Intermediate acid **I-3x** and EDC.HCl were dissolved in 10 ml DCM and stirred for 5 minutes. The appropriate phenol and DMAP were added, and the mixture stirred overnight and monitored by TLC. The reaction mixture was washed 3 times with water and the organic layer dried over magnesium sulfate and removed *in vacuo*. The crude solid was dissolved in the minimum amount of chloroform and precipitated with hexane or ethanol

Method B

Intermediate acid **I-4x** and DCC were dissolved in 10 ml DCM in an icebath and stirred for 30 minutes. The appropriate phenol was added, and the mixture stirred overnight and monitored by TLC. The reaction mixture was filtered and the solvent removed *in vacuo*. The crude solid was dissolved in the minimum amount of chloroform and precipitated with hexane or ethanol.

| | Method | EDC.HCl | DMAP | DCC | Acid | Phenol | Antisolvent | Isolated Yield |
|---|---|---|---|---|---|---|---|---|
| **E-0-1** | A | 67 mg, 0.3 mmol | 3 mg, 0.02 mmol | - | 76 mg, 0.2 mmol | 93 mg, 0.2 mmol | Hexane | 60 mg |
| **E-0-2** | B | | | 55mg 0.3 mmol | 100 mg, 0.3 mmol | 104 mg, 0.2 mmol | Ethanol | 52 mg |
| **E-1-1** | A | 94 mg, 0.5 mmol | 6 mg, 0.05 mmol | - | 100 mg, 0.3 mmol | 111 mg, 0.3 mmol | Ethanol | 43 mg |
| **E-1-2** | B | - | - | 31 mg, 0.2 mmol | 60 mg, 0.2 mmol | 59 mg, 0.1 mmol | Ethanol | 19 mg |
| **P-0-1** | A | 191 mg, 1 mmol | 7 mg, 0.05 mmol | - | 180 mg, 0.5 mmol | 191 mg, 0.4 mmol | Ethanol | 152 mg |
| **P-0-2** | B | - | - | 38 mg, 0.2 mmol | 79 mg, 0.2 mmol | 80 mg, 0.2 mmol | Ethanol | 12 mg |
| **P-1-1** | A | 62 mg, 0.3 mmol | 7 mg, 0.05 mmol | - | 80 mg, 0.2 mmol | 88 mg, 0.2 mmol | Ethanol | 68 mg |
| **P-1-2** | B | - | - | 71 mg, 0.3 mmol | 159 mg, 0.5 mmol | 150 mg, 0.3 mmol | Hexane | 95 mg |
| **B-0-1** | A | 46 mg, 0.2 mmol | 2 mg, 0.02 mmol | - | 46 mg, 0.1 mmol | 57 mg, 0.1 mmol | Hexane | 31 mg |
| **B-0-2** | B | - | - | 68 mg, 0.3 mmol | 99 mg, 0.3 mmol | 92 mg 0.2 mmol | Ethanol | 30 mg |
| **B-1-1** | A | 69 mg, 0.3 mmol | 8 mg, 0.06 mmol | - | 76 mg, 0.2 mmol | 80 mg, 0.2 mmol | Ethanol | 73 mg |
| **B-1-2** | B | - | - | 34 mg, 0.2 mmol | 80 mg, 0.2 mmol | 73 mg, 0.2 mmol | Ethanol | 27 mg |
| **Q-0-1** | A | 65 mg, 0.3 mmol | 3 mg, 0.02 mmol | - | 78 mg, 0.2 mmol | 83 mg, 0.2 mmol | Hexane | 44 mg |
| **Q-0-2** | B | - | - | 43 mg, 0.2 mmol | 80 mg, 0.2 mmol | 68 mg, 0.2 mmol | Ethanol | 85 mg |
| **Q-1-1** | A | 48 mg, 0.2 mmol | 3 mg, 0.02 mmol | - | 62 mg, 0.2 mmol | 69 mg, 0.2 mmol | Ethanol | 47 mg |
| **Q-1-2** | B | - | - | 28 mg, 0.1 mmol | 68 mg, 0.2 mmol | 60 mg, 0.1 mmol | Ethanol | 33 mg |

**E-0-1**

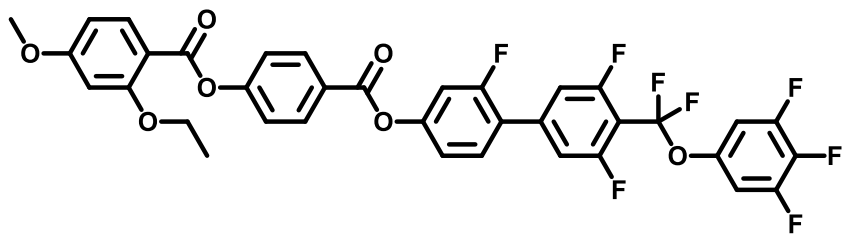

HRMS (ESI) m/z Calculated for $C_{36}H_{22}O_7F_8$:
$[M+H]^+$ theoretical mass: 719.13105, found 719.13037, difference -0.952 ppm.

$^{1}$H NMR (400 MHz, $CDCl_3$) δ = 8.26 (d, *J*=8.6, 2H), 8.08 (d, *J*=8.9, 1H), 7.50 (t, *J*=8.6, 1H), 7.39 (d, *J*=8.6, 2H), 7.27 – 7.14 (m, 4H), 7.04 – 6.96 (m, 2H), 6.57 (dd, *J*=8.9, 2.3, 1H), 6.53 (d, *J*=2.3, 1H), 4.15 (q, *J*=7.0, 2H), 3.89 (s, 3H), 1.50 (t, *J*=7.0, 3H).

$^{19}$F NMR (376 MHz, $CDCl_3$) δ = -61.82 (t, *J*=26.3, 2F), -110.44 (td, *J*=26.3, 11.0, 2F), -113.87 (t, *J*=9.8, 1F), -132.45 (dd, *J*=21.0, 8.0, 2F), -163.10 (tt, *J*=21.0, 5.8, 1F).

$^{13}$C NMR (101 MHz, $CDCl_3$) δ 165.22, 163.99, 163.08, 161.85, 161.24, 161.17, 160.80, 158.99, 158.93, 158.70, 158.64, 158.62, 158.61, 158.29, 155.92, 152.40, 152.29, 149.67, 144.53, 140.79, 139.71, 134.57, 131.87, 130.58, 130.54, 125.75, 122.44, 118.52, 118.48, 113.25, 113.23, 113.19, 113.01, 112.98, 112.96, 111.08, 110.83, 110.64, 107.60, 107.54, 107.41, 107.36, 104.99, 99.85, 64.64, 55.60, 14.66.

**E-0-2**

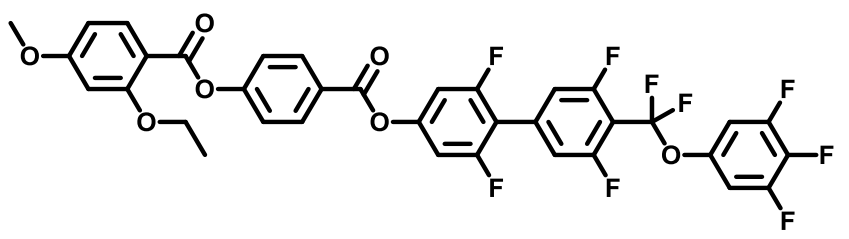

HRMS (ESI) m/z Calculated for $C_{36}H_{21}O_7F_9$:
$[M+H]^+$ theoretical mass: 737.12163, found 737.12174, difference 0.145 ppm.

$^{1}$H NMR (400 MHz, CDcl$_3$) δ = 8.25 (d, *J*=8.7, 2H), 8.08 (d, *J*=8.7, 1H), 7.40 (d, *J*=8.7, 2H), 7.18 (d, *J*=10.7, 2H), 7.01 (t, *J*=7.7, 4H), 6.58 (d, *J*=8.7, 1H), 6.53 (d, *J*=2.0, 1H), 4.20 – 4.10 (m, 2H), 3.90 (s, 3H), 1.54 – 1.46 (m, 3H).

$^{19}$F NMR (376 MHz, $CDCl_3$) δ = -61.99 (t, *J*=26.5, 2F), -110.63 (td, *J*=26.5, 10.7, 2F), -112.06 (d, *J*=8.8, 2F), -132.43 (dd, *J*=20.9, 8.3, 2F), -163.06 (tt, *J*=20.9, 5.9, 1F).

$^{13}$C NMR (101 MHz, cdcl$_3$) δ 165.24, 163.60, 163.02, 161.87, 160.99, 160.92, 158.49, 158.41, 156.10, 134.58, 131.93, 125.34, 122.52, 114.88, 114.64, 110.59, 107.65, 107.41, 106.97, 106.68, 105.00, 99.85, 64.64, 55.60, 14.66.

**E-1-1**

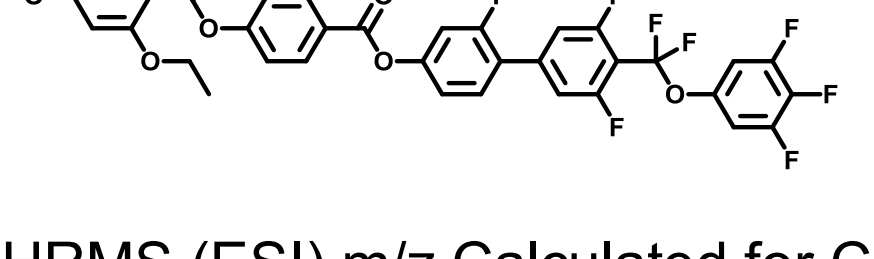

HRMS (ESI) m/z Calculated for $C_{36}H_{21}O_7F_9$:
$[M+H]^+$ theoretical mass: 737.12163, found 737.12200, difference 0.498 ppm.

$^{1}H$ NMR (400 MHz, $CDCl_3$) δ = 8.17 (t, *J*=8.3, 1H), 8.06 (d, *J*=8.8, 1H), 7.50 (t, *J*=8.5, 1H), 7.29 – 7.16 (m, 6H), 7.00 (t, *J*=7.0, 2H), 6.57 (d, *J*=8.8, 1H), 6.53 (d, *J*=2.5, 1H), 4.15 (q, *J*=7.1, 2H), 3.90 (s, 3H), 1.50 (t, *J*=7.1, 3H).

$^{19}F$ NMR (376 MHz, $CDCl_3$) δ = -61.83 (t, *J*=26.2, 2F), -104.30 – -104.42 (m, 1F), -110.42 (td, *J*=26.2, 10.9, 2F), -113.82 (t, *J*=9.8, 1F), -132.44 (dd, *J*=20.7, 8.3, 2F), -163.10 (tt, *J*=20.7, 5.9, 1F).

$^{13}C$ NMR (101 MHz, $CDCl_3$) δ 165.44, 164.23, 162.49, 161.98, 161.67, 161.63, 161.60, 161.23, 161.18, 160.77, 158.26, 156.84, 156.72, 152.32, 152.21, 152.02, 151.91, 149.77, 149.67, 140.93, 140.73, 140.63, 139.71, 137.31, 134.64, 133.35, 130.57, 130.53, 123.40, 120.14, 118.48, 118.45, 118.25, 118.21, 117.44, 114.25, 114.16, 113.23, 112.99, 111.69, 111.44, 111.06, 110.80, 110.11, 107.60, 107.53, 107.41, 107.36, 105.07, 99.80, 64.63, 55.62, 14.65.

**E-1-2**

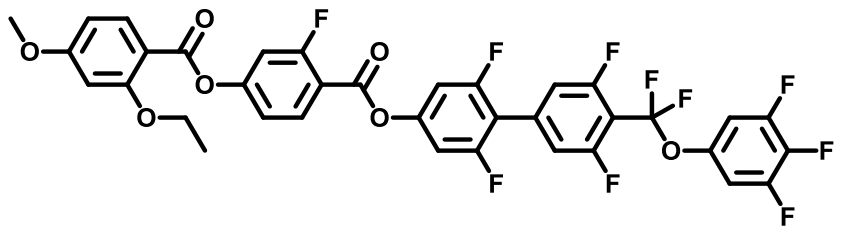

HRMS (ESI) m/z Calculated for $C_{36}H_{20}O_7F_{10}$:
$[M+H]^+$ theoretical mass: 755.11221, found 755.11244, difference 0.303 ppm.

$^{1}H$ NMR (400 MHz, $CDCl_3$) δ = 8.15 (t, *J*=8.2, 1H), 8.05 (d, *J*=8.8, 1H), 7.28 – 7.14 (m, 4H), 7.07 – 6.97 (m, 4H), 6.57 (dd, *J*=8.8, 2.3, 1H), 6.52 (d, *J*=2.7, 1H), 4.15 (q, *J*=7.0, 2H), 3.90 (s, 3H), 1.50 (t, *J*=7.0, 3H).

$^{19}F$ NMR (376 MHz, $CDCl_3$) δ = -61.99 (t, *J*=26.4, 2F), -104.10 (dd, *J*=11.2, 8.1, 1F), -110.61 (td, *J*=26.4, 10.6, 2F), -111.99 (d, *J*=8.9, 2F), -132.44 (dd, *J*=20.7, 7.1, 2F), -163.06 (tt, *J*=20.7, 5.8, 1F).

$^{13}C$ NMR (101 MHz, $CDCl_3$) δ 165.47, 164.26, 162.44, 162.00, 161.63, 161.27, 161.23, 160.97, 160.89, 158.47, 158.39, 157.04, 156.93, 152.32, 152.29, 152.23, 152.17, 151.80, 151.66, 151.52, 149.82, 149.77, 144.49, 139.87, 139.74, 134.65, 134.43, 134.36, 134.20, 133.37, 119.99, 118.32, 118.29, 114.88, 114.63, 113.86, 113.77, 113.02, 111.74, 111.49, 110.05, 107.64, 107.55, 107.47, 107.40, 106.95, 106.92, 106.87, 106.68, 106.66, 106.65, 105.08, 99.80, , 64.63, 55.62, 14.64.

**P-0-1**

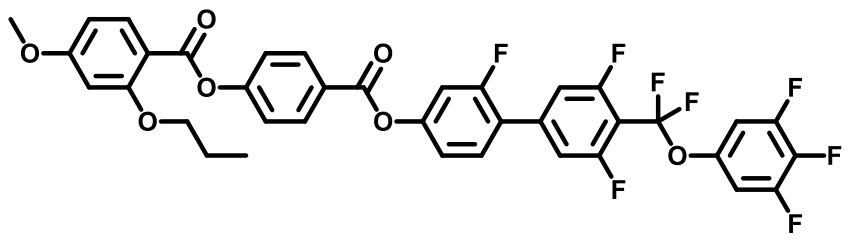

HRMS (ESI) m/z Calculated for $C_{37}H_{24}O_7F_8$:
$[M+H]^+$ theoretical mass: 733.14670, found 733.14596, difference -1.016 ppm.

$^{1}H$ NMR (400 MHz, $CDCl_3$) δ = 8.27 (d, *J*=8.0, 2H), 8.07 (d, *J*=8.7, 1H), 7.50 (t, *J*=8.6, 1H), 7.39 (d, *J*=8.0, 2H), 7.23 (d, *J*=10.8, 2H), 7.21 – 7.13 (m, 2H), 7.04 – 6.96 (m, 2H), 6.57 (d, *J*=8.7, 1H), 6.53 (s, 1H), 4.03 (t, *J*=6.4, 2H), 3.89 (s, 3H), 1.89 (h, *J*=7.1, 2H), 1.08 (t, *J*=7.4, 3H).

$^{19}F$ NMR (376 MHz, $CDCl_3$) δ = -61.82 (t, *J*=26.3, 2F), -110.44 (td, *J*=26.3, 10.9, 2F), -113.87 (t, *J*=9.9, 1F), -132.44 (dd, *J*=20.7, 8.3, 2F), -163.11 (tt, *J*=20.7, 5.8, 1F).

$^{13}C$ NMR (101 MHz, $CDCl_3$) δ 165.22, 163.99, 163.24, 161.93, 161.24, 161.18, 160.80, 158.67, 158.59, 158.29, 155.96, 152.41, 152.30, 152.22, 152.16, 149.83, 149.77, 149.72, 149.67, 144.58, 140.79, 139.70, 137.22, 134.64, 131.90, 130.58, 130.54, 125.74, 123.27, 123.14, 122.79, 122.43, 120.13, 120.12, 118.52, 118.48, 117.45, 113.26, 113.22, 113.01, 112.98, 112.95, 111.08, 110.82, 110.62, 107.60, 107.54, 107.43, 107.36, 104.95, 99.67, 70.40, 55.59, 22.51, 10.62.

**P-0-2**

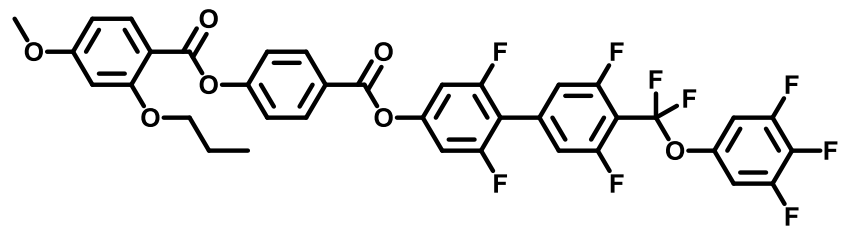

HRMS (ESI) m/z Calculated for $C_{37}H_{23}O_7F_9$:
$[M+H]^+$ theoretical mass: 751.13728, found 751.13738, difference 0.129 ppm.

$^1H$ NMR (400 MHz, $CDCl_3$) δ = 8.24 (d, *J*=8.7, 2H), 8.06 (d, *J*=8.8, 1H), 7.38 (d, *J*=8.7, 2H), 7.16 (d, *J*=10.7, 2H), 7.04 – 6.96 (m, 4H), 6.56 (dd, *J*=8.8, 2.3, 1H), 6.52 (d, *J*=2.3, 1H), 4.02 (t, *J*=6.5, 2H), 3.88 (s, 3H), 1.94 – 1.81 (m, 2H), 1.07 (t, *J*=7.4, 3H).

$^{19}F$ NMR (376 MHz, $CDCll_3$) δ = -61.99 (t, *J*=26.5, 2F), -110.63 (td, *J*=26.5, 10.7, 2F), -112.06 (d, *J*=8.7, 2F), -132.43 (dd, *J*=20.9, 7.9, 2F), -163.06 (tt, *J*=20.9, 5.8, 1F).

$^{13}C$ NMR (101 MHz, $CDCl_3$) δ 165.25, 163.60, 163.18, 161.95, 160.99, 160.91, 158.49, 158.41, 156.13, 134.65, 131.95, 125.34, 122.51, 114.87, 114.63, 110.56, 107.65, 107.41, 106.97, 106.94, 106.67, 104.96, 99.67,.40, 55.60, 22.51, 10.62.

**P-1-1**

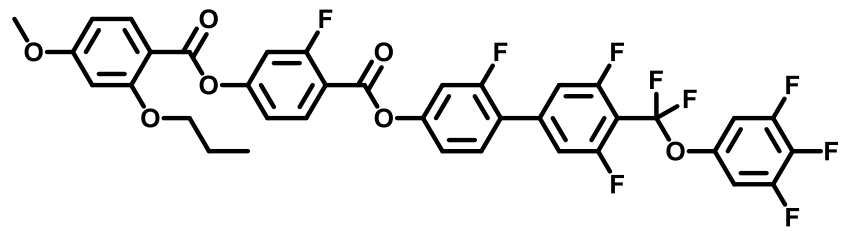

HRMS (ESI) m/z Calculated for $C_{37}H_{23}O_7F_9$:
$[M+H]^+$ theoretical mass: 751.13728, found 751.13811, difference 1.101 ppm.

$^1H$ NMR (400 MHz, $CDCl_3$) δ = 8.17 (t, *J*=8.5, 1H), 8.05 (d, *J*=8.8, 1H), 7.50 (t, *J*=8.6, 1H), 7.26 – 7.14 (m, 6H), 7.04 – 6.96 (m, 2H), 6.56 (dd, *J*=8.5, 2.3, 1H), 6.52 (d, *J*=2.3, 1H), 4.03 (t, *J*=6.4, 2H), 3.90 (s, 3H), 1.89 (h, *J*=7.1, 2H), 1.08 (t, *J*=7.4, 3H).

$^{19}F$ NMR (376 MHz, $CDCl_3$) δ = -61.83 (t, *J*=26.4, 2F), -104.34 (dd, *J*=11.3, 8.2, 1F), -110.42 (td, *J*=26.4, 10.7, 2F), -113.82 (t, *J*=9.8, 1F), -132.36 – -132.53 (m, 2F), -163.10 (tt, *J*=20.9, 5.9, 1F).

$^{13}C$ NMR (101 MHz, $CDCl_3$) δ 165.44, 164.24, 162.64, 162.07, 161.61, 160.77, 158.26, 156.86, 156.75, 152.27, 152.02, 151.76, 149.76, 149.67, 144.59, 140.74, 134.72, 133.39, 130.57, 130.54, 118.48, 118.44, 118.25, 114.26, 113.23, 112.99, 111.69, 111.44, 111.06, 110.80, 110.08, 107.60, 107.36, 105.04, 99.63, 70.40, 55.62, 22.49, 10.62.

**P-1-2**

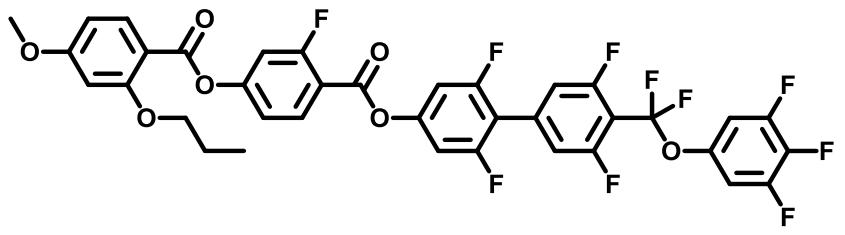

HRMS (ESI) m/z Calculated for $C_{37}H_{22}O_7F_{10}$:
$[M+H]^+$ theoretical mass: 769.12786, found 769.12802, difference 0.207 ppm.

$^1$H NMR (400 MHz, $CDCl_3$) δ = 8.15 (t, *J*=8.3, 1H), 8.05 (d, *J*=8.8, 1H), 7.23 – 7.14 (m, 4H), 7.02 (p, *J*=6.5, 4H), 6.56 (dd, *J*=8.8, 2.0, 1H), 6.52 (d, *J*=2.0, 1H), 4.03 (d, *J*=7.2, 2H), 3.89 (s, 3H), 1.89 (h, *J*=7.1, 2H), 1.08 (t, *J*=7.5, 3H).

$^{19}$F NMR (376 MHz, $CDCl_3$) δ = -61.99 (t, *J*=26.5, 2F), -104.08 (dd, *J*=11.3, 8.0, 1F), -110.61 (td, *J*=26.5, 10.8, 2F), -111.99 (d, *J*=8.8, 2F), -132.44 (dd, *J*=20.9, 8.2, 2F), -163.06 (tt, *J*=20.9, 5.9, 1F).

$^{13}$C NMR (101 MHz, $CDCl_3$) δ 165.47, 164.27, 163.33, 162.58, 162.09, 161.65, 161.27, 161.23, 161.03, 160.97, 160.89, 158.47, 158.39, 157.08, 156.96, 152.34, 152.25, 152.21, 152.16, 151.66, 149.83, 149.78, 149.67, 139.71, 134.72, 134.46, 134.36, 134.30, 133.40, 127.18, 120.01, 118.32, 118.29, 114.88, 114.63, 113.86, 113.77, 111.74, 111.49, 110.02, 107.64, 107.57, 107.46, 107.41, 106.95, 106.92, 106.67, 106.65, 105.05, 99.62, 70.39, 55.62, 22.49, 10.62.

**B-0-1**

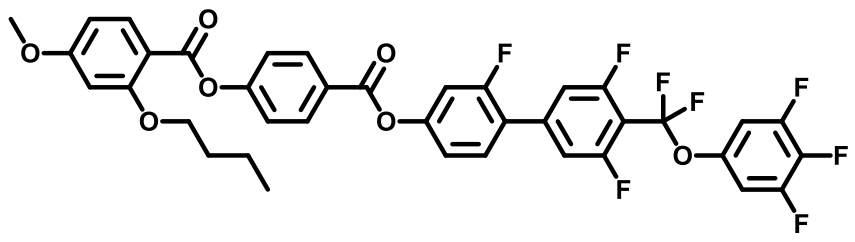

HRMS (ESI) m/z Calculated for $C_{38}H_{26}O_7F_8$:
$[M+H]^+$ theoretical mass: 747.162350, found 747.16190, difference -0.609 ppm.

$^1$H NMR (400 MHz, $CDCl_3$) δ = 8.27 (d, *J*=8.3, 2H), 8.07 (d, *J*=8.7, 1H), 7.50 (t, *J*=8.6, 1H), 7.38 (d, *J*=8.3, 2H), 7.28 – 7.13 (m, 4H), 7.04 – 6.96 (m, 2H), 6.57 (d, *J*=8.7, 1H), 6.53 (s, 1H), 4.07 (t, *J*=6.4, 2H), 3.90 (s, 3H), 1.85 (p, *J*=6.7, 2H), 1.61 – 1.47 (m, 2H), 0.96 (t, *J*=7.3, 3H).

$^{19}$F NMR (376 MHz, $CDCl_3$) δ = -61.82 (t, *J*=26.3, 2F), -110.43 (td, *J*=26.3, 10.9, 2F), -113.86 (t, *J*=9.8, 1F), -132.44 (dq, *J*=20.6, 8.0, 2F), -163.10 (tt, *J*=20.6, 5.8, 1F).

$^{13}$C NMR (101 MHz, $CDCl_3$) δ 165.21, 163.99, 163.20, 161.94, 161.23, 161.18, 160.79, 158.68, 158.29, 158.00, 155.96, 152.30, 134.64, 131.89, 130.54, 125.74, 122.43, 118.48, 113.23, 112.99, 111.08, 110.83, 110.63, 107.60, 107.36, 104.91, 99.67, 68.59, 55.60, 31.13, 19.20, 13.80.

**B-0-2**

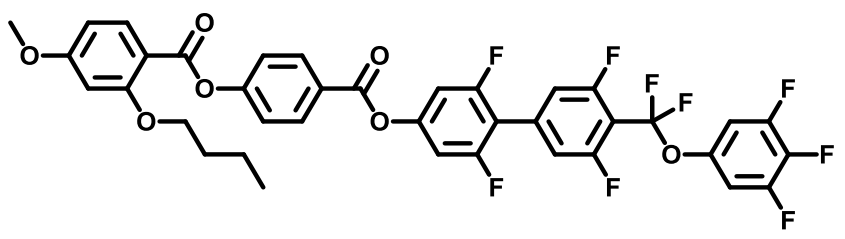

HRMS (ESI) m/z Calculated for $C_{38}H_{25}O_7F_9$:
$[M+H]^+$ theoretical mass: 765.15293, found 765.15322, difference 0.375 ppm.

$^1$H NMR (400 MHz, $CDCl_3$) δ = 8.25 (d, *J*=8.3, 2H), 8.07 (d, *J*=8.7, 1H), 7.39 (d, *J*=8.3, 2H), 7.18 (d, *J*=10.7, 2H), 7.01 (t, *J*=7.6, 4H), 6.57 (d, *J*=9.0, 1H), 6.53 (s, 1H), 4.07 (t, *J*=6.5, 2H), 3.90 (s, 3H), 1.84 (p, *J*=6.9, 2H), 1.62 – 1.47 (m, 2H), 0.96 (t, *J*=7.4, 3H).

$^{19}$F NMR (376 MHz, $CDCl_3$) δ = -61.99 (t, *J*=26.3, 2F), -110.63 (td, *J*=26.3, 10.8, 2F), -112.05 (d, *J*=8.9, 2F), -132.43 (dd, *J*=21.0, 8.3, 2F), -163.06 (tt, *J*=21.0, 6.0, 1F).

$^{13}$C NMR (101 MHz, $CDCl_3$) δ 165.24, 163.60, 163.15, 161.96, 161.00, 160.91, 158.49, 158.47, 158.37, 156.13, 152.20, 152.06, 149.77, 149.63, 134.65, 131.95, 125.33, 122.51, 114.88, 114.63, 110.56, 107.65, 107.41, 106.97, 106.68, 104.93, 99.66, 68.59, 55.60, 31.12, 19.20, 13.80.

**B-1-1**

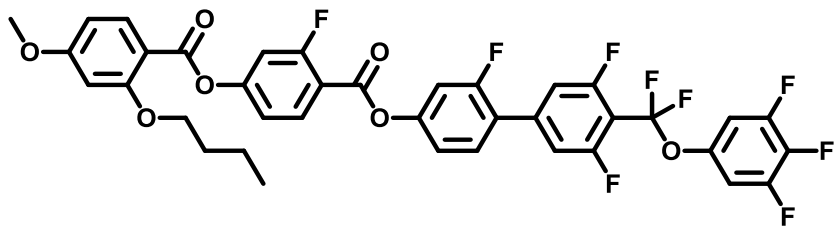

HRMS (ESI) m/z Calculated for $C_{38}H_{25}O_7F_9$:
$[M+H]^+$ theoretical mass: 765.15293, found 765.15317, difference 0.310 ppm.

$^1$H NMR (400 MHz, $CDCl_3$) δ = 8.23 (t, *J*=8.4, 1H), 8.11 (d, *J*=8.8, 1H), 7.56 (t, *J*=8.5, 1H), 7.34 – 7.21 (m, 6H), 7.06 (t, *J*=7.0, 2H), 6.62 (d, *J*=8.8, 1H), 6.59 (d, *J*=2.2, 1H), 4.13 (t, *J*=6.4, 2H), 3.96 (s, 3H), 1.91 (p, *J*=6.7, 2H), 1.67 – 1.54 (m, 2H), 1.03 (t, *J*=7.4, 3H).

$^{19}$F NMR (376 MHz, $CDCl_3$) δ = -61.82 (t, *J*=26.3, 2F), -104.35 (dd, *J*=11.4, 8.1, 1F), -110.42 (td, *J*=26.3, 10.8, 2F), -113.82 (t, *J*=9.9, 1F), -132.45 (dd, *J*=20.7, 8.2, 2F), -163.11 (tt, *J*=20.7, 5.8, 1F).

$^{13}$C NMR (101 MHz, $CDCl_3$) δ 165.44, 164.24, 162.60, 162.08, 161.63, 161.24, 160.77, 158.61, 156.87, 152.03, 151.92, 149.83, 144.57, 140.74, 139.95, 137.19, 134.71, 133.38, 130.57, 130.53, 123.40, 120.08, 118.48, 118.44, 118.23, 118.20, 117.37, 114.25, 114.16, 113.26, 113.23, 113.20, 113.01, 112.99, 112.95, 111.68, 111.43, 111.06, 110.80, 110.09, 107.60, 107.54, 107.42, 107.36, 105.01, 99.63, 68.59, 55.62, 31.11, 19.20, 13.79.

**B-1-2**

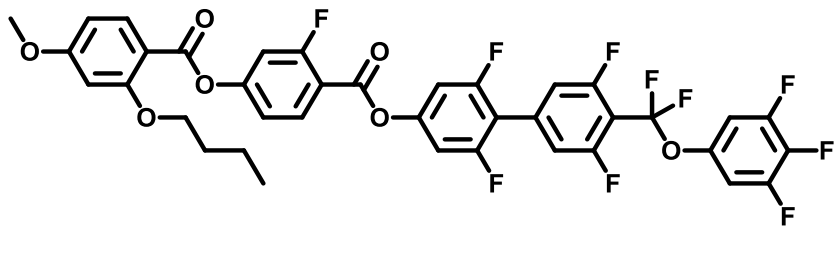

HRMS (ESI) m/z Calculated for $C_{38}H_{24}O_7F_{10}$:
$[M+H]^+$ theoretical mass: 783.14351, found 783.14392, difference 0.522 ppm.

$^1$H NMR (400 MHz, $CDCl_3$) δ = 8.14 (t, *J*=9.0, 1H), 8.04 (d, *J*=8.8, 1H), 7.22 – 7.13 (m, 4H), 7.07 – 6.96 (m, 4H), 6.56 (dd, *J*=8.9, 2.3, 1H), 6.52 (d, *J*=2.3, 1H), 4.06 (t, *J*=6.4, 2H), 3.89 (s, 3H), 1.90 – 1.78 (m, 2H), 1.60 – 1.47 (m, 2H), 0.96 (t, *J*=7.4, 3H).

$^{19}F$ NMR (376 MHz, $CDCl_3$) δ = -61.99 (t, *J*=26.5, 2F), -104.10 (t, *J*=9.8, 1F), -110.62 (td, *J*=26.5, 10.8, 2F), -111.99 (d, *J*=8.9, 2F), -132.44 (dd, *J*=20.8, 8.2, 2F), -163.07 (tt, *J*=20.8, 6.2, 1F).

$^{13}C$ NMR (101 MHz, $CDCl_3$) δ 165.46, 164.27, 162.64, 162.06, 161.64, 161.28, 161.24, 161.02, 160.90, 158.47, 158.37, 157.08, 156.96, 155.00, 152.25, 152.18, 134.73, 133.38, 118.31, 118.28, 114.91, 114.88, 114.66, 114.64, 113.86, 111.73, 111.48, 110.05, 107.65, 107.57, 107.45, 107.41, 106.95, 106.94, 106.68, 106.65, 105.03, 99.63, 68.93, 55.62, 28.77, 28.13, 22.37, 14.00.

**Q-0-1**

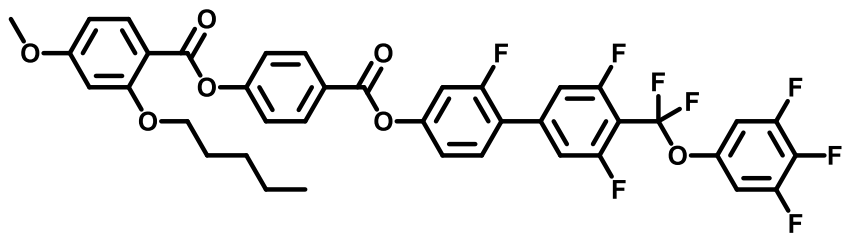

HRMS (ESI) m/z Calculated for $C_{39}H_{28}O_7F_8$:
$[M+H]^+$ theoretical mass: 761.17800, found 761.17749, difference -0.677 ppm.

$^1H$ NMR (400 MHz, $CDCl_3$) δ = 8.27 (d, *J*=8.3, 2H), 8.07 (d, *J*=8.8, 1H), 7.50 (t, *J*=8.5, 1H), 7.38 (d, *J*=8.3, 2H), 7.28 – 7.14 (m, 4H), 7.00 (t, *J*=7.0, 2H), 6.57 (d, *J*=8.7, 1H), 6.53 (s, 1H), 4.06 (t, *J*=6.6, 2H), 3.89 (s, 3H), 1.86 (p, *J*=6.9, 2H), 1.53 – 1.43 (m, 2H), 1.36 (h, *J*=7.5, 2H), 0.89 (t, *J*=7.3, 3H).

$^{19}F$ NMR (376 MHz, $CDCl_3$) δ = -61.82 (t, *J*=26.3, 2F), -110.43 (td, *J*=26.3, 10.8, 2F), -113.87 (t, *J*=10.0, 1F), -132.44 (dd, *J*=20.9, 8.2, 2F), -163.10 (tt, *J*=20.9, 5.7, 1F).

$^{13}C$ NMR (101 MHz, $CDCl_3$) δ 165.21, 163.99, 163.28, 161.90, 160.79, 158.65, 158.29, 155.96, 152.30, 149.70, 134.65, 131.88, 130.54, 125.74, 122.43, 118.48, 113.23, 112.99, 111.08, 110.83, 110.65, 107.60, 107.37, 104.93, 99.68, 68.93, 55.60, 28.79, 28.13, 22.39, 14.00.

**Q-0-2**

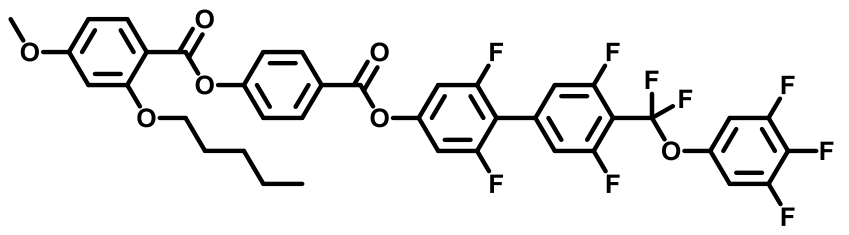

HRMS (ESI) m/z Calculated for $C_{39}H_{27}O_7F_9$:
$[M+H]^+$ theoretical mass: 779.16858, found 779.16885, difference 0.343 ppm.

$^1H$ NMR (400 MHz, $CDCl_3$) δ = 8.25 (d, *J*=7.8, 2H), 8.06 (d, *J*=8.7, 1H), 7.39 (d, *J*=7.8, 2H), 7.18 (d, *J*=10.7, 2H), 7.05 – 6.97 (m, 4H), 6.57 (dd, *J*=8.7, 2.6, 1H), 6.52 (d, *J*=2.6, 1H), 4.06 (t, *J*=6.6, 2H), 3.89 (s, 3H), 1.86 (p, *J*=7.2, 2H), 1.49 (p, *J*=7.4, 2H), 1.37 (p, *J*=7.5, 2H), 0.89 (t, *J*=6.9, 3H).

$^{19}F$ NMR (376 MHz, $CDCl_3$) δ = -61.99 (t, *J*=26.3, 1F), -110.63 (dt, *J*=26.3, 10.8, 1F), -112.06 (d, *J*=8.9, 1F), -132.43 (dd, *J*=20.9, 8.3, 1F), -163.06 (tt, *J*=20.9, 5.9, 1F).

$^{13}C$ NMR (101 MHz, $cdcl_3$) δ 165.24, 163.60, 163.23, 161.92, 160.98, 160.91, 158.49, 158.41, 156.13, 152.22, 152.06, 151.93, 149.86, 149.73, 142.53, 141.40, 139.73, 134.66,

134.43, 134.39, 131.94, 125.34, 122.51, 114.87, 114.63, 110.57, 107.65, 107.41, 106.97, 106.68, 104.94, 99.67, , 68.93, 55.60, 28.79, 28.13, 22.39, 14.00.

**Q-1-1**

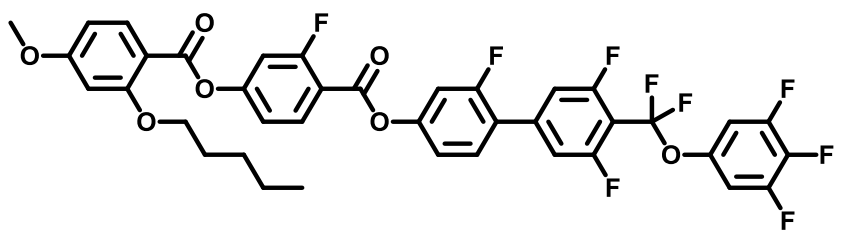

HRMS (ESI) m/z Calculated for $C_{39}H_{27}O_7F_9$:
$[M+H]^+$ theoretical mass: 779.16858, found 779.16901, difference 0.548 ppm.

$^1$H NMR (400 MHz, $CDCl_3$) δ = 8.17 (t, *J*=8.3, 1H), 8.04 (d, *J*=8.7, 1H), 7.50 (t, *J*=8.5, 1H), 7.28 – 7.14 (m, 6H), 7.00 (t, *J*=7.1, 2H), 6.56 (d, *J*=8.7, 1H), 6.52 (s, 1H), 4.06 (t, *J*=6.4, 2H), 3.89 (s, 3H), 1.87 (p, *J*=6.9, 2H), 1.53 – 1.43 (m, 2H), 1.43 – 1.30 (m, 2H), 0.90 (t, *J*=7.4, 3H).

$^{19}$F NMR (376 MHz, $CDCl_3$) δ = -61.82 (t, *J*=26.3, 2F), -104.36 (dd, *J*=11.4, 8.1, 1F), -110.42 (td, *J*=26.3, 10.8, 2F), -113.82 (t, *J*=9.8, 1F), -132.44 (dd, *J*=20.7, 8.2, 2F), -163.10 (tt, *J*=20.7, 5.8, 1F).

$^{13}$C NMR (101 MHz, $CDCl_3$) δ 165.43, 162.68, 162.04, 161.61, 134.72, 133.37, 130.57, 130.53, 118.48, 118.45, 118.20, 114.24, 114.15, 113.27, 113.23, 113.00, 112.99, 111.96, 111.68, 111.43, 111.06, 110.80, 110.12, 107.60, 107.52, 107.42, 107.36, 105.02, 99.64, 68.93, 55.62, 28.78, 28.13, 22.38, 14.00.

**Q-1-2**

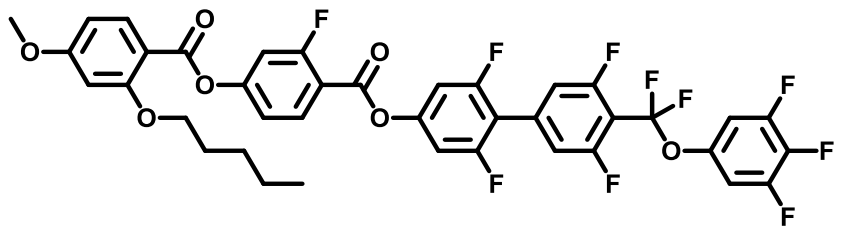

HRMS (ESI) m/z Calculated for $C_{39}H_{26}O_7F_{10}$:
$[M+H]^+$ theoretical mass: 797.15916, found 797.15962, difference 0.575 ppm.

$^1$H NMR (400 MHz, $CDCl_3$) δ = 8.14 (t, *J*=8.8, 1H), 8.03 (d, *J*=8.8, 1H), 7.23 – 7.13 (m, 4H), 7.06 – 6.96 (m, 4H), 6.56 (dd, *J*=8.8, 2.3, 1H), 6.51 (d, *J*=2.3, 1H), 4.05 (t, *J*=6.5, 2H), 3.89 (s, 3H), 1.91 – 1.80 (m, 2H), 1.52 – 1.42 (m, 2H), 1.41 – 1.31 (m, 2H), 0.90 (t, *J*=7.2, 3H).

$^{19}$F NMR (376 MHz, $CDCl_3$) δ = -61.99 (t, *J*=26.5, 2F), -104.10 (t, *J*=9.8, 1F), -110.62 (td, *J*=26.5, 10.8, 2F), -111.99 (d, *J*=8.9, 2F), -132.44 (dd, *J*=20.8, 8.2, 2F), -163.07 (tt, *J*=20.8, 6.2, 1F).

$^{13}$C NMR (101 MHz, $CDCl_3$) δ 166.26, 165.46, 164.27, 162.64, 162.06, 161.64, 161.29, 161.24, 160.97, 160.87, 158.47, 158.40, 157.08, 134.73, 133.38, 118.31, 118.28, 114.88, 114.65, 111.73, 111.48, 110.05, 107.65, 107.58, 107.47, 107.41, 106.95, 106.92, 106.67, 106.65, 105.03, 99.63, 68.93, 55.62, 28.77, 28.13, 22.37, 14.00.

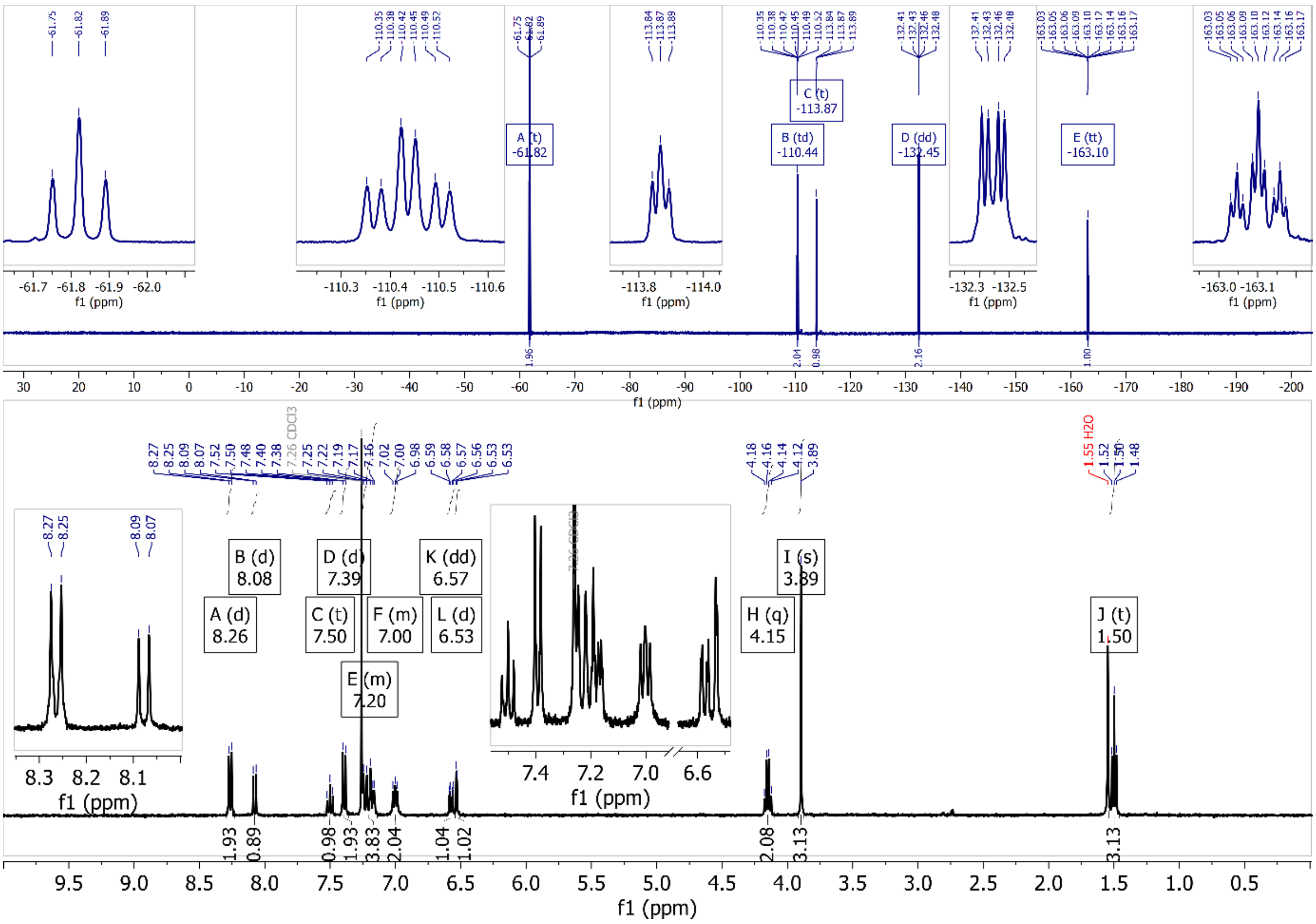

Figure S4. [19]F and [1]H NMR spectra of **E-0-1** in $CDCl_3$.

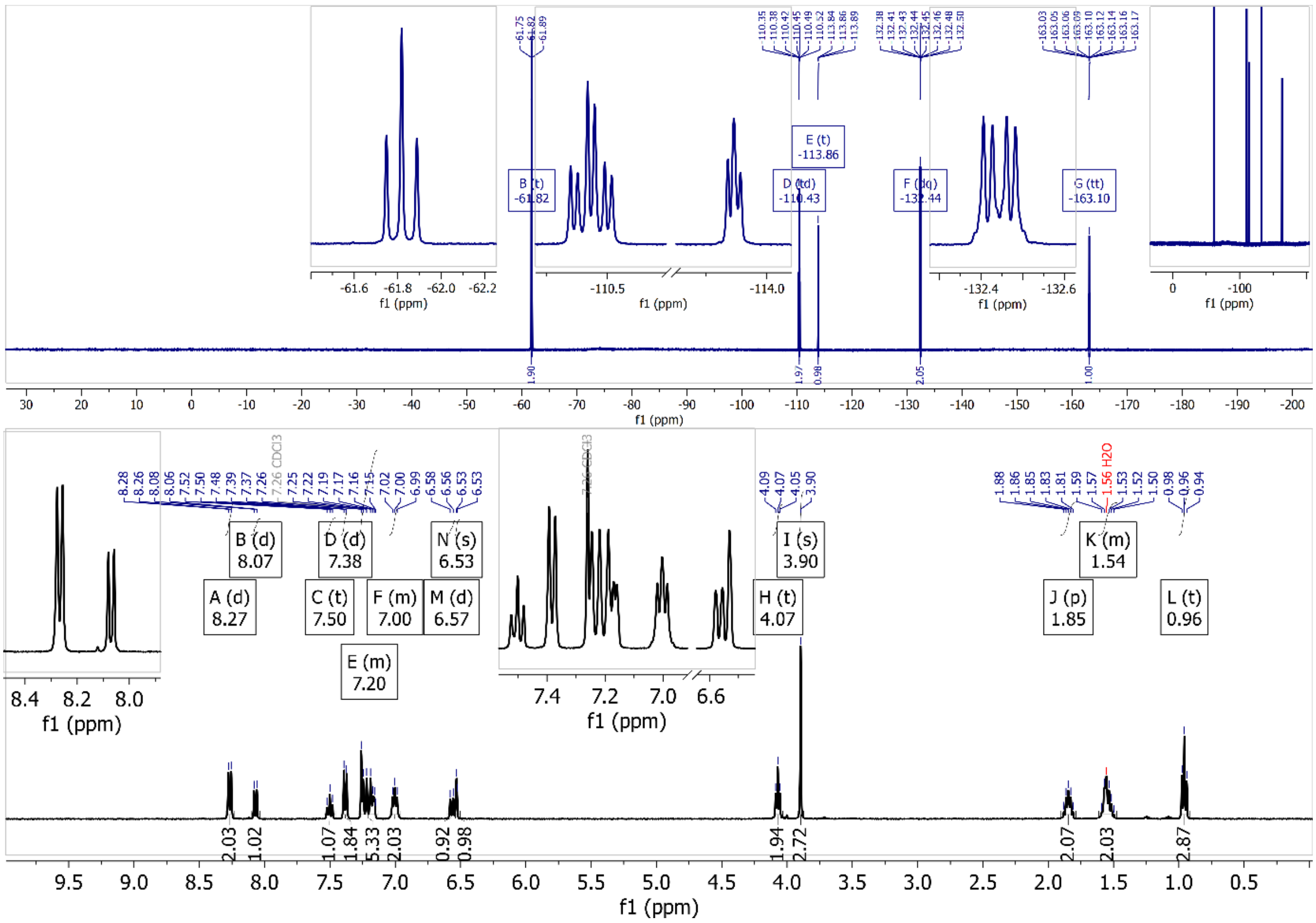

Figure S5. [19]F and [1]H NMR spectra of **P-0-1** in $CDCl_3$.

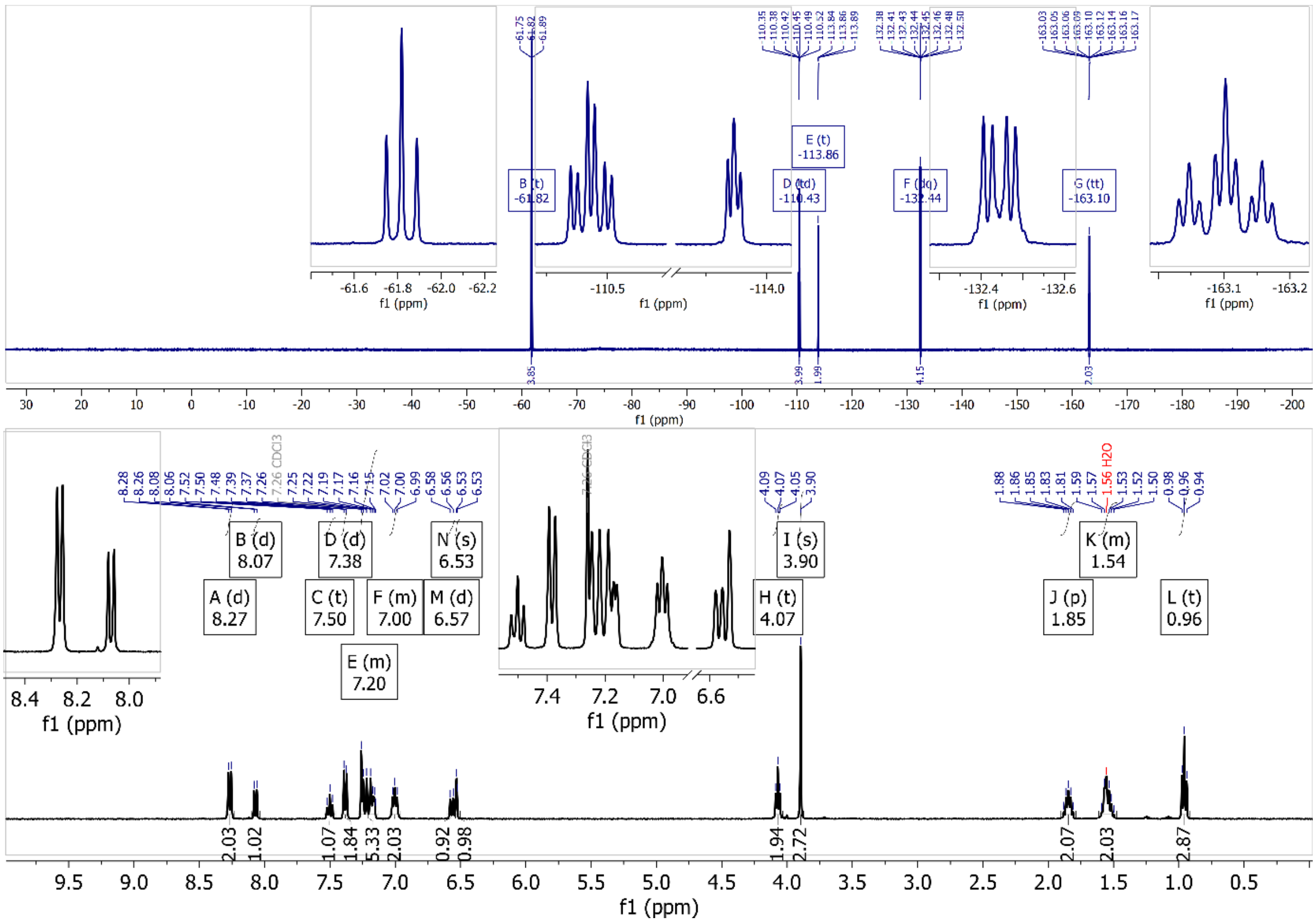

Figure S6. $^{19}$F and $^{1}$H NMR spectra of **B-0-1** in $CDCl_3$.

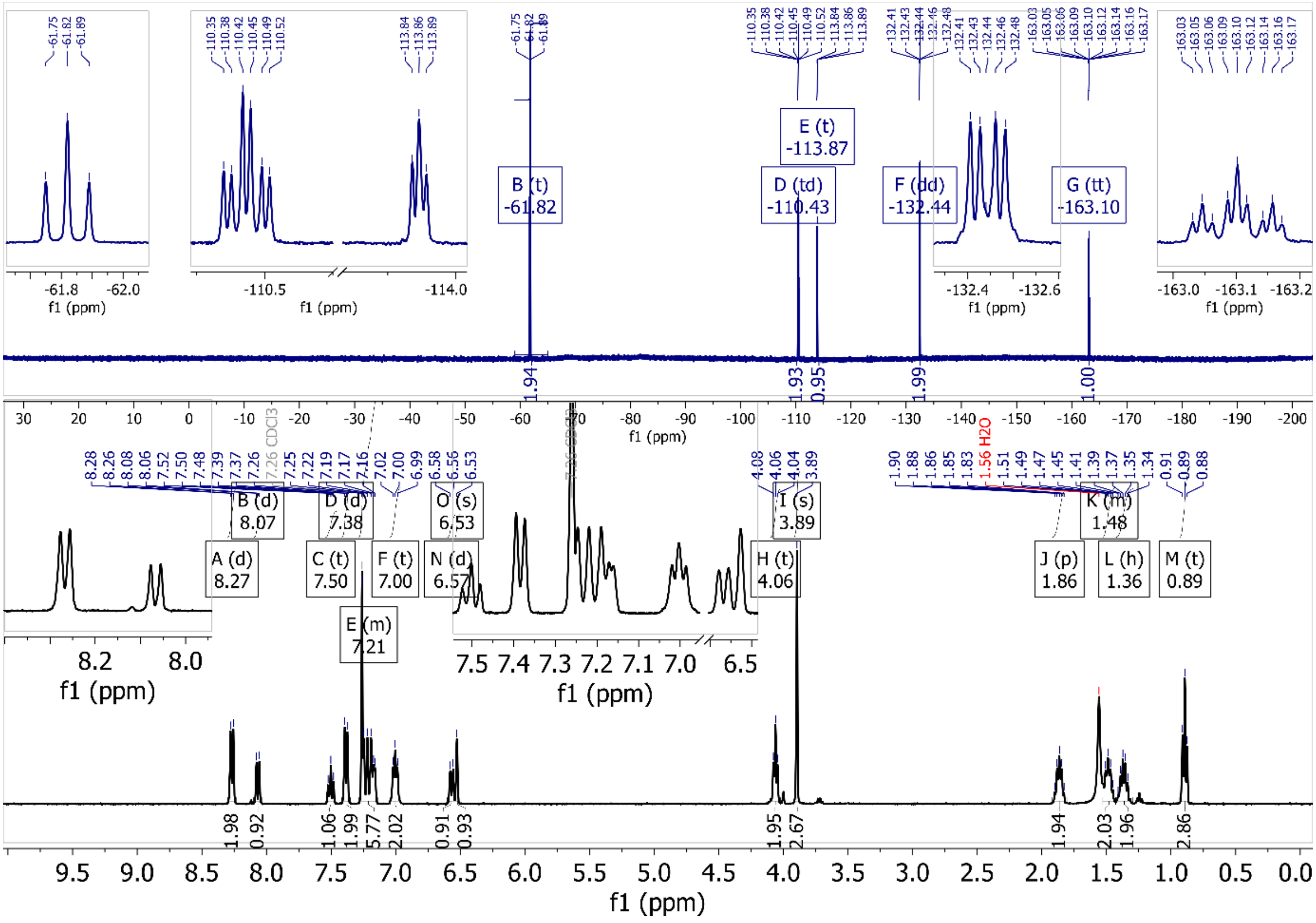

Figure S7. $^{19}$F and $^{1}$H NMR spectra of **Q-0-1** in $CDCl_3$.

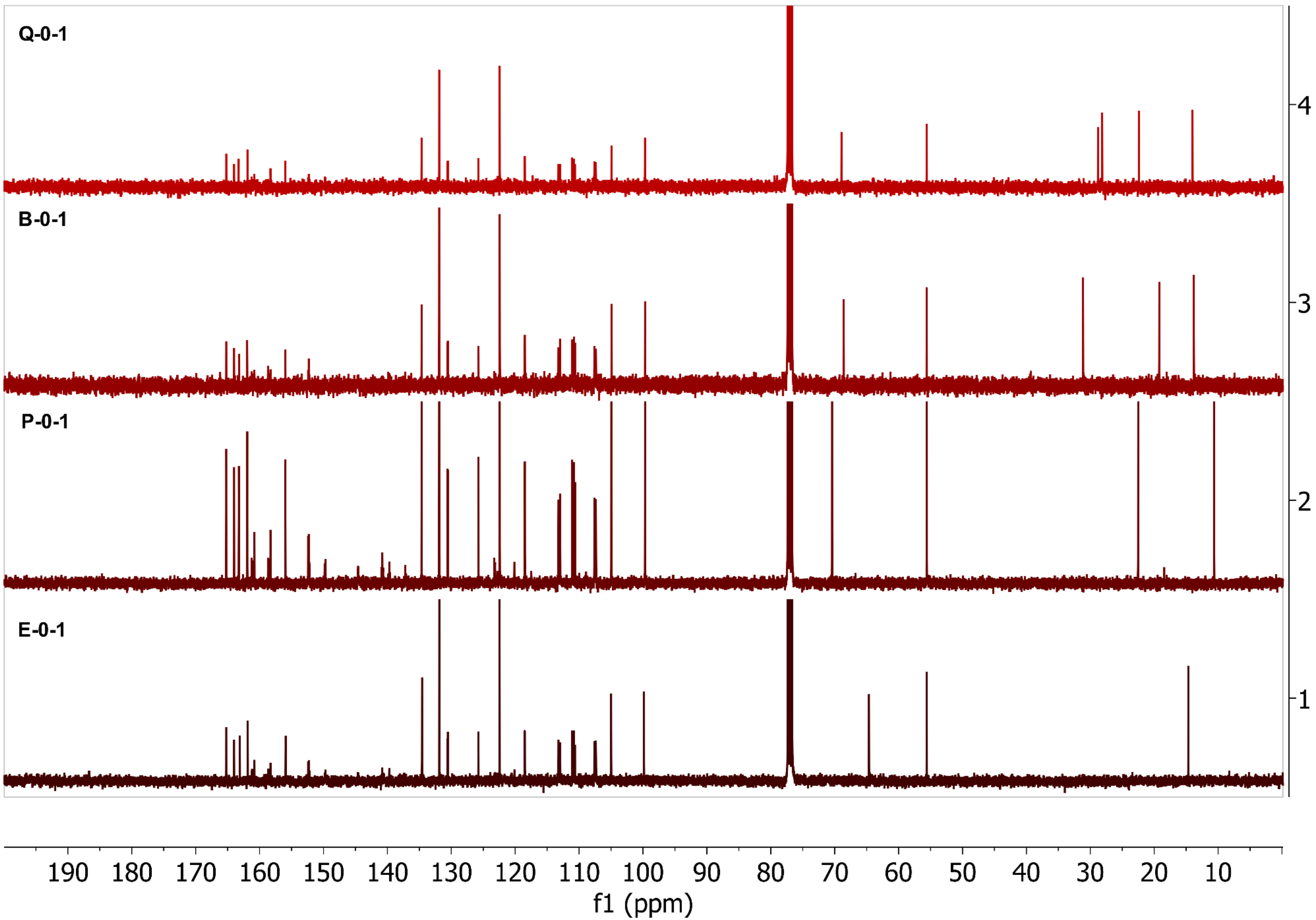

Figure S8. Stacked $^{13}C$ NMR spectra of **E-0-1**, **P-0-1**, **B-0-1** and **Q-0-1** in $CDCl_3$.

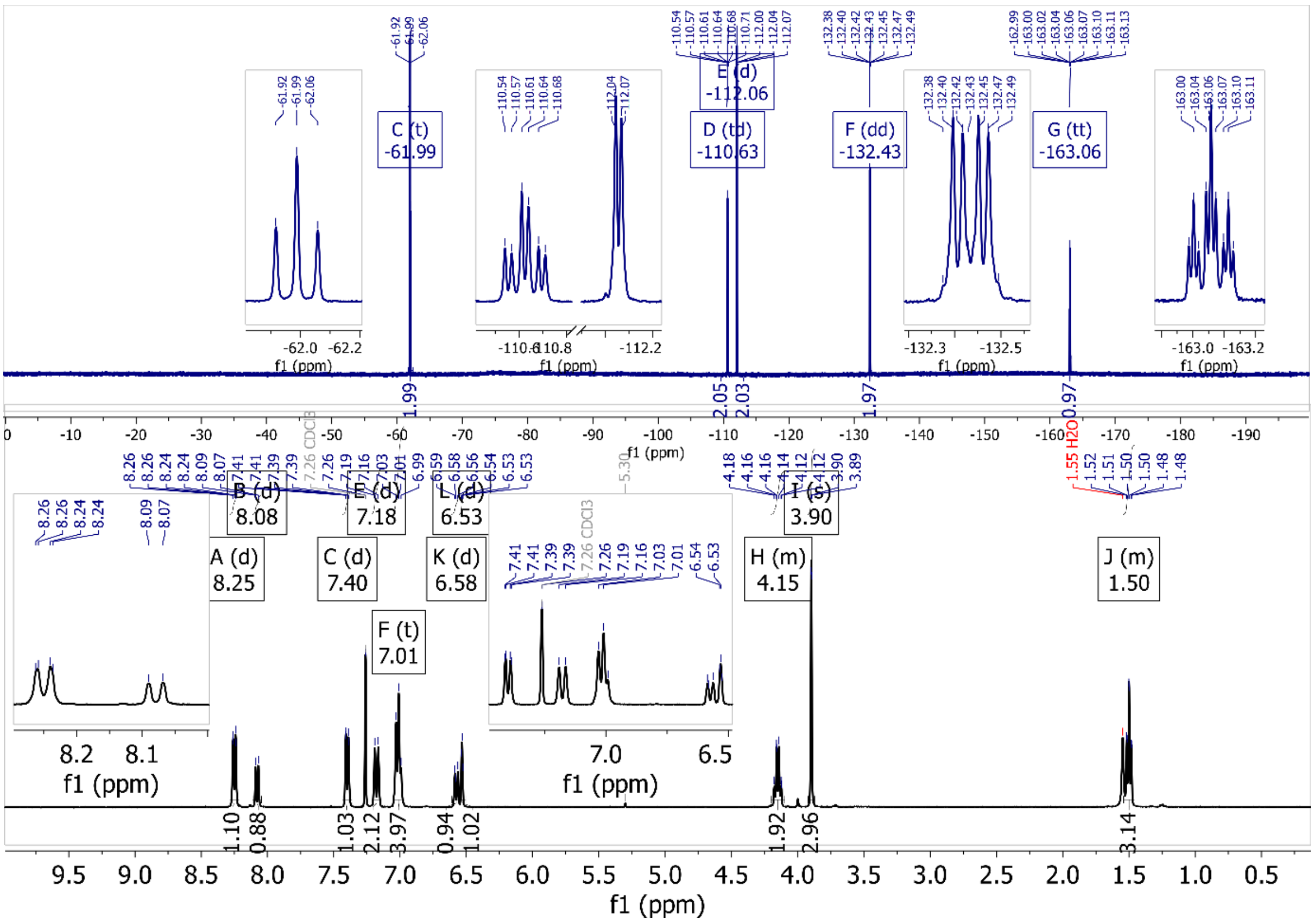

Figure S9. $^{19}$F and $^{1}$H NMR spectra of **E-0-2** in $CDCl_3$.

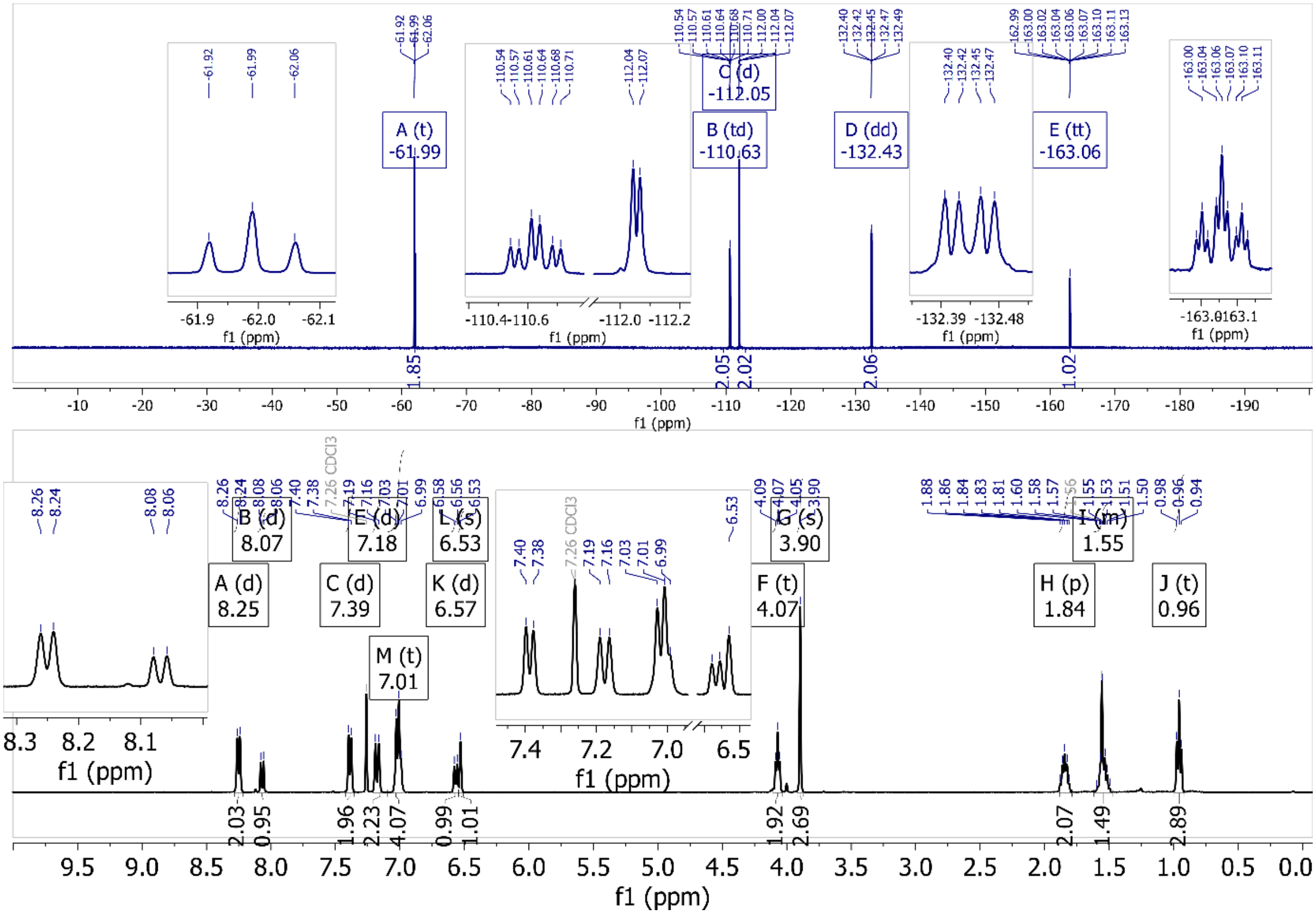

Figure S10. $^{19}$F and $^{1}$H NMR spectra of **P-0-2** in $CDCl_3$.

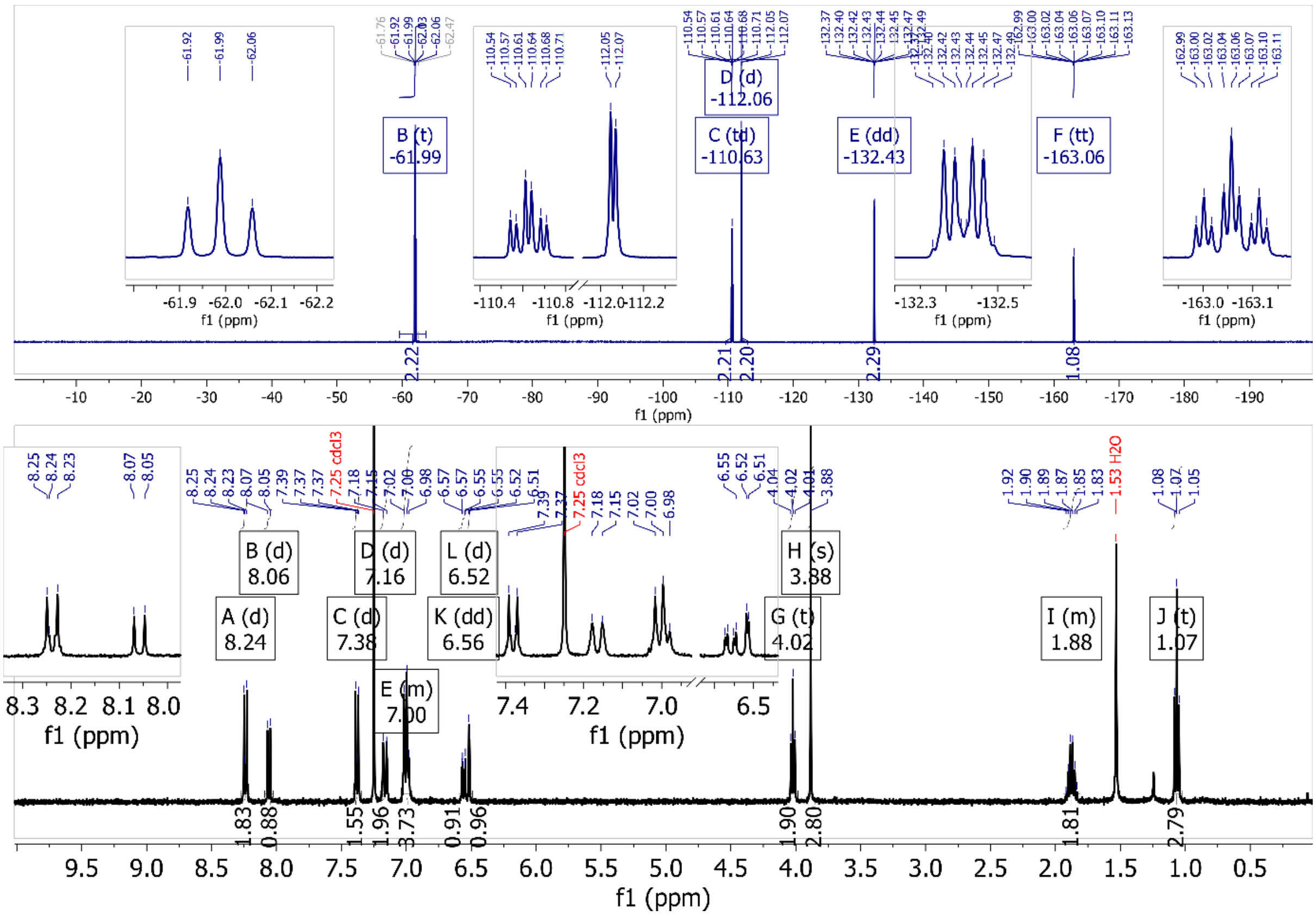

Figure S11. [19]F and [1]H NMR spectra of **P-0-2** in $CDCl_3$.

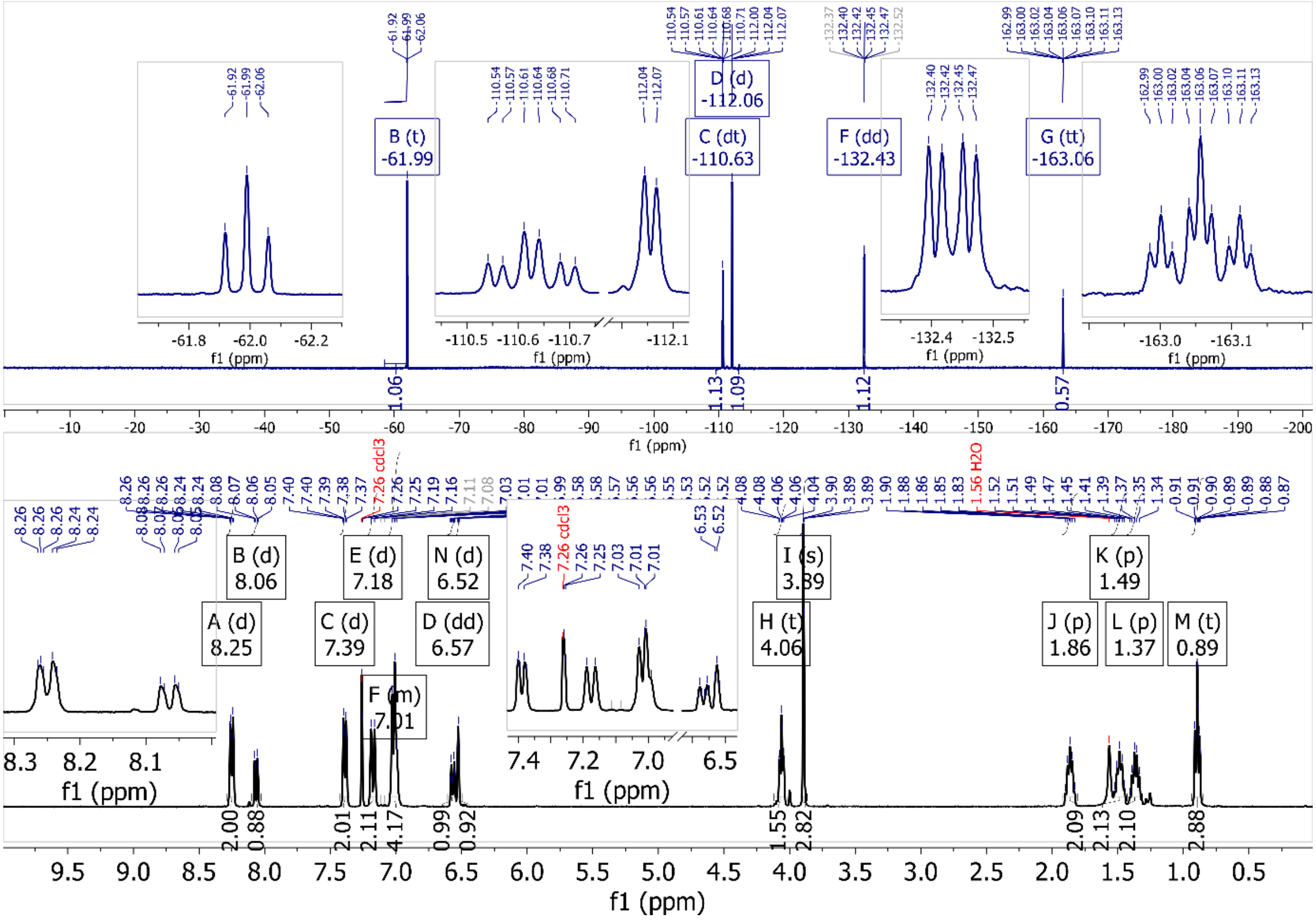

Figure S12. [19]F and [1]H NMR spectra of **Q-0-2** in $CDCl_3$.

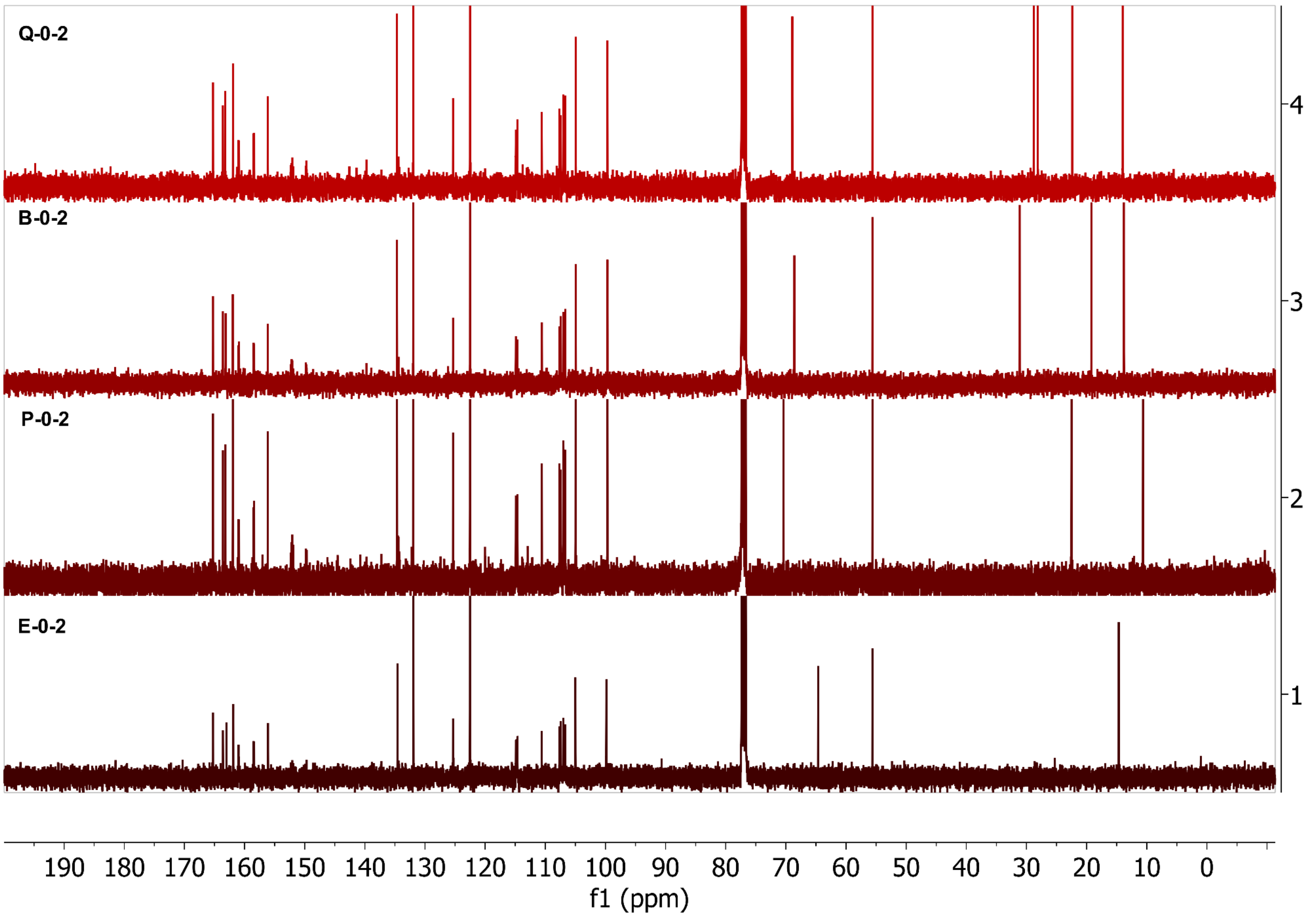

Figure S13. Stacked $^{13}C$ NMR spectra of **E-0-2**, **P-0-2**, **B-0-2** and **Q-0-2** in $CDCl_3$

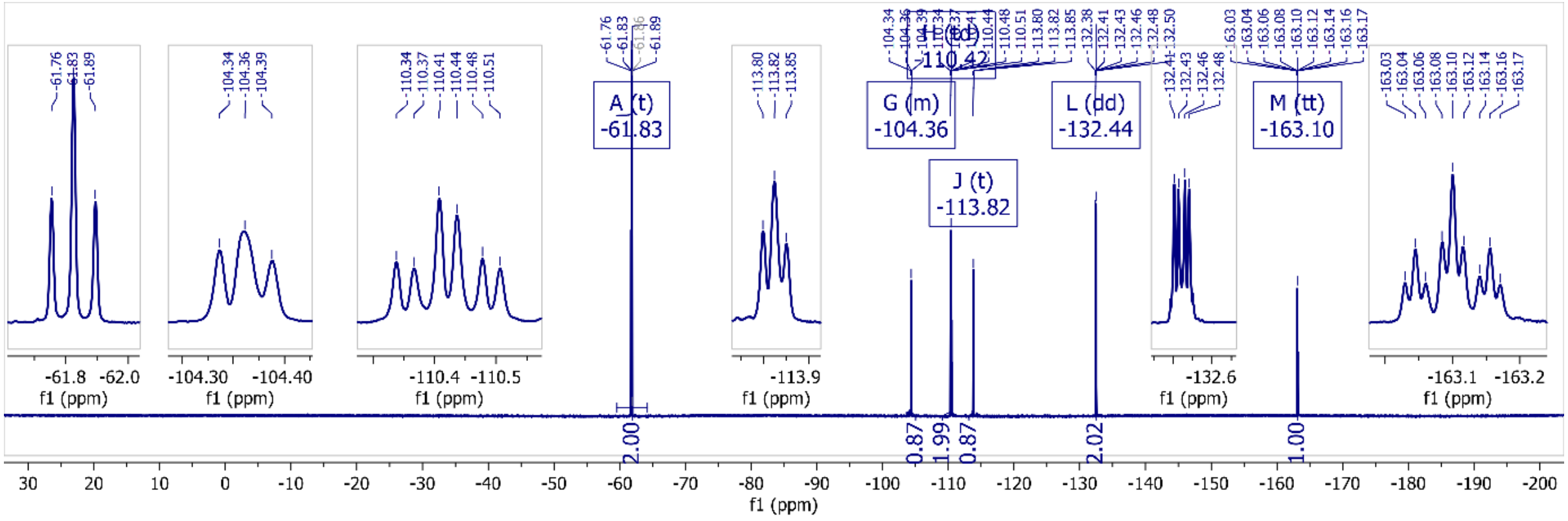

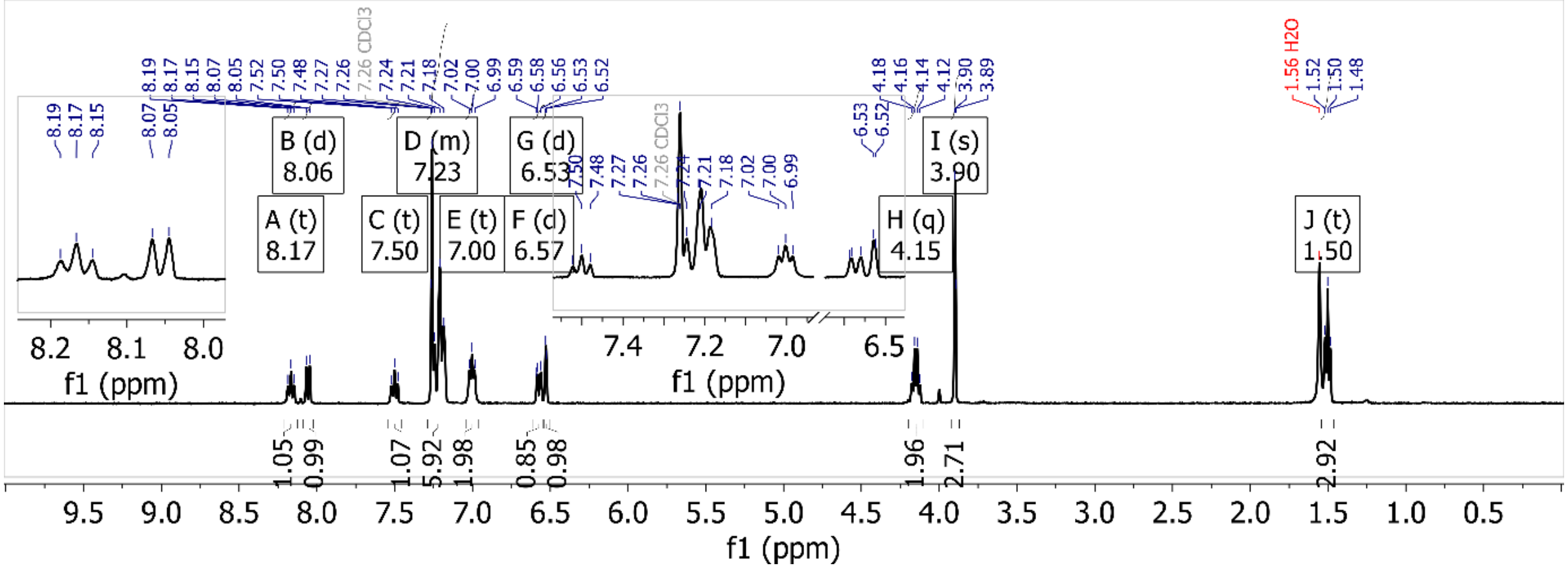

Figure S14. $^{19}F$ and $^{1}H$ NMR spectra of **E-1-1** in $CDCl_3$.

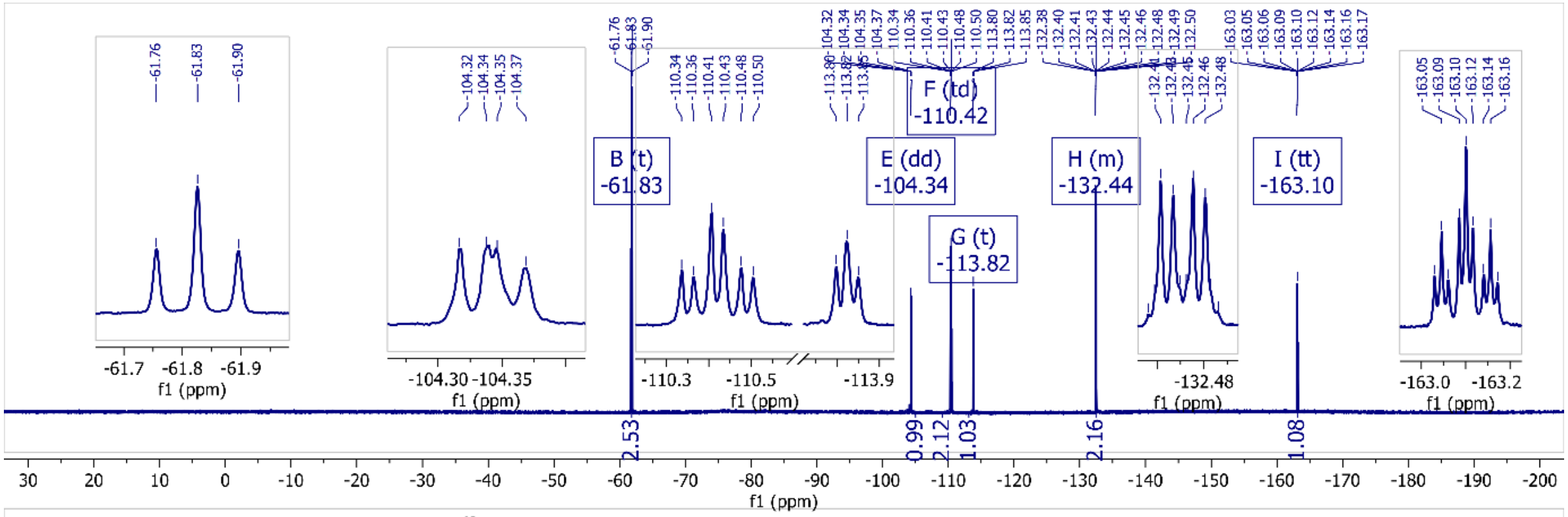

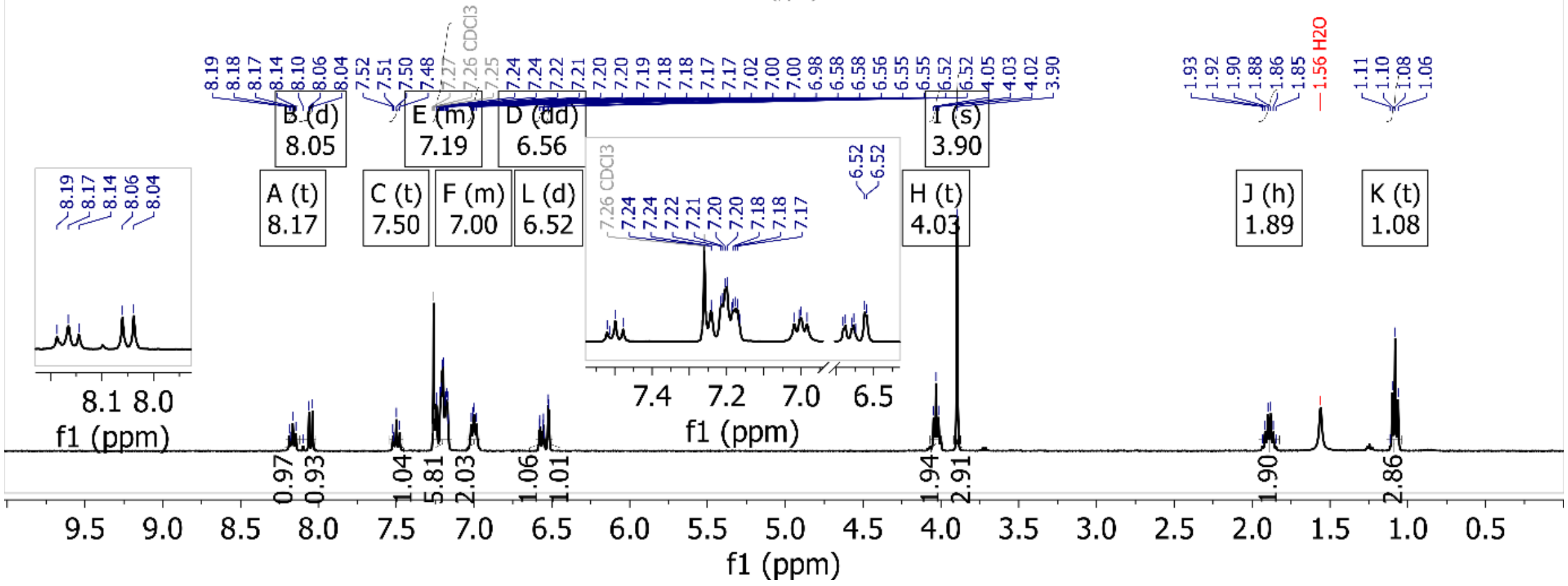

Figure S15. $^{19}F$ and $^{1}H$ NMR spectra of **P-1-1** in $CDCl_3$.

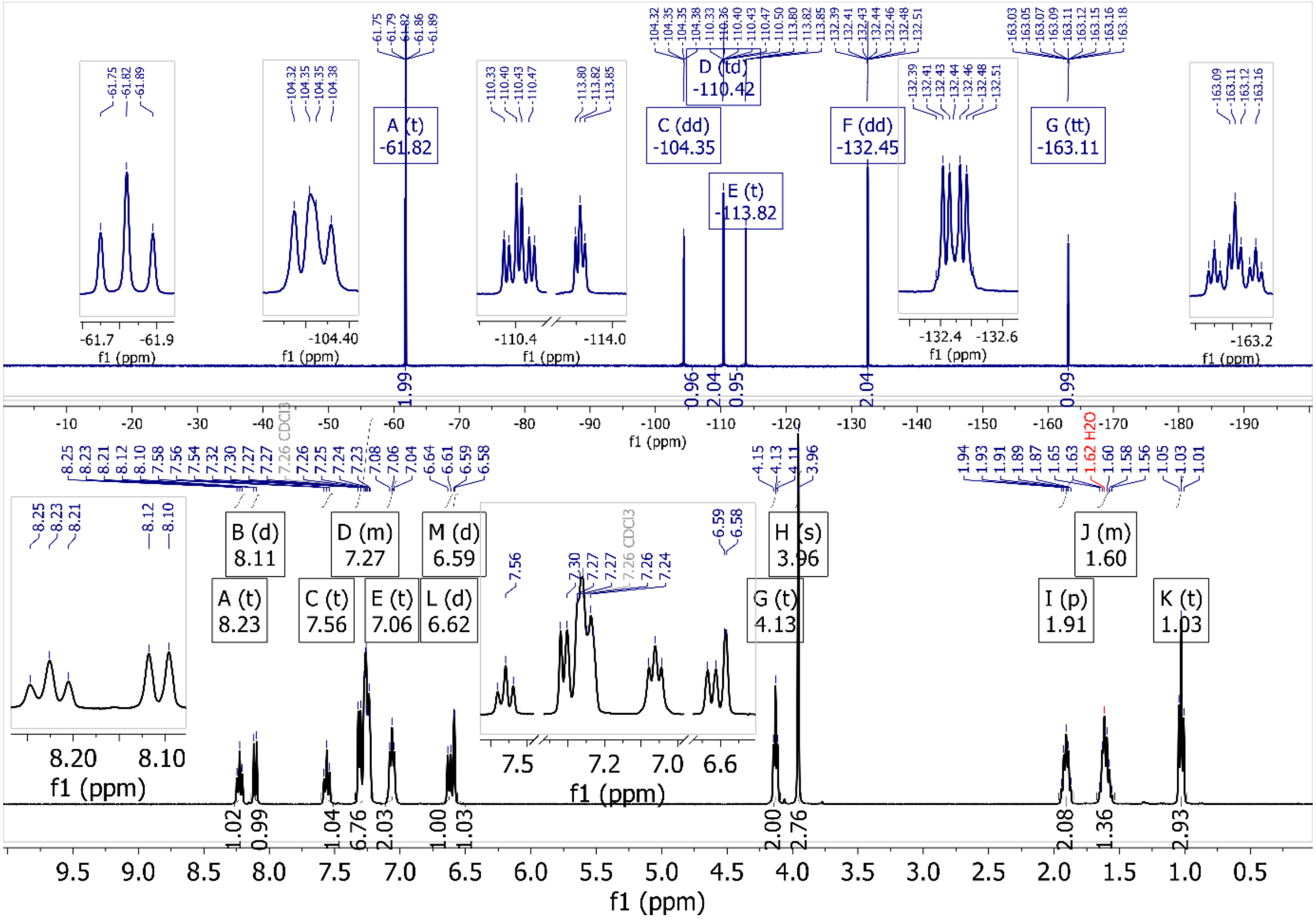

Figure S16. $^{19}$F and $^{1}$H NMR spectra of **B-1-1** in $CDCl_3$.

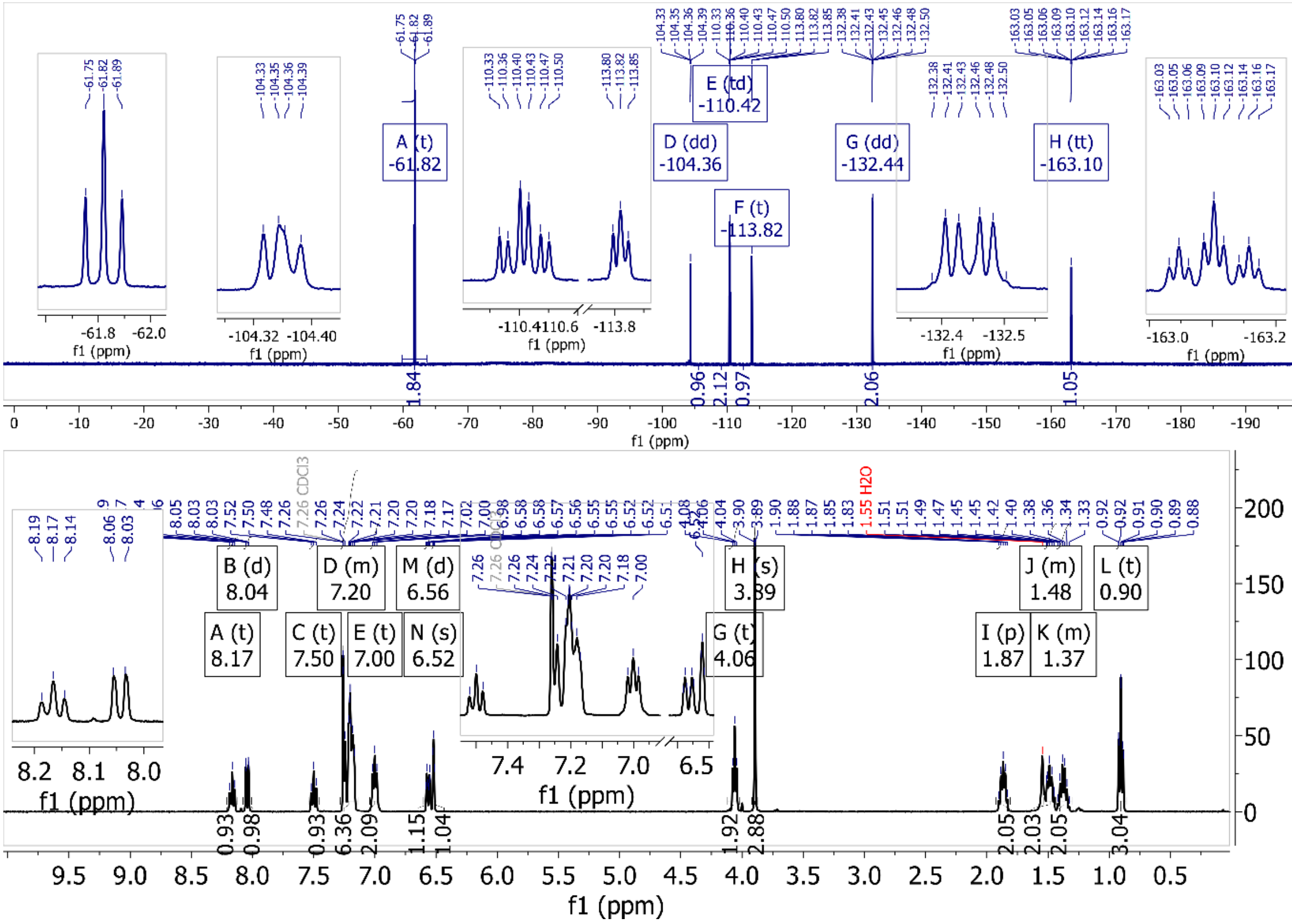

Figure S17. $^{19}$F and $^{1}$H NMR spectra of **Q-1-1** in $CDCl_3$.

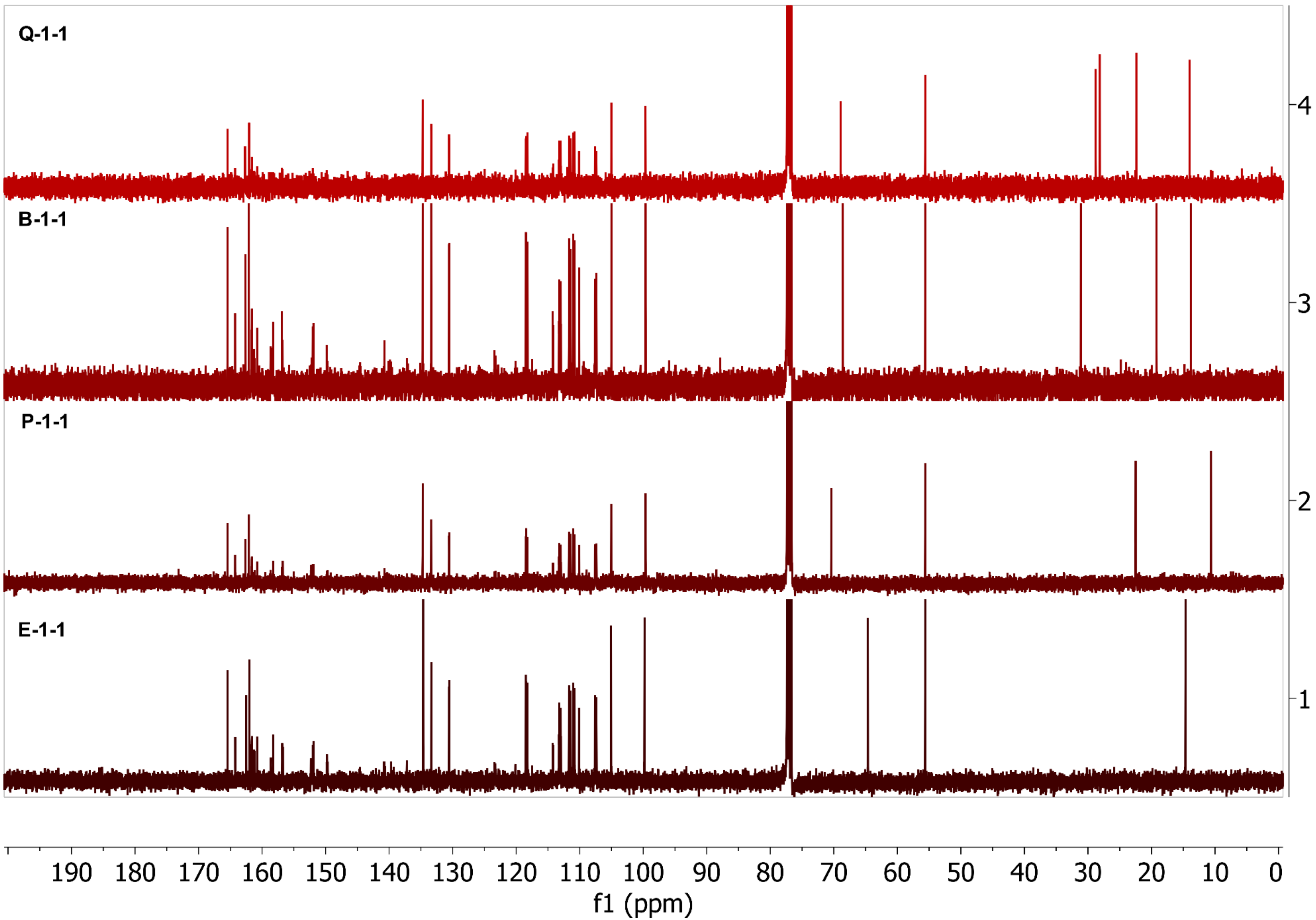

Figure S18. Stacked $^{13}$C NMR spectra of **E-1-1**, **P-1-1**, **B-1-1** and **Q-1-1** in $CDCl_3$.

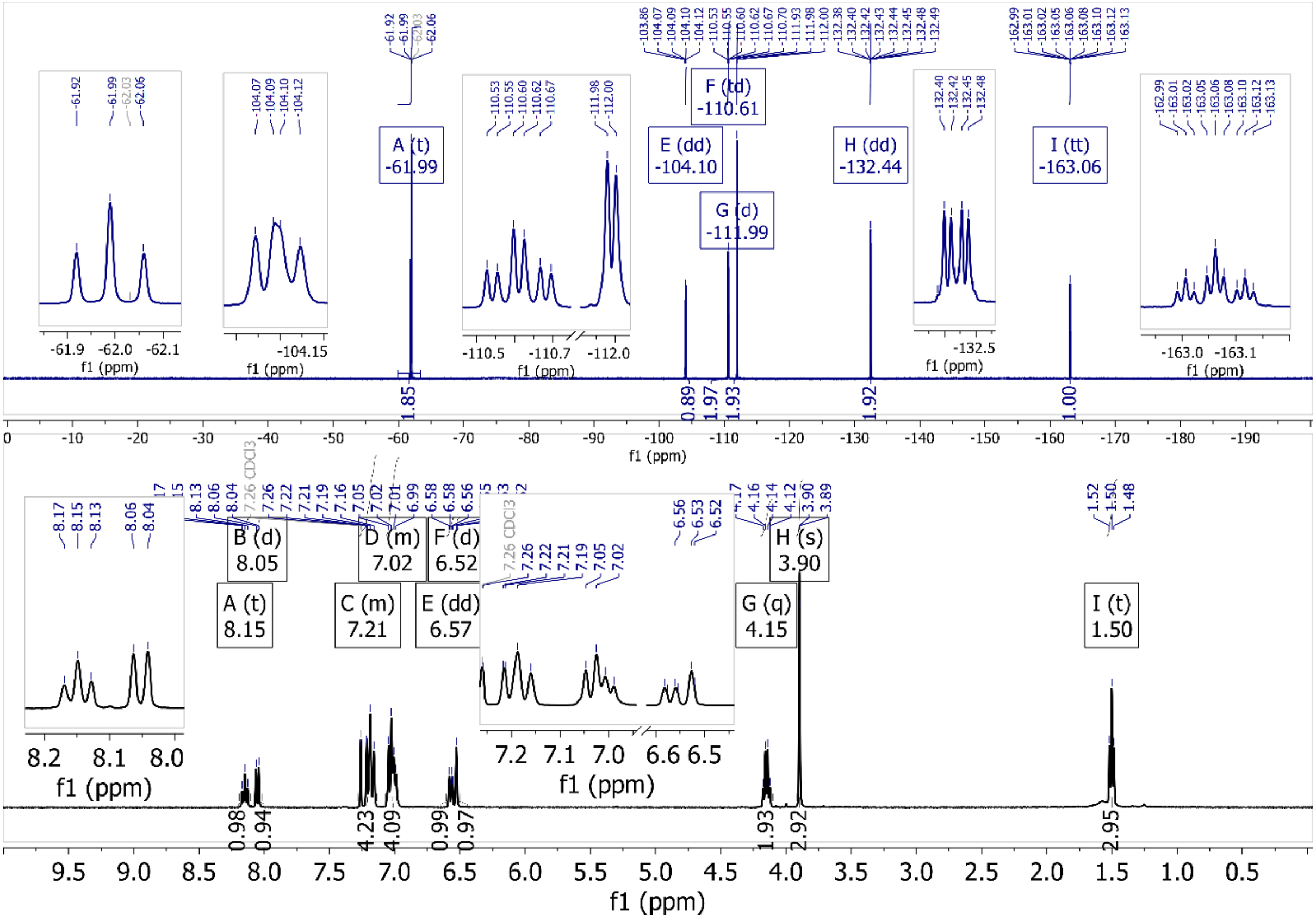

Figure S19. $^{19}$F and $^{1}$H NMR spectra of **E-1-2** in $CDCl_3$.

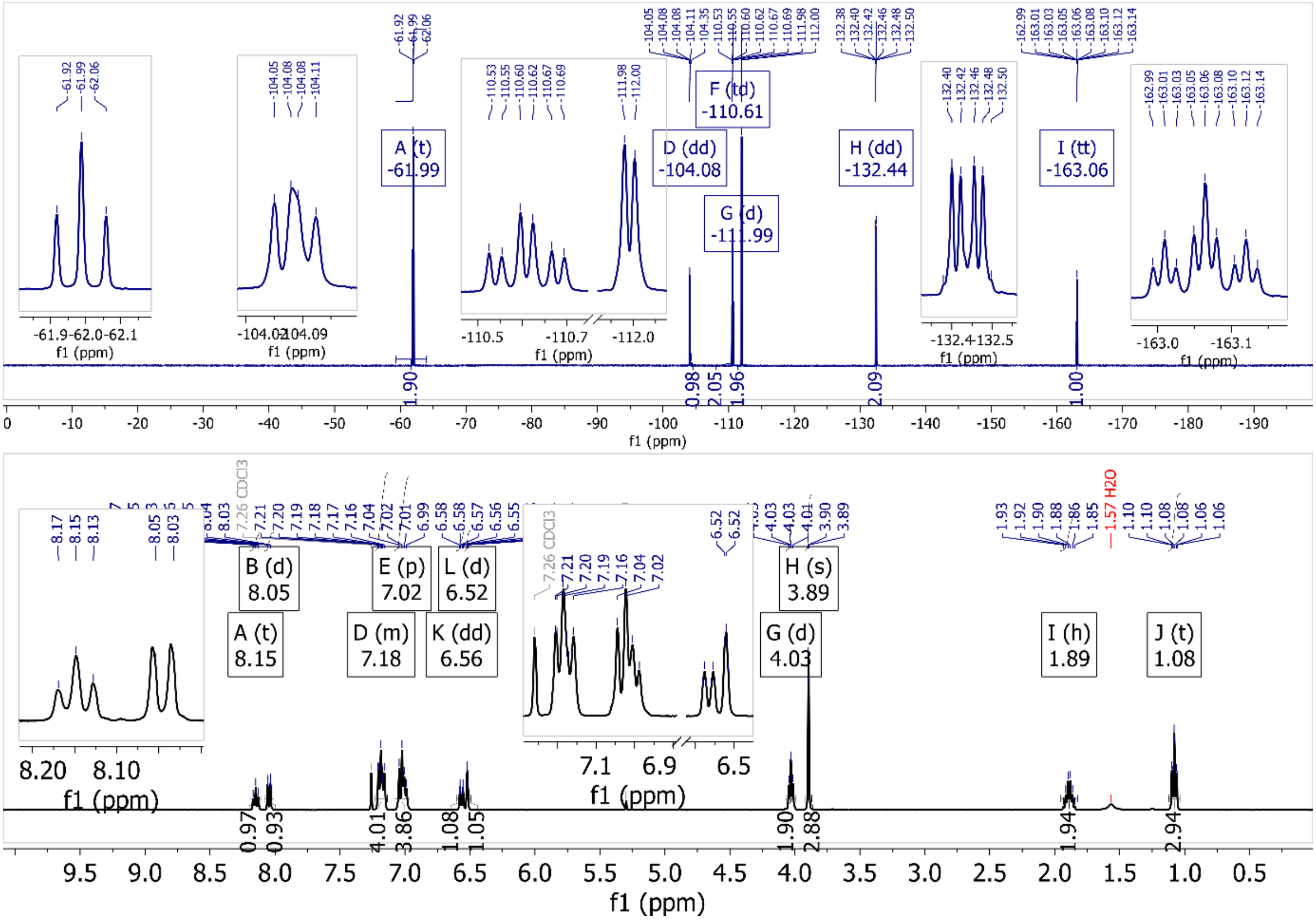

Figure S20. $^{19}$F and $^{1}$H NMR spectra of **P-1-2** in $CDCl_3$.

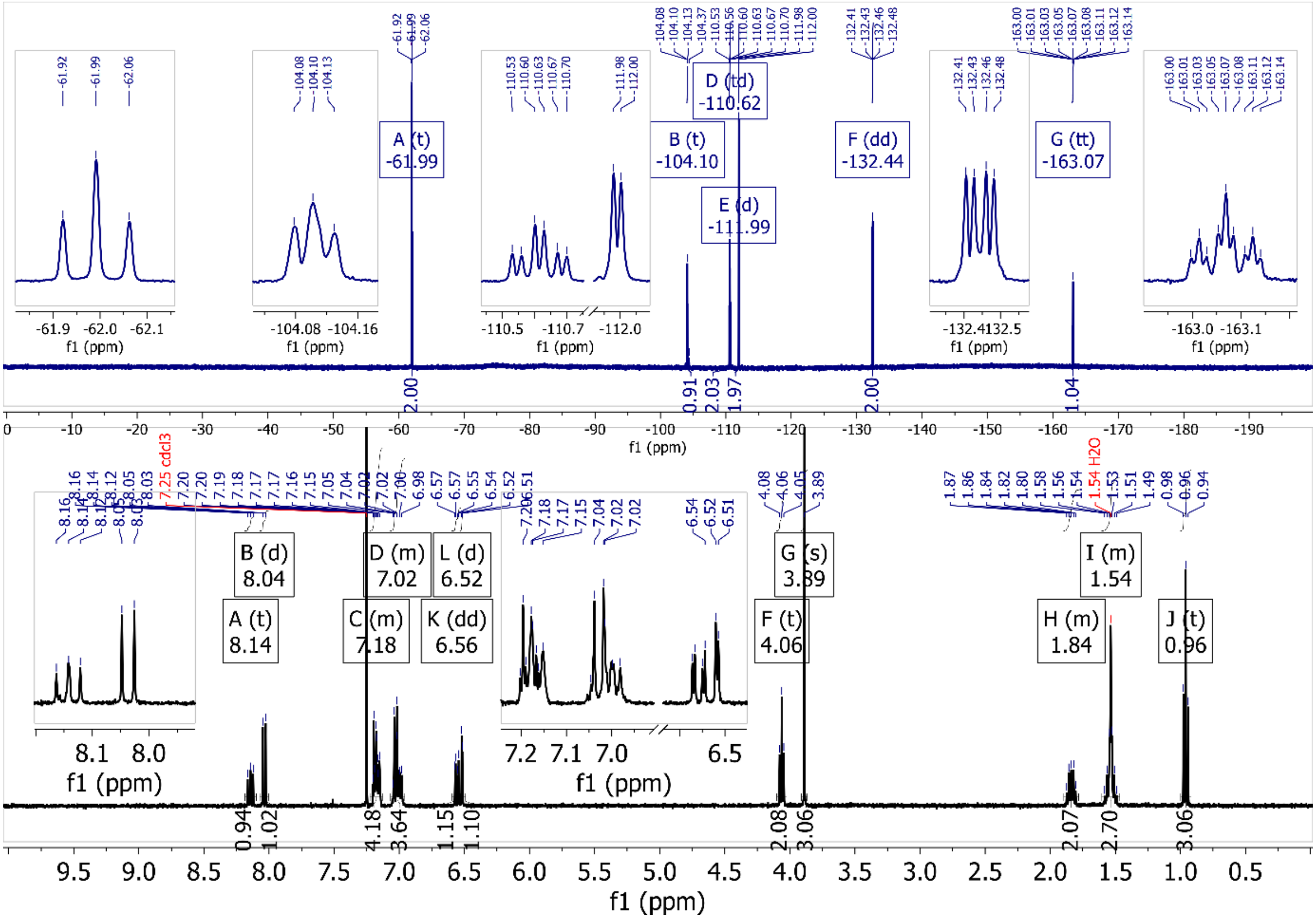

Figure S21. $^{19}$F and $^{1}$H NMR spectra of **B-1-2** in $CDCl_3$.

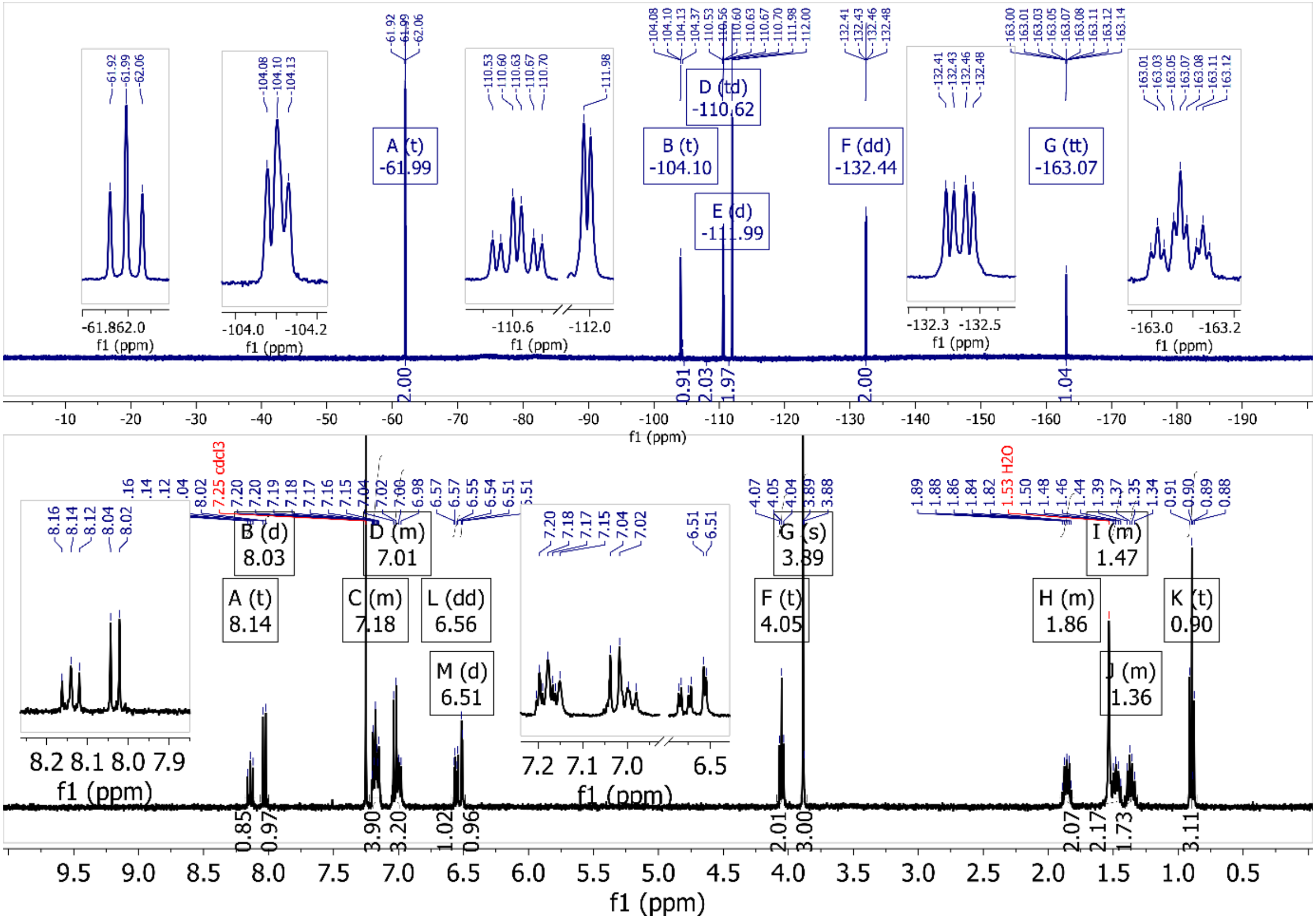

Figure S22. $^{19}$F and $^{1}$H NMR spectra of **Q-1-2** in $CDCl_3$.

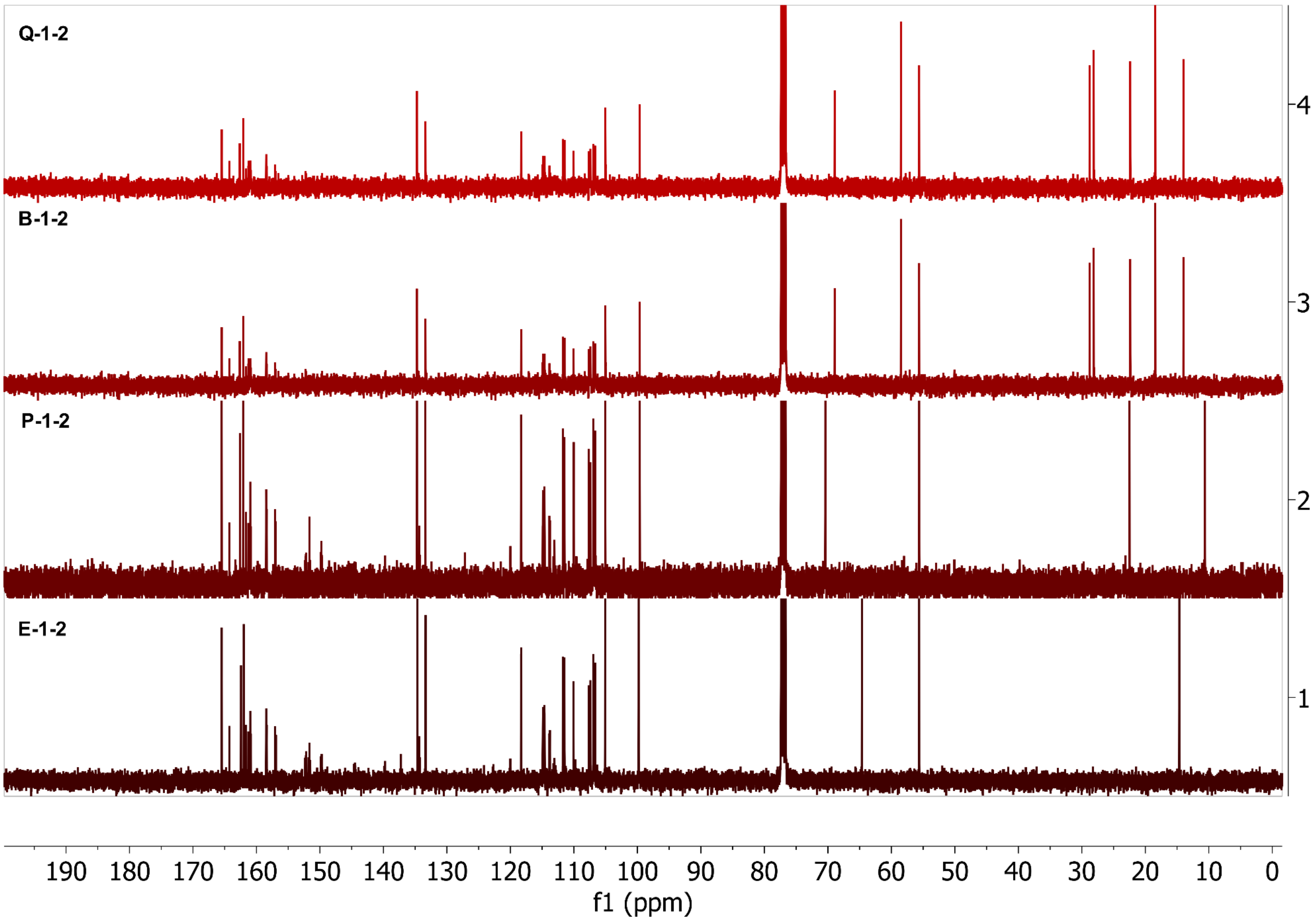

Figure S23. Stacked $^{13}C$ NMR spectra of **E-1-2**, **P-1-2**, **B-1-2** and **Q-1-2** in $CDCl_3$. **B-1-2** and **Q-1-2** contain additional peaks from ethanol at 18 and 58 ppm.

**Supplemental References**